\documentclass[
    aps,
    prx,
    superscriptaddress,
    twocolumn,
    a4paper,
    floatfix,
    longbibliography
]{revtex4-2}

\usepackage[american]{babel}

\usepackage{amsmath}
\usepackage{amssymb}
\usepackage{graphicx}
\usepackage[caption=false]{subfig}
\usepackage{color}
\usepackage[percent]{overpic}
\usepackage{bm}
\usepackage{rotating}

\usepackage{hyperref}

\usepackage{pbox}
\usepackage{array}
\usepackage{physics}
\usepackage{mathtools}
\usepackage{enumerate}
\usepackage{leftidx}
\usepackage{comment}
\usepackage{booktabs}

\usepackage{xcolor}

\usepackage[normalem]{ulem}

\usepackage{float}

\usepackage{cleveref}

\DeclareFontFamily{U}{BOONDOX-calo}{\skewchar\font=45 }
\DeclareFontShape{U}{BOONDOX-calo}{m}{n}{
  <-> s*[1.05] BOONDOX-r-calo}{}
\DeclareFontShape{U}{BOONDOX-calo}{b}{n}{
  <-> s*[1.05] BOONDOX-b-calo}{}
\DeclareMathAlphabet{\mathcalboondox}{U}{BOONDOX-calo}{m}{n}
\SetMathAlphabet{\mathcalboondox}{bold}{U}{BOONDOX-calo}{b}{n}
\DeclareMathAlphabet{\mathbcalboondox}{U}{BOONDOX-calo}{b}{n}

\def\approxprop{%
  \def\p{%
    \setbox0=\vbox{\hbox{$\propto$}}%
    \ht0=0.6ex \box0 }%
  \def\s{%
    \vbox{\hbox{$\sim$}}%
  }%
  \mathrel{\raisebox{0.7ex}{%
      \mbox{$\underset{\s}{\p}$}%
    }}%
}

\newcommand{\ii}{\mathrm{i}}

\usepackage{suffix}

\usepackage[export]{adjustbox}% http://ctan.org/pkg/adjustbox

\DeclareMathOperator{\e}{\mathrm{e}}

\DeclareMathOperator{\E}{\textbf{E}}
\DeclareMathOperator{\B}{\textbf{B}}

\DeclareMathOperator{\rr}{\textbf{\textit{r}}}

\DeclareMathOperator{\kk}{\textbf{\textit{k}}}

\DeclareMathOperator{\p}{\textbf{p}}

\DeclareMathOperator{\aop}{\hat{\textit{a}}}

\newcommand{\re}{\bm r_\mathrm{e}}

\newcommand{\ye}{\bm y_\mathrm{e}}
\newcommand{\ze}{z_\mathrm{e}}
\newcommand{\erone}{\partial_{(y_1)} }
\newcommand{\ertwo}{\partial_{(y_2)} }

\newcommand{\Smear}{\bm F_{\bm m,\bm s \bm s'}}

\newcommand\inner[2]{\left\langle #1, #2 \right\rangle}

\DeclareFontFamily{OMS}{oasy}{\skewchar\font48 }
\DeclareFontShape{OMS}{oasy}{m}{n}{%
         <-5.5> oasy5     <5.5-6.5> oasy6
      <6.5-7.5> oasy7     <7.5-8.5> oasy8
      <8.5-9.5> oasy9     <9.5->  oasy10
      }{}
\DeclareFontShape{OMS}{oasy}{b}{n}{%
       <-6> oabsy5
      <6-8> oabsy7
      <8->  oabsy10
      }{}
\DeclareSymbolFont{oasy}{OMS}{oasy}{m}{n}
\SetSymbolFont{oasy}{bold}{OMS}{oasy}{b}{n}

\DeclareMathSymbol{\smallleftarrow}     {\mathrel}{oasy}{"20}
\DeclareMathSymbol{\smallrightarrow}    {\mathrel}{oasy}{"21}
\DeclareMathSymbol{\smallleftrightarrow}{\mathrel}{oasy}{"24}

\DeclareMathAlphabet\mathbfcal{OMS}{cmsy}{b}{n}

\begin{document}
	\title{Dimensional Reduction in Quantum Optics \\ \normalfont{ \small{(This article has been published in \href{https://doi.org/10.1103/PhysRevResearch.6.013285}{Phys. Rev. Res. \textbf{6}, 013285 (2024)})}}}

	\author{Jannik Str\"ohle}
 \email{jannik.stroehle@uni-ulm.de}
	\affiliation{Institut f{\"u}r Quantenphysik and Center for Integrated Quantum
    Science and Technology (IQST), Universit{\"a}t Ulm, Albert-Einstein-Allee 11, D-89069 Ulm, Germany}

%\author{others}
%	\affiliation{Institut f{\"u}r Quantenphysik and Center for Integrated Quantum
%    Science and Technology (IQST), Universit{\"a}t Ulm, Albert-Einstein-Allee 11, D-89069 Ulm, Germany}

    \author{Richard Lopp}
    \email{richard.lopp@uni-ulm.de}
	\affiliation{Institut f{\"u}r Quantenphysik and Center for Integrated Quantum
    Science and Technology (IQST), Universit{\"a}t Ulm, Albert-Einstein-Allee 11, D-89069 Ulm, Germany}

\begin{abstract}
One-dimensional quantum optical models usually rest on the intuition of large-scale separation or frozen dynamics associated with the different spatial dimensions, for example when studying quasi one-dimensional atomic dynamics, potentially resulting in the violation of ($3+1$)-dimensional Maxwell's theory. 
Here, we provide a rigorous foundation for this approximation by means of the light-matter interaction. We show how the quantized electromagnetic field can be decomposed exactly into an infinite number of \textit{subfields} living on a lower-dimensional subspace and containing the entirety of the spectrum when studying axially symmetric setups, such as with an optical fiber, a laser beam, or a waveguide. The \textit{dimensional reduction} approximation then corresponds to a truncation in the number of such subfields that in turn, when considering the interaction with for instance an atom, corresponds to a modification to the atomic spatial profile. We explore under what conditions the standard approach is justified and when corrections are necessary in order to account for the dynamics due to the neglected spatial dimensions. In particular we examine what role vacuum fluctuations and structured laser modes play in the validity of the approximation.
\end{abstract}
	
\maketitle

\section{Introduction}
\label{Intro}

The success of theoretical quantum optics describing the tremendous experimental progress in the past decades is largely due to effective models which provide very accurate statements to physical problems via  simple models that are valid in limited parameter regimes only~\cite{Rokaj,Blaha,NCS1,NCS2,Lopp3,McKay,Fleming,Blaha,Meiser,Krimer,Lechner}. 
For instance, lower-dimensional effective models are ubiquitously used to describe the dynamics of Bose-Einstein condensates (BECs)~\cite{Pit,Salas,Gorlitz,Blackie2,Edmons,Edler,Tononi}, semiconductor devices~\cite{Painter,Zhou}, quantum dots~\cite{Johnson,Reimann,Shapt,Ullmo} or quantum state engineering~\cite{SchleichBabula, Akulin}.   
Care needs to be applied, however, when quantum models are based on different spatial dimensions as significantly different predictions may occur; this is evident for instance in  thermalization~\cite{Ooguri, Arrechea}, communication~\cite{Jonsson, Blasco} or scattering processes~\cite{Salas,Gorlitz,PBOEGEL}.

Cavities, in particular, are of special interest as  effects become experimentally relevant which are highly restricted in free space, like power enhancement, spatial filtering and more accurate beam profiles~\cite{BOOZER,Heinzen1,Heinzen2,Bochove}. Based hereon is the concept of lasers and masers~\cite{Renk,Svelto,Kogelnik,Scully4} and cavity matter-wave interferometry~\cite{CAVAI1,CAVAI2,CAVAI3,Greve}, but also photonic BECs~\cite{PBEC,Kirton} and quantum dots~\cite{Childress,Najer,Hennessy,Hohenester} benefit from the effects offered by cavity quantum electrodynamics (QED). All together, this reveals cavity QED as a powerful area of research in quantum optics~\cite{Walther,Mabuchi,Haroche}.

Nonetheless, the mode structure of the cavity, which depends on the cavity geometry, can be almost arbitrarily complex.
To this end there is despite exact three dimensional (3D) models containing the full spectrum~\cite{Lopp3,Friedrich,KAKAZUSE,KAKAZU} usually an underlying, simpler low-dimensional model which, however, is nothing but a physically motivated educated guess.
In this context, lower-dimensional models  are typically prescribed  \textit{ad hoc} in a wide range of applications. For instance one might consider an atom in a cavity of \mbox{length $L$} interacting dipolarly with the cavity field via a \mbox{coupling $g$} ~\cite{Scully,Childress}:
\begin{align}
\hat{H}^{\text{di}} =\! \hbar\! \sum_l \!\left[\omega^{\vphantom{\dagger}}_l \hat{a}^\dagger_l \hat{a}^{\vphantom{\dagger}}_l + g (\hat\sigma^\dagger-\hat\sigma)(\bar{u}^{\vphantom{\dagger}}_l(x)\hat{a}_l^\dagger - u^{\vphantom{\dagger}}_l(x)\hat{a}^{\vphantom{\dagger}}_l)\!\right],
\end{align}   
where $\omega_l= c |k_l|$ are the 1D cavity field's frequencies with $\hat{a}^{(\dagger)}_l$ being the creation and annihilation operators, and $\hat\sigma^{(\dagger)}$ being the atomic ladder operators. The modes $u_l$ of the field are usually scalar quantities, respectively evaluated at the atomic 1D position $x$, e.g., \mbox{$u(x) \sim \cos( k_l x)$}.
Subsequent approximations lead to the 1D versions of the Jaynes-Cummings model~\cite{Childress,Deb,VANENK,VANENK2,Larson,Blais,Zhu}, or for an ensemble of atoms to the Dicke model~\cite{Albrecht,Paulisch,GARRAWAY,Garcia} and the Tavis-Cummings model~\cite{TC3}.  
One might also add a semiclassical pumping that drives the cavity with strength $\eta$ or the atom with the Rabi frequency $\Omega$ via~\cite{Scully,Steck,Deb}
\begin{align}
   \hat H^{\text {pump}} =\ii \hbar\eta\left(\hat a_l^{\dagger}-\hat a^{\vphantom{\dagger}}_l\right)+\ii \hbar \Omega \cos( k_l x)\left(\hat \sigma^{\dagger}-\hat \sigma\right) .
\end{align}
Furthermore effective dispersive interactions in the large-detuning limit can be obtained, e.g.,~\cite{Zhang2,Szirmai,Brenecke}
\begin{align}
 \hat{H}^{\text{disp}}= \hbar U_0|u^{\vphantom{\dagger}}_l(x)|^2 \hat a_l^{\dagger} \hat a^{\vphantom{\dagger}}_l.
\end{align}
The common feature of these (and many more) Hamiltonians is that they arise by considering electromagnetic fields as scalars living in 1D, i.e.,~\cite{Deutsch2,BLOW,Southall}
\begin{align}
    \hat E(x)=\ii \sum_{l}\sqrt{\frac{\hbar \omega_{l} }{2 \varepsilon_0 \mathcal{A}_l}} (u_l(x) \hat a_l-\bar{u}_l(x)\hat a^\dagger_l),
    \label{1DField}
\end{align}
with the (effective) cross sectional area $\mathcal{A}_l$ being the only remnant of the higher dimensions.  Besides the lack of  two polarizations, Maxwell's equations do force three spatial dimensions~\cite{Ehrenfest}. Various approaches have been made to reduce the dimensions of electromagnetism when starting from 3D, both at the classical level, e.g., Refs.~\cite{Maggi, Schippa, Angelone,Hammer,Benson}, as well as for quantized fields, see e.g., Refs.~\cite{Chang,Southall,Edery}; often employing the method of \textit{Hadamard's decent}. That is, lower-dimensional models are obtained by assuming that the model is constant in certain degrees of freedom~\cite{Hadamard}.

Here, however, we do not rely on \textit{descent} conditions nor impose any restrictions on our original model in order to reduce the dimensions (apart from a symmetry consideration for analytical purposes).
Instead we show, starting from the Helmholtz equation, how to rigorously realize a lower-dimensional model from 3D cavity QED. To that end, we consider axially symmetric geometries (such as optical fibers or Fabry-Pérot cavities) which manifest in a separation of the cavity modes. By projecting an appropriate ancilla basis onto the 3D modes, the resulting reduced modes only depend on the complementary spatial coordinates and are solutions of a reduced Helmholtz equation. This extends Ref.~\cite{Lopp1}, where this problem has been studied with a scalar version of the light-matter interaction, by actively accounting for the vector nature of electromagnetism and the nontrivial coupling of the different components due to the polarization induced by Maxwell's equations. 
Naturally, as we show, the common light-matter interactions  as well as setups with laser beams  can be incorporated. 

In the process, the electromagnetic fields decompose into an infinite collection of vector-valued \textit{subfields} which live on the remaining dimensions but encode geometrical information of the original model. 
This allows us to answer the question under which circumstances a 3D cavity can be treated as a, e.g., 1D problem which is usually justified by having some length scales of the cavity or the matter system much larger than the remaining ones, or the dynamics is assumed to be frozen in some dimensions. 
Here, a dimensionally reduced simple model can be achieved via a single- or few-mode approximation on those subfields. Due to corrections that arise from the 3D model, it is not equivalent to the usual way of prescribing this approximation \textit{ad hoc}. As we will show, this is also not generally the case in the common regime of having a very long but narrow fiber. Finally, we highlight the difference between the role of vacuum fluctuations and strongly excited modes, such as for a laser, in order to reconstruct the full 3D dynamics.

This paper is organized as follows: In Sec.~\ref{DIMREDELFIELDS}
we establish the formalism for the dimensional reduction of ideal cavities. In particular we will discuss how the 3D electromagnetic modes decompose into lower-dimensional sectors, each governed by its independent dynamics in the absence of interactions. In Sec.~\ref{DIMREDSEC3LASER} the dimensional reduction is applied to the Hamiltonian dynamics, including interactions with matter, showing that the common quantum optical models can be treated in this framework. We identify the typical approach of dimensional reduction as a truncation in the number of the lower-dimensional fields.   The validity of such a number-of-subfield approximation is then investigated for different parameter regimes of a waveguide and an optical cavity in Sec.~\ref{NUMERICALESTIMATON}. Lastly, in Sec.~\ref{LASERSEC} we provide an extension of the dimensional reduction to structured laser beams.

\section{Dimensional Reduction of the Electromagnetic Fields}
\label{DIMREDELFIELDS}

 We start with  the free, second-quantized electromagnetic fields inside an ideal, i.e., perfectly conducting, cavity of volume $V$.
The mode decomposition of the Heisenberg fields may be written in the form 
\begin{subequations} 
\begin{align}
\hat{\E}(\rr,t) &=   \sum_{\bm j, \mu} \Big( A_{\bm j,\mu} \aop_{\bm j, \mu} (t)  \bm u_{\bm j,\mu}(\rr) + \mathrm{H.c.}\Big),\label{QEF}\\
\hat{\textbf{B}}(\rr,t) &=   \sum_{\bm j, \mu}  \Big( C_{\bm j,\mu} \aop_{\bm j, \mu} (t)  \bm v_{\bm j,\mu}(\rr) + \mathrm{H.c.}\Big),\hspace{.2cm}
\label{QBF}
\end{align}
\label{QEBF}%
\end{subequations}\noindent
with frequency $\omega_{\bm j,\mu}$, the electric and magnetic field modes $\bm u_{\bm j,\mu} (\bm r)$ and $\bm v_{\bm j,\mu} (\bm r)$, and the creation and the annihilation operators in the Heisenberg picture $\hat{a}^\dagger_{\bm j,\mu} (t)$ and $\hat{a}_{\bm j,\mu} (t)$. 
The index \mbox{$\bm j$} denotes the tuple of unspecified mode numbers, and $\mu \in \{\mu_1,\mu_2\}$ the two polarizations. The field amplitudes read 
\begin{align}
A_{\bm j,\mu} = \ii  \sqrt{\frac{\hbar \omega_{\bm j,\mu} }{2 \varepsilon_0 }}, \quad C_{\bm j,\mu} = \sqrt{\frac{\hbar \omega_{\bm j,\mu}}{2 \varepsilon_0 c^2}}.
\end{align}

It follows directly from Maxwell's equations and the dispersion relation, $\omega_{\bm j,\mu}= c |\bm k_{\bm j,\mu}|$,
that the electric and magnetic modes defined in Eqs.~\eqref{QEBF} are solutions of the unsourced Helmholtz equation, i.e., 
\begin{subequations}
\begin{align}
\left(\bm \Delta + \bm k_{\bm j,\mu}^2 \right)  
        \bm u_ {\bm j,\mu} (\rr)= \bm  0,
        \label{HELMHOLTZEQ}
\end{align}
with the boundary conditions for the electric modes being 
\begin{align}
\bm n \times \bm u_{\bm j,\mu} (\rr) \Big|_{\rr \in \partial V} &=  \bm 0,
\label{CONDEFIELD}\\
\bm \nabla \cdot \bm u_{\bm j,\mu} (\rr)\Big|_{\rr \in \partial V} &= 0, 
\label{CONDBFIELD}
\end{align}
where $\bm n$ is the normal vector to the cavity surface.
Whereas the Dirichlet type boundary condition~\eqref{CONDEFIELD} is obeyed by the electric field on a perfectly conducting cavity wall, the Neumann type condition~\eqref{CONDBFIELD} is physically motivated by Gauss' law~\cite{Hanson}.
With Faraday's law  the magnetic field modes can be expressed in terms of the electric field modes via
\begin{align}
\bm v_{\bm j,\mu} (\rr) =   |\bm k_{\bm j,\mu}|^{-1}\bm \nabla \times \bm u_{\bm j,\mu} (\rr)  .
\label{ID1}
\end{align}
For the boundary conditions of the magnetic modes one obtains then 
\begin{align}
\bm n \cdot \bm v_{\bm j,\mu} (\rr) \Big|_{\rr \in \partial V} &= 0,
\label{CONDBBFIELD}\\
\bm n \times \left[\bm \nabla \times \bm v_{\bm j,\mu} (\rr)\right]\Big|_{\rr \in \partial V} &= \bm 0. 
\label{CONDBBBFIELD}
\end{align}
\label{FULLHEHOEQ}%
\end{subequations}
The Helmholtz equation~\eqref{HELMHOLTZEQ} together with the boundary conditions~\eqref{CONDEFIELD}-\eqref{CONDBFIELD}, or~\eqref{CONDBBFIELD}-\eqref{CONDBBBFIELD} respectively (also known as the \textit{short-circuit} boundary conditions), form a self-adjoint boundary problem (cf.~\cite{Hanson}, Th. 4.4.6) defined on the Hilbert space of square integrable functions $\textbf{L}^2 (V)$ on the cavity volume $V$, which is considered simply connected enclosed by a sufficiently smooth boundary surface. 
Therefore, magnetic modes $\bm v_{\bm j, \mu}(\bm r)$ and electric modes $\bm u_{\bm j,\mu}(\bm r)$ form an orthonormal mode basis with respect to the $L^2$ inner product on $V$, e.g., 
\begin{align}
\inner{ 
        \bm u_{\bm j,\mu} (\rr)}{ \bm u_{\bm j',\mu'} (\rr)}_V &=  \int_V  \dd^3  r \, \bm u^\dagger_{\bm j,\mu} (\rr)  \cdot  \bm u_{\bm j',\mu'} (\rr) \nonumber\\ &=  \delta_{\bm j,\bm j'} \delta_{\mu,\mu'}.
\label{FIRTSL2NORM}
\end{align}
To perform the dimensional reduction of the 3D model we restrict ourselves subsequently to geometries which exhibit axial symmetry such that the model becomes separable in the longitudinal and transverse degrees of freedom (note, we connote these terms with respect to the symmetry axis and not the wave vector, cf. Fig.~\ref{CAVLOPP}). This covers a wide area of common cavity QED setups, and later we will see how the formalism can be extended to setups including laser beams (see Sec.~\ref{LASERSEC}).

\subsection{Separability of the Modes under Axial Symmetry}
\label{SeparabilityofModesunderAxialSymmetry}
 
\subsubsection*{General Idea: Mapping onto Ancilla Bases}

\label{BASISPROJECTIONCHAP}
\begin{figure}
\begin{center}
\includegraphics[width=6.5cm]{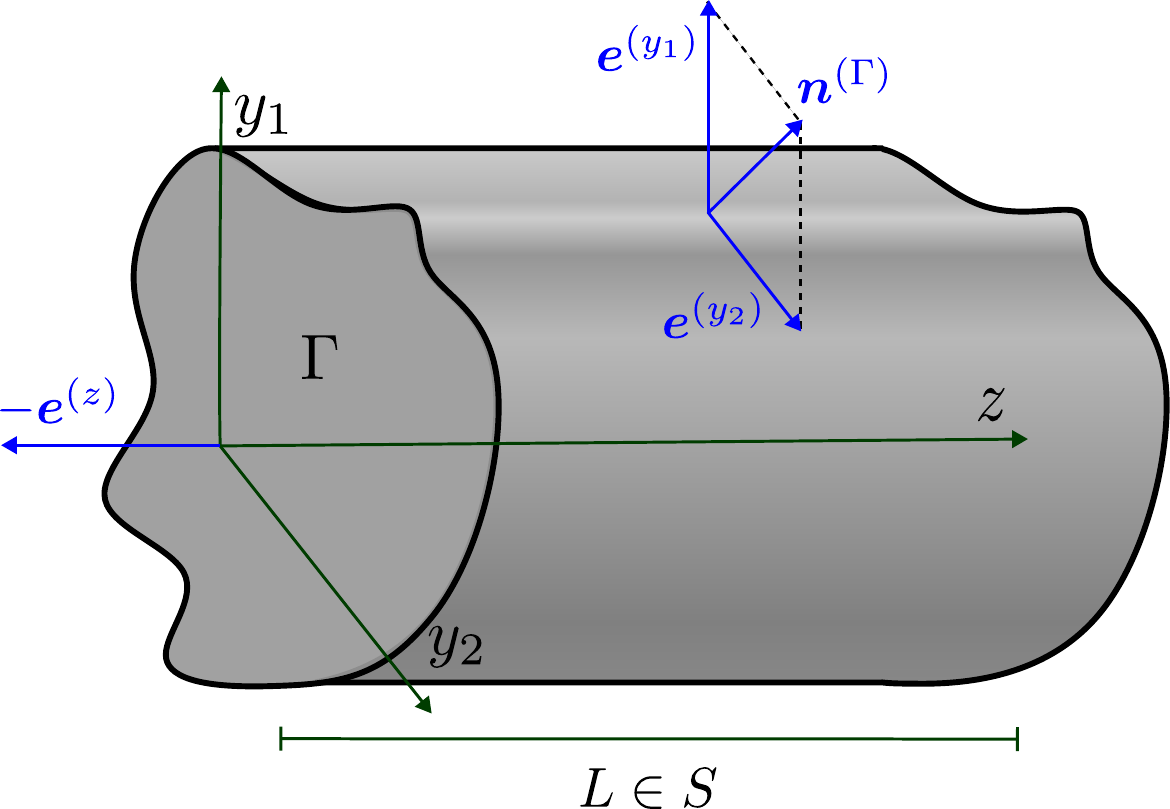}
\end{center}
\caption{Axially symmetric cavity of length $L$ with arbitrary but integrable cross section $\Gamma$ and sufficiently smooth boundary $\partial \Gamma$. The cavity is spanned by the basis $\mathcal{C}$: the symmetry axis defines the longitudinal direction $\bm e^{(z)}$, and the cross section is spanned by $\bm e^{(y_1)}$ and $\bm e^{(y_2)}$ which are the transversal directions. The normal vectors of the cross section are $\pm \bm e^{(z)}$, and  $\bm n^{(\Gamma)}$ for the lateral surface.   }
\label{CAVLOPP}
\end{figure}
We provide first a short mathematical motivation of the dimensional reduction.
In detail, let us consider a separable cavity of simply connected volume $V = \Gamma \times S$ which is spanned by an arbitrary cross section \mbox{$\Gamma \ni \bm y = (y_1, y_2)$} and a longitudinal section $S \ni z$, assuming a sufficiently smooth boundary surface $\partial V$. Thus, one can for example consider a rectangular, circular, or more exotic cross sections as shown in Fig.~\ref{CAVLOPP}.   
We can accordingly define an orthonormal basis with respect to this geometry via
\begin{align}
\mathcal{C}=\{\bm e^{(y_1)}, \bm e^{(y_2)}, \bm e^{(z)} \}.    \label{BASIS}
\end{align}
To stress that the cross section is treated differently from the longitudinal
direction, we introduce the tuple of mode numbers  \mbox{$\bm m =  (m_1,m_2)$}, which is associated with the cross section, while the mode number $l$ is associated with the longitudinal degrees of freedom, i.e., \mbox{$\bm j=(\bm m, l)$}.

Mathematically speaking a separable cavity geometry implies that the modes factorize into a set of orthonormal basis modes for the cross section $\Gamma$ and  one for the longitudinal section $S$ (cf. Th. 2.17 in~\cite{Hanson}). However, these modes couple via the polarization induced by Maxwell's equations in a nontrivial way.

For a separation of the equations of motion it is required that the eigenvalues of the 3D Laplace operator in the Helmholtz equation separate. 
Indeed, for a separable geometry  the wave vectors decompose as
\begin{align}
\kk_{\bm j,\mu}^2 = \kk_{\bm m,\mu}^2  + \kk_{l}^2, 
\label{SEPMODE}
\end{align}
where $\kk_{\bm m,\mu}$ is the wave vector spanned by the transverse basis vectors of $\bm y$, and $\kk_{l}$ corresponds to the longitudinal $z$ direction.
Since we keep the cavity cross section arbitrary, $\kk_{\bm m,\mu}$ may depend on the polarization $\mu$. 
Moreover, we do not impose $\Gamma$ to be separable in its individual degrees of freedom $y_1$ and $y_2$. In that case it is not possible to give the individual components of the wave vector associated with those coordinates. An example of such a cavity is a cylinder~\cite{KAKAZUSE}, which we study in detail subsequently as an example of how the dimensional reduction can be implemented, see Sec.~\ref{NUMERICALESTIMATON}.

The idea in order to obtain a lower-dimensional model from the originating 3D model is to project the ancilla mode basis, which  spans  the cavity cross section  (if one wishes an 1D model associated with the longitudinal space $S$)  and has not been endowed with a polarization structure, onto the 3D mode basis such that  \mbox{$\textbf{L}^2(V) \rightarrow \textbf{L}^2(S)=\bigoplus_{\bm m} \textbf{L}^2_{\bm m} (S)$}.
This results in an infinite set of Hilbert spaces $\textbf{L}^2_{\bm m} (S)$ in 1D where each of these, which we will call \textit{longitudinal-mode spaces}, is associated with fixed transversal mode numbers $\bm m$. 
Alternatively, one could consider a two-dimensional (2D) model: go from the 3D model to the 2D cross section $\Gamma$ via a mapping onto the ancilla basis associated with the longitudinal degrees of freedom, i.e., $\textbf{L}^2(V) \rightarrow \textbf{L}^2 (\Gamma)  = \bigoplus_l \textbf{L}_l^2(\Gamma)$. 
Again, an infinite set of spaces $\textbf{L}^2_l(\Gamma)$ is found for each longitudinal-mode number $l$. We will denote these subspaces as \textit{transverse-mode spaces}.
\begin{figure*}
\begin{center}
\includegraphics[width=16.8cm]{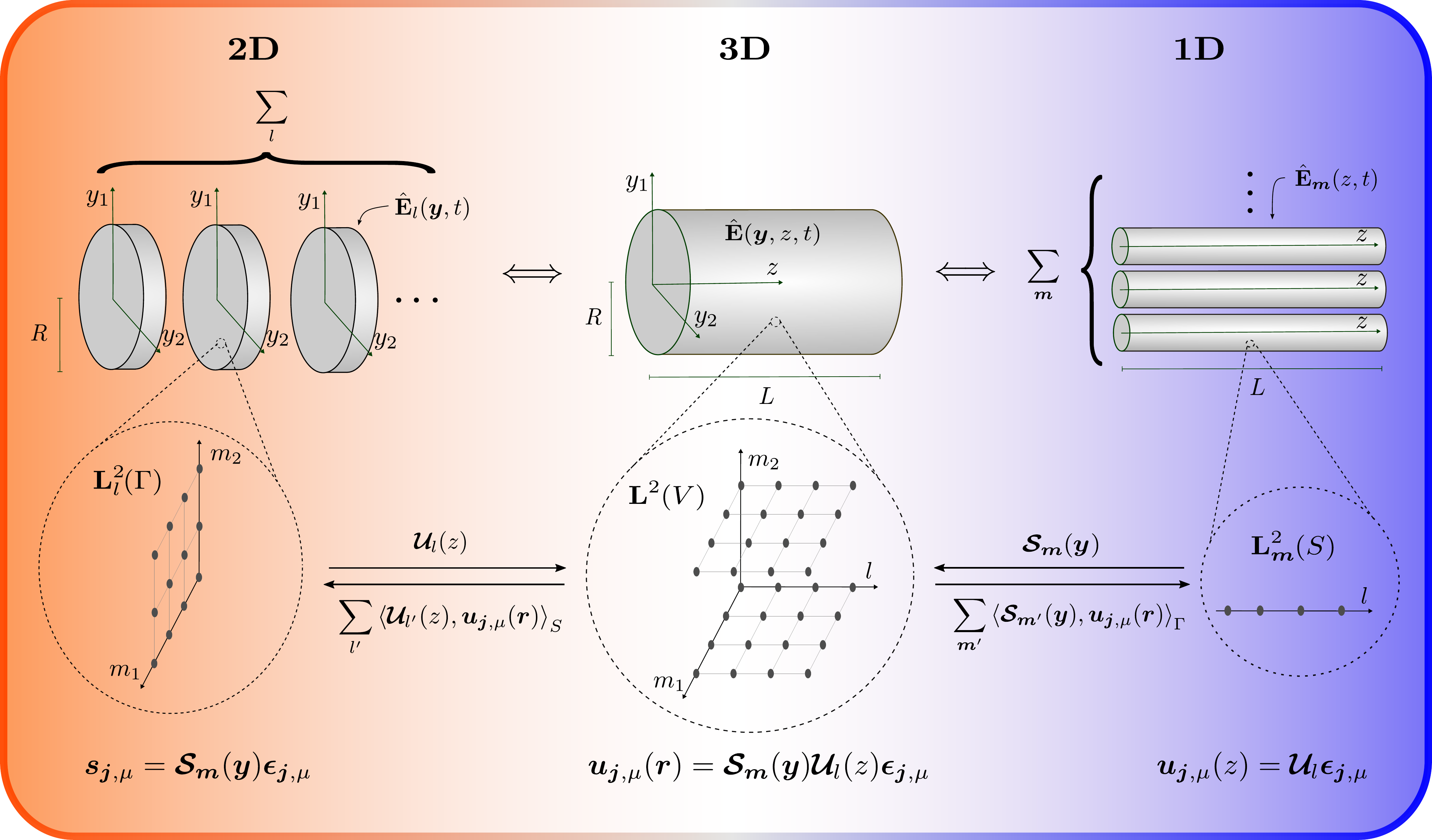}
\end{center}
\caption{Principle of the dimensional reduction of the electric field $\hat\E(\bm y,z,t)$ and its dynamics in an arbitrary axially symmetric 3D cavity of volume $V$ (here example of cylinder of length $L$ and radius $R$). The electric modes are represented by the grids with each dot describing a different tuple of quantum numbers $\bm j=(\bm m, l)$, where $\bm m$ is associated with the cross section $\Gamma$ and $l$ with the longitudinal direction $S$, and polarization $\mu$. The 3D modes  $\bm u_{\bm j,\mu}(\bm r)$ can be written as a product of matrices $\mathbfcal{S}_{\bm m}(\bm y)$ and $\mathbfcal{U}_{l} (z)$, serving as ancilla bases on $\Gamma$ and $S$ respectively, and polarization $\bm \epsilon_{\bm j,\mu}$.   
The reduction to 1D on $S$ is performed by projection onto the ancilla basis $\mathbfcal{S}_{\bm m}(\bm y)$. One obtains an infinite set of subfields $\hat\E_{\bm m} (z,t)$, each spanned by the set of modes $\bm u_{\bm j,\mu} (z)$. Every $\hat\E_{\bm m} (z,t)$ can be seen as corresponding to an independent sector, visualized by the set of 1D cavities. The collection of all of them re-comprises the original 3D dynamics.
The back transformation is applied by multiplying each of the $\bm u_{\bm m l,\mu} (z)$ with their corresponding $\mathbfcal{S}_{\bm m} (\bm y)$. Likewise, a 2D model on $\Gamma$ with subfields $\hat\E_{l} (\bm y,t)$ built from modes $\bm s_{\bm j,\mu} (\bm y)$ is realized via  projection onto ancilla modes $\mathbfcal{U}_{l} (z)$. The same follows for the magnetic field and extends to the common quantum optical interactions with matter.}
\label{FIGFIELDS}
\end{figure*}
Each transverse and longitudinal-mode space is spanned by its own set of basis modes, cf. Fig~\ref{FIGFIELDS}. However, it is important to note that the modes corresponding to different longitudinal and transverse-mode spaces, i.e., for different  $\bm m$ and for different $l$, respectively, have a non-zero overlap. Thus, the different subspaces are in principle coupled to each other. Nonetheless, as we will see in Sec.~\ref{DIMREDSEC3LASER}, for common quantum optical interactions with matter these couplings will not be of relevance.
Accordingly, the dimensional reduction  results  in an exact decomposition of the 3D model where the non-orthogonal, lower-dimensional Hilbert spaces completely decouple. 
For more general dynamics, an extension to the procedure would be required and shall not be the focus here.
Additionally, the modes living on the reduced mode spaces  are the solution of a lower-dimensional Helmholtz equation defined on the respective domain, appended with dimensionally reduced boundary conditions.

The dimensional reduction that we will construct in the following can always be implemented for cavity QED with axial symmetry, e.g., an optical fiber or a Fabry-Pérot cavity. More generally, the dimensional reduction is also applicable to cavities with infinite extension (such as an open-ended cavity), and to setups where the radiation field decays sufficiently fast in some direction (for instance in a laser beam).

\subsubsection*{Mode Decomposition using Ancilla Bases}
\label{BASICS}

To arrive at a lower-dimensional model of the electromagnetic theory, we re-cast the electric  and magnetic modes  as a  product of components associated with the different spatial degrees of freedom. 
To that end, we define for the electric  modes the matrix $\mathbfcal{S}_{\bm m}(\bm y)$ for the cavity cross section $\Gamma$, the matrix $\mathbfcal{U}_{l}(z)$ associated with the longitudinal direction $S$, and the polarization vectors $\bm \epsilon_{\bm j,\mu}$ such that
\begin{subequations}
\begin{align} 
\bm u_{\bm j,\mu} (\rr) &= \mathbfcal{S}_{\bm m} (\bm y) \mathbfcal{U}_{l}(z) \bm \epsilon_{\bm j,\mu}.
\label{IMPORTANTCONDITION}
\end{align}
A similar decomposition can be applied to the magnetic modes. Defining a suitable matrix $\mathbfcal{T}_{\bm m}(\bm y)$ for $\Gamma$, a matrix $\mathbfcal{V}_{l} (z)$ for $S$, and the polarization vector $\bm \kappa_{\bm j,\mu}$, the magnetic modes decompose as
\begin{align}
 \bm v_{\bm j,\mu} (\rr) =\mathbfcal{T}_{\bm m} (\bm y) \mathbfcal{V}_{l}(z)  \bm \kappa_{\bm j,\mu}.
 \label{IMPORTANTCONDITIONVMODES}
\end{align}
\label{IMPORTANTCONDITIONSUANDV}%
\end{subequations}
As we show in App.~\ref{SecB} such decompositions are always possible when considering axially symmetric geometries. In particular,  the transversal components $\mathbfcal{S}_{\bm m} (\bm y)$ and $\mathbfcal{T}_{\bm m} (\bm y)$ are given without polarization index even though $\bm k^2_{\bm m,\mu}$ generally depends on $\mu$, and  the orthonormality of the polarization is preserved, i.e., \mbox{$\bm \epsilon^\dagger_{\bm j,\mu}  \cdot \bm \epsilon^{\vphantom{\dagger}}_{\bm j,\mu'} = \bm \kappa^\dagger_{\bm j,\mu}\cdot \bm \kappa^{\vphantom{\dagger}}_{\bm j,\mu'} = \delta_{\mu,\mu'}$}.

Without specifying the cross section, the matrices $\mathbfcal{U}_{l}(z)$ and $\mathbfcal{V}_{l}(z)$ can be written explicitly for any axially symmetric cavity of finite length $L$. Thus, assuming that \mbox{$z \in S = [0,L]$}, the longitudinal degrees of freedom of the electric and magnetic modes read
\begin{subequations}
\begin{align}
\mathbfcal{U}_{l} (z) 
=  \sqrt{\frac{2}{L}}  \mathrm{diag} \left[ 
\sin \left( k_l z \right), 
\sin \left( k_l z \right), 
\cos \left( k_l z \right)  \right], \\
~\mathbfcal{V}_{l} (z) = \sqrt{\frac{2}{L}} \mathrm{diag}   \left[ 
\cos \left( k_l z \right), 
\cos \left( k_l z \right), 
\sin \left( k_l z \right)  \right],
\end{align}
\label{LONGFIELDMODESSINCOS}%
\end{subequations}
where $k_l=l \pi/L$. The matrices $\mathbfcal{S}_{\bm m}(\bm y)$ and $\mathbfcal{T}_{\bm m}(\bm y)$ for the transversal degrees of freedom, on the other hand, cannot be explicitly written without specifying the cross section $\Gamma$; the same is true for the polarizations $ \bm \epsilon_{\bm j,\mu}$ and $ \bm \kappa_{\bm j,\mu}$. We can, nonetheless, provide a simple reconstruction in terms of a differential operator acting on the eigenbasis of the scalar Helmholtz equation,  cf. App.~\ref{SecB}. This also serves as a bridge to previous literature where the dimensional reduction has been applied to a scalar version of the light-matter interaction~\cite{Lopp1}.
As worked out in the following, the matrices $\mathbfcal{S}_{\bm m} (\bm y)$ and $\mathbfcal{U}_l(z)$ (and analogously for the magnetic field) serve as ancilla modes without polarization structure and allow to generate dimensionally reduced field modes living on a lower-dimensional subspace that satisfy the expected orthonormalization properties and corresponding lower-dimensional Helmholtz equations. Indeed, the dimensional reduction corresponds then to a projection onto orthonormalized basis modes of a certain subspace one wishes to integrate out, and the reduced modes evolve independently from the other degrees of freedom. In Fig.~\ref{FIGFIELDS} we provide a visualization for the procedure.

That the decomposition of the field modes also implies a complete decoupling of the dynamics of the dimensionally reduced problem is not guaranteed, \textit{a priori}, due to the polarization coupling the different degrees of freedom. Despite being common in the literature, leading to celebrated and successful models in many regimes of quantum optics (e.g., 1D versions of the Jaynes-Cummings and Dicke model), decoupling of the field modes can in general not be obtained by brute force distilling out the degrees of freedom suitable for the lower-dimensional model. On the contrary a more general way to realize lower-dimensional models, which is achieved without any approximation, is the mapping of the modes onto a lower-dimensional space. 

\subsection{lower-dimensional Modes}
\label{SecC}
\subsubsection*{1D Dynamics}
\label{1DDynamics}
 
We  begin with the dimensional reduction to the longitudinal degrees of freedom which in an intuitive picture would correspond to a very long and narrow cavity with negligible cross section. Note that, however, that at this point we impose no approximations on the fields. In detail, the reduction from 3D to the 1D  subspace $S$ can be defined solely in terms of the  transverse ancilla modes $\mathbfcal{S}_{\bm m}$ or $\mathbfcal{T}_{\bm m}$ via the $L^2$ inner product on $\Gamma$. Hence,  the  reduced, normalized electric modes of the 1D system are given by 
\begin{subequations}
\begin{align}
\bm u_{\bm j,\mu} (z) &= \sum_{\bm m'} \inner{\mathbfcal{S}_{\bm m'} (\bm y)}{\bm u_{\bm j,\mu}(\rr)}_\Gamma 
\label{SUREANOTHER} 
=  \mathbfcal{U}_{l}(z)  \bm \epsilon_{\bm j,\mu}.\\
\intertext{Analogously the magnetic  modes~\eqref{IMPORTANTCONDITIONVMODES} reduce to:}
\bm v_{\bm j,\mu} (z) &= \sum_{\bm m'} \inner{\mathbfcal{T}_{\bm m'} (\bm y)}{\bm v_{\bm j,\mu}(\rr)}_\Gamma 
=  \mathbfcal{V}_{l}(z)  \bm \kappa_{\bm j,\mu}.
\label{YETANOTHER}
\end{align}
\label{PROJECTIONSUANDV}%
\end{subequations}
 The orthonormality conditions
 \begin{subequations}
 \begin{align}
  \inner{\mathbfcal{S}_{\bm m} (\bm y)}{ \mathbfcal{S}_{\bm m'} (\bm y)}_\Gamma &=  \delta_{\bm m,\bm m'}\openone,
 \label{ORTHOGSS}\\
 \inner{\mathbfcal{T}_{\bm m} (\bm y)}{ \mathbfcal{T}_{\bm m'} (\bm y)}_\Gamma&= \delta_{\bm m,\bm m'}\openone,
 \label{ORTHOGTT}
 \end{align}
 \label{STID1}%
 \end{subequations}
 with $\openone$ being the identity matrix, emerge naturally from the orthonormality of the 3D modes in Eq.~\eqref{FIRTSL2NORM}.
 Accordingly, the dimensional reduction can be viewed as an orthogonal (with respect to transverse-mode numbers $\bm m$) projection onto the transversal ancilla basis on $\Gamma$, integrating out those degrees of freedom. 
It is due to the infinite number of transverse ancilla modes that one obtains an infinite set of longitudinal modes $\bm u_{\bm j,\mu} (z)$ and $\bm v_{\bm j,\mu} (z)$, each pair corresponding to a subspace $\textbf{L}^2_{\bm m} (S) \subset \textbf{L}^2(V)$ with distinct transverse-mode numbers $\bm m$.

With the longitudinal 1D modes~\eqref{SUREANOTHER} and~\eqref{YETANOTHER} at hand, we need to verify that their dynamics completely separates from the dynamics of the transversal degrees of freedom. Therefore, one has to not only split the 3D Helmholtz equation~\eqref{HELMHOLTZEQ} in two independent equations for transverse and longitudinal degrees of freedom but also the boundary conditions of Eq.~\eqref{FULLHEHOEQ}. While the Helmholtz equation can be split by a simple separation ansatz~\cite{Hassani}, it is shown in App.~~\ref{BOUNDARYCONDITIONSFIELDMODES} that the boundary-value problem for the electric 1D modes becomes 
\begin{align}
\begin{split}
(\partial^2_z + k_l^2) \bm u_{\bm j,\mu} (z) &= \bm 0, 
\\
\bm e^{(z)} \cdot\partial_z   \bm u_{\bm j,\mu} (z)\Big|_{z \in \partial S} &=0, \\
\bm e^{(z)} \times  \bm u_{\bm j,\mu} (z)\Big|_{z \in \partial S}&= \bm 0.
\end{split}
\label{INITU1}
\end{align}
For the 1D magnetic modes we have analogously
\begin{align}
\begin{split}
(\partial^2_z + k_l^2) \bm v_{\bm j,\mu} (z) &= \bm 0, \\
\bm e^{(z)} \partial_z \times  \bm v_{\bm j,\mu} (z)\Big|_{z \in \partial S} &= \bm 0, \\ 
\bm e^{(z)} \cdot  \bm v_{\bm j,\mu} (z)\Big|_{z \in \partial S} &= 0.
\end{split}
\label{INITV1}
\end{align}
We further show in App.~\ref{SELFADJOINTNESS1D} that these boundary-value problems correspond to self-adjoint, dimensionally reduced Laplacian operators with the corresponding $L^2$ norm on $\textbf{L}_{\bm m}^2(S)$ (cf. \cite{Hanson}, Def. 3.14).
In other words, each longitudinal-mode space $\textbf{L}_{\bm m}^2(S)$ is equipped with an orthonormal basis of the dimensionally reduced longitudinal modes~\eqref{PROJECTIONSUANDV}. 
They obey orthonormality conditions, for fixed cross sectional mode numbers $\bm m$,
\begin{subequations}
\begin{align}
\inner{\bm u_{\bm m l,\mu}(z)}{\bm u_{\bm m l',\mu'}(z)}_S = \delta_{l,l'} \delta_{\mu,\mu'}
\label{normCondlongmodespace},\\
\inner{\bm v_{\bm m l,\mu}(z)}{\bm v_{\bm m l',\mu'}(z)}_S = \delta_{l,l'} \delta_{\mu,\mu'}.
\end{align}
\end{subequations}
Lastly, the longitudinal modes are imbued with the polarization vectors $\bm \epsilon_{\bm j,\mu}$, respectively $\bm \kappa_{\bm j,\mu}$, of the 3D problem. As we see in detail later, this induces the lower-dimensional model to still contain information on the original 3D model. 
Furthermore,  the 1D electric and magnetic modes obey the usual orthogonality condition (see App.~\ref{APPREDTO1D})
\begin{align}
\inner{\bm u_{\bm j, \mu} (z)}{ \bm v_{\bm j, \mu}(z)}_S = 0.
\label{UVSTR}
\end{align}

\subsubsection*{2D Dynamics}
\label{2DDynamics}
Conversely, the same procedure can also be applied to the transverse degrees of freedom by projecting the longitudinal ancilla basis onto the 3D modes. To that end, we switch the role of the longitudinal and transverse components. The orthonormality condition for the longitudinal ancilla modes, for the electric and magnetic modes respectively, read (cf. Eqs.~\eqref{LONGFIELDMODESSINCOS})
\begin{subequations}
\begin{align}
\inner{\mathbfcal{U}_{l} (z)}{ \mathbfcal{U}_{l'}(z)}_S &= \delta_{ll'}\openone, \label{COMPONENTWISEULZ}\\
\inner{ \mathbfcal{V}_{l} (z) }{\mathbfcal{V}_{l'}(z)}_S &= \delta_{ll'}\openone.
\end{align}    
\label{ORTHONORMALITYUANDVMATRICES}%
\end{subequations}
%Therefore, the maps to $\Gamma$ become
%\begin{subequations}
%\begin{align}
%\mathbfcal{P}^{(\Gamma)} [\bullet] &= \sum_{l}  \int_S \dd z \, \mathbfcal{U}^\dagger_l (z) \, \bullet,\\
%\mathbfcal{Q}^{(\Gamma)} [\bullet]&= \sum_{l}  \int_S \dd z \, \mathbfcal{V}^\dagger_l (z) \bullet.
%\label{PGAMMA}
%\end{align}
%\label{PROJECTIONSOPERATORSUV}%
%\end{subequations}
Continuing analogously, one can define dimensionally reduced modes  depending solely on the transversal coordinates $\bm y$: 
\begin{subequations}
\begin{align}
\bm s_{\bm j,\mu} (\bm y) &= \sum_{l'}\inner{\mathbfcal{U}_{l'} (z)}{\bm u_{\bm j,\mu} (\rr)}_S
= \mathbfcal{S}_{\bm m} (\bm y) \bm \epsilon_{\bm j,\mu},
\label{NONONOTANOTHER}\\
\bm t_{\bm j,\mu} (\bm y) &= \sum_{l'}\inner{\mathbfcal{V}_{l'} (z)}{\bm v_{\bm j,\mu}(\rr)}_S
= \mathbfcal{T}_{\bm m} (\bm y) \bm \kappa_{\bm j,\mu}.
\label{SEEMSMYWORDHASNOWEIGHTHERE}
\end{align}
\label{2DMODEVECTORSSANDT}%
\end{subequations}
Combined with the orthonormality conditions~\eqref{STID1} this implies that the dimensionally reduced transverse modes are $L^2$ normalized on their respective subspace $\textbf{L}^2_l (\Gamma)$ (for fixed longitudinal-mode number $l$):
\begin{subequations}
\begin{align}
\inner{ \bm s_{\bm m l,\mu} (\bm y)}{ \bm s_{\bm m'l,\mu'} (\bm y)}_\Gamma  = \delta_{\bm m, \bm m'}\delta_{\mu,\mu'},\\
\inner{ \bm t_{\bm m l,\mu} (\bm y)}{ \bm t_{\bm m'l,\mu'} (\bm y)}_\Gamma  = \delta_{\bm m, \bm m'}\delta_{\mu,\mu'}.
\end{align}
\label{COND15}%
\end{subequations}
We remark that due to the polarization-independent map applied in Eqs.~\eqref{NONONOTANOTHER} and~\eqref{SEEMSMYWORDHASNOWEIGHTHERE}, similarly to the map to 1D, all information of the 3D fields' polarizations is conserved, resulting in the usual orthogonality of the modes $\bm s_{\bm j,\mu} (\bm y)$  and $\bm t_{\bm j, \mu} (\bm y)$ (see App.~\ref{APPREDTO2D}):
\begin{align}
 \inner{ \bm s_{\bm j, \mu} (\bm y) }{ \bm t_{\bm j, \mu} (\bm y)}_\Gamma  = 0.
\label{COND15EXTENDED}
\end{align}
Moreover, via the same separation ansatz that led to Eq.~\eqref{INITU1}, we obtain the transverse Helmholtz equation for the transverse electric modes on $\Gamma$ i.e., 
\begin{align}
\begin{split}
\left(\Delta_{\Gamma} + \kk_{\bm m,\mu}^2 \right) \bm s_{\bm j,\mu}(\bm y) = \bm 0,
\\
\bm n_{\Gamma}\times   \bm s_{\bm j,\mu} (\bm y)  \Big|_{\bm y \in \partial \Gamma } = \bm 0, 
\\ 
\bm  \nabla_\Gamma \cdot  \bm s_{\bm j,\mu} (\bm y) \Big|_{\bm y \in \partial \Gamma } = 0.
\end{split}
\label{INITM1}
\end{align}
Here $\bm n_\Gamma$ is the normal vector of the surface $\partial \Gamma$ (cf. Fig.~\ref{CAVLOPP}), $\bm \nabla_\Gamma$ the transverse nabla operator (both spanned by the cavity basis vectors of the cross section), and $\Delta_\Gamma = \Delta - \partial_z^2$ is the Laplacian acting only on the transverse coordinates.  
Analogously, one finds for the dimensionally reduced, transverse magnetic modes 
\begin{align}
\begin{split}
\left(\Delta_{\Gamma} + \kk_{\bm m,\mu }^2 \right) \bm t_{\bm j,\mu}(\bm y) &= \bm 0,
\\
\bm n_{\Gamma}\cdot   \bm t_{\bm j,\mu} (\bm y)  \Big|_{\bm y \in \partial \Gamma } &= 0, 
\\
\bm n_{\Gamma}\times \left[ \bm  \nabla_\Gamma \times \bm t_{\bm j,\mu} (\bm y)  \right] \Big|_{\bm y \in \partial \Gamma } &= \bm 0.
\end{split}
\label{INITN1}    
\end{align}
A derivation of the boundary conditions is shown in App.~\ref{BOUNDARYCONDITIONSFIELDMODES}; a proof of the self-adjointness of the boundary-value problem on the underlying transverse-mode space $\textbf{L}^2_{l} (\Gamma)$ can be found in App.~\eqref{SELFADJOINTNESS2D}.
Accordingly, $\bm s_{\bm m, \mu} (\bm y)$ and $\bm t_{\bm m, \mu} (\bm y)$ are both eigenmodes of a self-adjoint boundary-value problem with respect to the inner product~\eqref{COND15} and thus form a complete orthonormal eigenbasis for  fixed $l$.
We want to emphasize again that the separation into two independent boundary-value problems of longitudinal~\eqref{INITU1} and transverse~\eqref{INITM1} degrees of freedom for the electric modes (similarly for longitudinal~\eqref{INITV1} and transverse~\eqref{INITN1} degrees of freedom in case of the magnetic modes) assumed only a separable cavity geometry, as discussed in Sec.~\ref{BASISPROJECTIONCHAP}. 

Hence, we have shown that by mapping the electromagnetic 3D modes onto  a certain subspace thereof, we obtain solutions to the Helmholtz equation associated with the complement space. This directly  defines a dimensional reduction procedure such that the original cavity modes after this  are no longer dependent on the spatial coordinates associated with the ancilla basis, which in turn acts as a means to view the cavity as a lower-dimensional problem. 
In App.~\ref{EXAMPLEAPP} we provide an example for the reduction of the modes of a cylindrical cavity to 2D as well as 1D.

\subsection{Dimensional Reduction of the Quantum Fields}
\label{DIMREDQF}

To understand how the quantum fields themselves behave under the dimensional reduction, we consider --  without loss of generality -- from now on the dimensional reduction to 1D.
Therefore, we employ the reduced longitudinal 1D modes from Eq.~\eqref{SUREANOTHER} for the electric field and~\eqref{YETANOTHER} for the magnetic field respectively. 

Since the electromagnetic fields are Hermitian, they have to be reduced in a way which does not violate Hermiticity. Accordingly, we decompose the fields into polarization independent positive and negative (frequency) field components. The positive field components of electric and magnetic field read  in our notation
\begin{subequations}
\begin{align}
\hat{\mathbfcal{E}}^{(+)}_{\bm j} (\rr ,t) &= \sum_\mu A_{\bm j,\mu}  \aop_{\bm j, \mu} (t) \mathbfcal{S}_{\bm m} (\bm y)  \mathbfcal{U}_{l}(z)  \bm \epsilon_{\bm j,\mu},
\label{POSNEGFIELDSMATRIXE}\\
\hat{\mathbfcal{B}}^{(+)}_{\bm j} (\rr ,t) &= \sum_\mu C_{\bm j,\mu}  \aop_{\bm j, \mu} (t) \mathbfcal{T}_{\bm m} (\bm y) \mathbfcal{V}_{l}(z)  \bm \kappa_{\bm j,\mu}.
\label{POSNEGFIELDSMATRIXB}
\end{align}
\label{POSNEGELBFIELD}%
\end{subequations}
Therefore one has to map the positive field components via the Hermitian conjugate of the transverse ancilla modes, i.e., $\mathbfcal{S}_{\bm m}^\dagger (\bm y)$ for the electric and $\mathbfcal{T}_{\bm m}^\dagger (\bm y)$ for the magnetic field respectively. 
The same is done analogously for the negative field components which are the Hermitian conjugate of~\eqref{POSNEGELBFIELD}.
Starting with the electric field, 
a dimensionally reduced field which only depends on the longitudinal coordinate $z$ is realized as 
\begin{align}
\hat{\E}(z,t) &=   \sum_{\bm j,\bm m'}  \inner{\mathbfcal{S}_{\bm m'} (\bm y)}{ \hat{\mathbfcal{E}}^{(+)}_{\bm j} (\rr,t) }_\Gamma  +  \text{H.c.} \nonumber\\
&= \sum_{\bm m} \hat{\E}_{\bm m}(z,t),
\label{DIMREDE}
\end{align}
where we defined the electric field mapped onto the $\bm m$-th transverse ancilla mode, i.e the electric modes live on the subspace $\textbf{L}^2_{\bm m}(S)$, as
\begin{align}
\begin{split}
\hat{\textbf{E}}_{\bm m} (z,t) &= \sum_{l}  \left( \hat{\mathbfcal{E}}^{(+)}_{\bm j} (z,t) + \hat{\mathbfcal{E}}^{(-)}_{\bm j} (z,t) \right)\\ 
&= \sum_{l,\mu}  \Big( A_{\bm j,\mu} \aop_{\bm j, \mu} (t)  \bm u_{\bm j,\mu}(z) + \mathrm{H.c.} \Big).
\end{split}
\label{RQEF}
\end{align}
The field modes $\bm u_{\bm j,\mu} (z)$ are the dimensionally reduced modes of Eq.~\eqref{SUREANOTHER}. 
Analogously, one finds for the quantized magnetic fields 
via the map $\mathbfcal{Q}^{(S)}$
\begin{align}
\hat{\B}(z,t)  &=  \sum_{\bm j,\bm m'}  \inner{\mathbfcal{T}_{\bm m'} (\bm y)}{ \hat{\mathbfcal{B}}^{(+)}_{\bm j} (\rr,t)}_\Gamma  +   \text{H.c.} \nonumber\\
&= \sum_{\bm m} \hat{\B}_{\bm m}(z,t),
\label{DIMREDB}
\end{align}
where the mapping of the magnetic field onto the $\bm m$-th mode of the transverse ancilla basis gives 
\begin{align}
\begin{split}
\hat{\textbf{B}}_{\bm m} (z,t) &= \sum_{l} \left( \hat{\mathbfcal{B}}^{(+)}_{\bm j} (z,t) + \hat{\mathbfcal{B}}^{(-)}_{\bm j} (z,t) \right)\\
&=  \sum_{l,\mu} \Big( C_{\bm j,\mu} \aop_{\bm j, \mu} (t) \bm v_{\bm j,\mu}(z) + \mathrm{H.c.}\Big),
\end{split}
\label{RQBF}
\end{align}
with $\bm v_{\bm j,\mu} (z)$ being the dimensionally reduced modes of Eq.~\eqref{YETANOTHER}.
Here, we defined $\hat{\mathbfcal{E}}^{(+/-)}_{\bm j} (z,t)$ and $\hat{\mathbfcal{B}}^{(+/-)}_{\bm j}(z,t)$ as the positive/negative frequency components of the reduced 1D fields, i.e.,
\begin{align}
\begin{split}
\hat{\mathbfcal{E}}^{(+)}_{\bm j} (z,t) 
&= \inner{\mathbfcal{S}_{\bm m'} (\bm y)}{ \hat{\mathbfcal{E}}^{(+)}_{\bm j} (\rr,t)}_\Gamma =  \hat{\mathbfcal{E}}^{(-)\dagger}_{\bm j} (z,t),\\
\hat{\mathbfcal{B}}^{(+)}_{\bm j} (z,t)
&= \inner{\mathbfcal{T}_{\bm m'} (\bm y)}{\hat{\mathbfcal{B}}^{(+)}_{\bm j} (\rr,t) }_\Gamma= \hat{\mathbfcal{B}}^{(-)\dagger}_{\bm j} (z,t).
\end{split}
\label{POSNEGLDIM}
\end{align}

Hence, the 1D electric~\eqref{DIMREDE} and 1D magnetic field~\eqref{DIMREDB} decompose into an infinite set of fields $\hat{\textbf{E}}_{\bm m}(z,t)$ and $\hat{\textbf{B}}_{\bm m}(z,t)$ for each given set of transverse-mode numbers $\bm m$ (this is analogous to the modes' decomposition in Sec.~\ref{1DDynamics}). 
In return, the 3D fields can be fully and without approximation reconstructed (by definition of a mode decomposition) from the longitudinal 1D subfields via the back transformation 
\begin{align}
\begin{split}
\hat{\textbf{E}}(\rr,t) =\sum_{\bm j}  &\left( \mathbfcal{S}_{\bm m} (\bm y) \hat{\mathbfcal{E}}^{(+)}_{\bm j}  (z,t) + \hat{\mathbfcal{E}}^{(-)}_{\bm j}  (z,t)\mathbfcal{S}_{\bm m}^\dagger  (\bm y)\right) ,\\
\hat{\textbf{B}}(\rr,t)  = \sum_{\bm j}  &\left( \mathbfcal{T}_{\bm m} (\bm y)  \hat{\mathbfcal{B}}^{(+)}_{\bm j}  (z,t) +  \hat{\mathbfcal{B}}^{(-)}_{\bm j} (z,t) \mathbfcal{T}_{\bm m}^\dagger (\bm y) \right).
\end{split}
\label{BACKTRANSFORM}
\end{align}
We want to emphasize that the operators $\hat{\textbf{E}}  (z,t)$ and $\hat{\textbf{B}} (z,t)$ are the dimensionally reduced fields  and are thus not to be confused with the physical 3D observables $\hat{\textbf{E}}(\bm r,t)$ and $\hat{\textbf{B}}(\bm r,t)$. Since the dimensionally reduced  fields describe the dynamics corresponding to exactly one specific longitudinal-mode space $\textbf{L}_{\bm m}^2(S)$ each, we call them~subfields (in App.~\ref{EXAMPLEAPP} we provide at the example of the cylinder the  explicit construction of the subfields for 1D as well as 2D).

\subsubsection*{Comparison to Ad-Hoc 1D Fields and Maxwell's Equations }

The collection of all subfields encodes information about the geometry of the cross section which is retained even in the lower-dimensional models (cf. Eqs.~\eqref{RQEF} and~\eqref{RQBF}).
In particular, the wave vector has neither been restricted to point in longitudinal direction nor are the frequencies then just 1D in nature. Contrast this with the commonly found 1D field structure of Eq.~\eqref{1DField}. 
In detail, with our choice of basis (see App.~\ref{SecB}), one of the polarizations indeed corresponds to the TE mode  one would prescribe \textit{ad hoc} to a 1D electric field on $z$ (and equivalently the TM mode for the magnetic field). However, the second polarization then necessarily contains a correction in the longitudinal $z$ component that is due to the transversal part of the 3D wave vector:
\begin{align}
\bm \epsilon_{\bm j,\mu_1}  = 
\begin{pmatrix}
 &1/\sqrt{2}\\
 -&1/\sqrt{2}\\
 &0
\end{pmatrix},\,
\bm \epsilon_{\bm j,\mu_2}   = \frac{1}{| \bm k_{\bm j,\mu_2}|}  \begin{pmatrix}
-k_l/\sqrt{2}\\
-k_l/\sqrt{2}\\
|\bm k_{\bm m,\mu_2}|
\end{pmatrix}.
\end{align}
That entry can never be identical to zero unless one considers the continuum limit for the cross section (corresponding to a Fabry-Pérot cavity with infinite sized mirrors). In that case, though, there is no energy gap to the lowest level transversal mode which cannot be expected to be the sole contributor to the dynamics, accordingly.  
In the opposite regime, and one where one would intuitively expect to use a 1D model, with the cross section much smaller in extension (with characteristic scale $R$) than the length $L$ of the cavity, the transversal modes dominate the wave vector since \mbox{$|\bm k_{\bm m, \mu}|\propto 1/R, \text{ whereas } |k_l|\propto 1/L$.}

Furthermore, we showed that the lower-dimensional modes satisfy dimensionally reduced Helmholtz equations with the analog boundary conditions from the original 3D theory on their respective mode space. 
In contrast, one could have also tried reducing the dimensions of Maxwell's equations to achieve  lower-dimensional Helmholtz equations~\cite{Maggi,Schippa}. This can be for instance done via the method of \textit{Hadamard's decent}~\cite{Hadamard,Angelone}, where spatial degrees which one wishes to reduce are assumed constant. The dimensional reduction procedure used here -- where no assumptions besides an axially symmetric cavity are applied -- does in general not commute with the spatial operations of Maxwell's equations. Thus, a dimensional reduction acting directly on the 3D Helmholtz equation~\eqref{HELMHOLTZEQ} but not on the 3D modes themselves is not obtained trivially by the maps defined in Sec.~\ref{BASISPROJECTIONCHAP}.
Naturally, the subfields~\eqref{RQEF} and~\eqref{RQBF} violate not only the 3D Maxwell's equations as well as the 3D Helmholtz equation~\eqref{HELMHOLTZEQ} but already the Coulomb gauge condition. In summary this gives rise to the subfields being an effective description as they do not represent physical fields but a convenient representation to perform the dimensional reduction.

The emergence of an effective theory is obviously given by the non-invertible mapping from 3D to the lower dimensions.
Here, the information that the eigenmodes of the transverse Helmholtz equation, or the longitudinal Helmholtz equation respectively, contribute to the mode structure is lost. However, their eigenvalues remain, with each subfield being associated with one of these eigenvalues. More precisely: The subfields do still contain information of the full cavity spectrum as the eigenvalues to the Helmholtz equation are preserved by the dimensional reduction. Through these eigenvalues, conclusions can again be drawn about the 3D cavity geometry from where the mode structure of the transverse degrees of freedom may be reconstructed~\cite{Lopp1}.
Nevertheless, the subfield decomposition itself is  exact  which we show explicitly when deriving the dynamics for the subfields in the next section. Conversely, assigning a single one of these subfields physical significance -- as is commonly the case in the literature -- can only be done in conjunction with quantifying the error introduced to the physical observables. The details of which will be discussed in Sec.~\ref{MEASUREOFVALIDITY}. In essence, the subfields provide physical results only in the same way that the few-mode approximation in quantum optics provides accurate results. As such one does not necessarily need the whole 3D model to back up a specific physical result but a sufficient convergence in error with respect to some metric.

\section{Dimensional Reduction at the Hamiltonian Level}
\label{DIMREDSEC3LASER}

\subsection{The Free-Field Dynamics}
\label{DIMREDFIELDHAMILTONIAN}

Analogously to the fields, the free-field Hamiltonian decomposes into an infinite number of effective 1D Hamiltonians  (see App.~\ref{FIELDHAMAPP}): 
\begin{align}
\begin{split}
\hat{H}^{\mathrm{field}} 
&= \frac{\varepsilon_0}{2} \int_V \dd^3 r \left[ | \hat{\textbf{E}}(\rr,t)|^2 + c^2  | \hat{\textbf{B}}(\rr,t)|^2 \right]= \sum_{\bm m} \hat{h}_{\bm m}^{\mathrm{field}}.
\end{split}
\label{Hh}
%\label{FIELDHAM}
\end{align}
Each Hamiltonian $\hat{h}_{\bm m}^\mathrm{field}$ is the 1D Hamiltonian analog in terms of the 1D electromagnetic subfields~\eqref{RQEF} and~\eqref{RQBF} for a given transversal mode set $\bm m$, i.e.,
\begin{align}
\hat{h}_{\bm m}^{\mathrm{field}} =& \frac{\varepsilon_0}{2}  \int_S \dd z   \left[  |\hat{\textbf{E}}_{\bm m}(z,t)|^2 + c^2    |\hat{\textbf{B}}_{\bm m}(z,t)|^2  \right]\nonumber\\
=&  \hbar \sum_{l,\mu} \,\,  \omega_{\bm j,\mu} \hat{a}^\dagger_{\bm j,\mu} \aop^{\vphantom{\dagger}}_{\bm j,\mu},
\label{h1D}
\end{align}
where the whole spectrum of the longitudinal modes and polarizations contribute.
Accordingly, all of these subfield Hamiltonians taken together completely comprise the original 3D free-field dynamics. Importantly, even though the modes corresponding to different subspaces $\textbf{L}_{\bm m}^2(S)$   have non-vanishing overlap the Hamiltonians~\eqref{h1D} commute, and  the free dynamics does not couple these subspaces.
Note: Since the subfield Hamiltonians are realized without any approximation from the 3D model they differ from the usual 1D theories discussed in Sec.~\ref{Intro}.

\subsection{The Light-Matter Interaction}
\label{DIMREDH}
Next we examine the dynamics of the fields induced by the interaction  with matter.  Without loss of generality, we restrict ourselves to the electric dipole Hamiltonian; extensions to magnetic atoms, higher-order multipoles or atomic center of mass delocalization follow analogously; in App.~\ref{MULTIPOLEAPP} we  explicitly show how this applies to the electric dipole interaction for an atom with a quantized center of mass.

Let us consider as a concrete example a general hydrogen-like, stationary (atomic motion can be included in the formalism in a straightforward manner) atom interacting with the electromagnetic field within the cavity.
We assume that the atom is an effective one-particle system with a classical nucleus that is much heavier than the electron. The atomic Hamiltonian may be written as
$
\hat{H}^\mathrm{A} = \sum_{\bm s} E_{\bm s} \dyad{\bm s}{\bm s}.
%\label{ATH}
$
The dipole interaction can then be written -- when prescribed from the field's quantization frame and in the joint atom-field interaction picture -- as~\cite{Lopp2, Vukics}
\begin{align}
\begin{split}
\hat{H}^\mathrm{I} (t)= &\chi(t)  \sum_{\bm s>\bm s'}\int_V \dd^3  r \, \hat{\bm d}_{\bm s\bm s'} (\re(\bm r), t) \cdot  \hat{\E} (\bm r,t),
\end{split}
\label{HDIPG}
\end{align}
being in position representation and the electronic position \mbox{$\bm r_\mathrm{e} (\bm r)$} accounts for the field's  (stationary) quantization frame to be potentially different from the atom's center of mass frame. 
The dipole operator can be expressed as 
\begin{align}
\hat{\bm{d}}_{\bm{s}\bm{s}'}(\re, t)&=e\bm F_{\bm s \bm s'}(\re) e^{\ii \Omega_{\bm s\bm s'} t}\dyad{\bm{s}}{\bm{s}'} + \text{H.c.}   
\end{align}
in terms of the spatial smearing vector
\begin{align}
\bm F_{\bm s\bm s'} (\re) =    \re \psi^{\ast}_{\bm s} (\re) \psi_{\bm s'} (\re),
\label{SMEARNORMAL}
\end{align} 
where we inserted the atomic wave functions \mbox{$\braket{\re }{\bm s}  = \psi_{\bm s}(\re)$}, and we defined the atomic energy spacing $\hbar \Omega_{\bm s \bm s'}= E_{\bm s}- E_{\bm s'}$.
The spatial smearing accounts for the atomic spatial profile associated with the internal states for any considered transition process.
We also introduced a switching function $\chi(t)$ which encodes the possibly time-dependent coupling between the atom and the field inside the cavity.
To achieve the dimensionally reduced dipole Hamiltonian we define the spatial smearing vector associated with the $\bm m$-th subfield, reading 
\begin{align}
\Smear (\ze (z)) = 
\inner{   \mathbfcal{S}^\dagger_{\bm m}(\bm y)}{ \bm F_{\bm s\bm s'} (\re(\bm r))}_\Gamma,
\label{KNMSS}
\end{align}
i.e., it is the mapping of the atomic profile onto the field's transversal ancilla basis.
Employing Eq.~\eqref{BACKTRANSFORM} yields
\begin{align}
\hat{H}^\mathrm{I} (t) = &\chi(t) \, e\sum_{\bm j} \sum_{\substack{\bm s,\bm s'\\ \bm s \neq \bm s'}}   \int_{S} \dd z   \Smear (\ze(z)) \hat{\mathbfcal{E}}_{\bm j}^{(+)}( z,t) \nonumber \\ 
&\times  \e^{\ii \Omega_{\bm s\bm s'}t}  \dyad{\bm s }{\bm s'}
+ \mathrm{H.c.} \label{HDEE}\\
=&\sum_{\bm m} \hat{h}^\mathrm{I}_{\bm m}(t)\nonumber,
\end{align}
where in the last step we defined the subfield-dipole Hamiltonians $\hat{h}_{\bm m}^{\mathrm{dipole}}$ for fixed transverse-mode numbers $\bm m$. Therefore, the projected smearing vector $\Smear (\ze(z))$ accounts for how much the atom couples to the corresponding subfield $\hat{\E}_{\bm m}(z,t)$. 
Analogously to the free-field Hamiltonian, the interaction Hamiltonian too decomposes  into an infinite number of dimensionally reduced  Hamiltonians, each prescribing the atom-light interaction on its respective Hilbert space.

\subsection{The Number-of-Subfield Approximation}
\label{MEASUREOFVALIDITY}

In a similar fashion, the decompositions of the Hamiltonians~\eqref{Hh} and~\eqref{HDEE}  could have been also attained in terms of a reduction to a 2D problem, that is via the mapping onto the longitudinal-mode spaces presented in Sec.~\ref{2DDynamics}. Since we have performed no approximation hitherto, both descriptions can be used to reconstruct the dynamics induced by the interactions discussed so far. Therefore, they correspond merely to a different basis expansion and thus correspond to different subspaces, cf. Sec.~\ref{SeparabilityofModesunderAxialSymmetry}. 
Both cases are, however, intuitively connected
to different cavity shapes and thus to different mode structures inside a cavity.

To illustrate this, take two opposite regimes of a cylindrical cavity: a thin disk-shaped cavity ($R\gg L$) and a long fiber-shaped  cavity ($R\ll L$), with radius $R$ and length $L$. For the disk-shaped cavity, the frequency difference between two modes with neighboring transverse-mode numbers $\bm m$ but equal longitudinal-mode number $l$  is much larger than for equal $\bm m$ but neighboring $l$. Thus, if we choose a fixed polarization and neglect the polarization index for compactness, we obtain the condition (in reference to frequency  $\omega_{\bm m,l}$)
\begin{subequations}
\begin{align}
&R \gg L: \, \min \{ \omega_{\bm m+ (1,0),l},\, \omega_{\bm m + (0,1),l} \} \gg \omega_{\bm m,l+1}. 
\label{DISKCAVITY}\\
\intertext{In contrast to this we have for the fiber-shaped cavity}
&R \ll L: \,\max \{ \omega_{\bm m + (1,0) ,l},\, \omega_{\bm m + (0,1) ,l} \} \ll \omega_{\bm m,l+1}. 
\label{THINWIRECAVITY}
\end{align}
\label{IDENTITIESWIRERESONATOR}%
\end{subequations}
Let us consider how lower-dimensional models can intuitively be achieved when considering an atom near resonance to one cavity mode, i.e., \mbox{$\Omega_\mathrm{A} \approx \omega_{\bm m^{\mathrm{res}},l^{\mathrm{res}}}$}. For a disk-shaped cavity with  Eq.~\eqref{DISKCAVITY} one would intuitively pick  the single subfield from the dimensionally reduced 2D model with longitudinal-mode number $l^\mathrm{res}$.  However, in case of the thin fiber, one would pick instead the resonant subfield  corresponding to the 1D model with transverse-mode number $\bm m^{\mathrm{res}}$.

The two limits are also well known when describing a quantum-mechanical particle in an axially symmetric harmonic trap. There, two cases can be distinguished based on the axial and perpendicular harmonic-oscillator length $a_z$ and $a_\bot$, respectively. When $a_z \ll a_\bot$, the system reduces to a two-dimensional disk, whereas for $a_\bot \ll a_z$, a one-dimensional tube is obtained. In both cases the frequency spacing along the strongly confined direction is too large to allow for any excitation, effectively freezing the system in the corresponding ground state of that direction, fully analogously to Eq.~\eqref{IDENTITIESWIRERESONATOR}.

Furthermore, in a BEC with interacting atoms, the picture is more complicated due to additional intrinsic length scales imprinted by the interaction. Here, the spectrum separates in collective and single-particle excitations at the healing length $\xi$. If in addition to above criteria either $a_z \ll \xi$ (disk) or $a_\bot \ll \xi$ (tube), the collective excitations of the BEC in the strongly confined direction are frozen out. Such circumstances allow for the description of the dynamics with a dimensionally reduced Gross-Pitaevskii equation~\cite{Pit, Salas}. Note that the local scattering between two atoms still retains its 3D character until the oscillator length reaches the order of the 3D scattering length~\cite{Adhi,Shlap}. This is in analogy to the models presented here where the 3D character due to the polarization of the field is preserved locally.

What we have achieved so far is to decompose -- without approximation -- the full light-matter interaction  in terms of subfields of the electromagnetic field and associated spatial profiles of the atom coupling to those subfields. To formulate an approximation scheme for dimensional reduction, and to connect this procedure with the \textit{ad hoc} prescription in the literature, we have to discuss how many of the subfields are necessary so as to have an accurate representation of the full 3D dynamics. Then a subfield approximation based on Eq.~\eqref{DIMREDE} is nothing but a number-of-modes approximation, equivalent to a single- or few-mode approximation.

For that truncation, an observable has to be chosen. A natural choice for this are atomic transition probabilities that can be attained via an experiment. That is to say, we address the question of how many subfields we need to take into account in order to only deviate slightly from the full 3D statistics.
In particular, we see that the atomic transition probabilities  can also be decomposed into a sum of independent terms each governing the transition probability induced by the appropriate subfield.

To calculate the transition probabilities we use a Dyson series~\cite{Scully} in the weak-coupling regime to first order which is valid as long as the relevant parameters are sufficiently small.
We assume that the initial state of the joint system is a product state of the field in the
vacuum state $\ket{\Omega}= \bigotimes_{\bm j,\mu} \ket{0}_{\bm j,\mu}$  and the atom in an arbitrary energy eigenstate $\ket{ \bm s  }$.
Thus, the transition probability between an initial atomic level $\ket{ \bm s  }$ and the arbitrary (yet distinct) level $\ket{ \bm s'  }$ can be, to leading order, written as 
\begin{align}
P_{\bm s\rightarrow \bm s'}  
&\approx 
\sum_{\bm j', \mu'} \sum_{n=1}^\infty \left|\matrixel{ n_{\bm j',\mu'};  \bm s' }{ \sum_{\bm m} \left(\frac{\ii}{\hbar}  \int_\mathbb{R} \dd t \,\, \hat{h}^\mathrm{I}_{\bm m}(t)  \right)}{\Omega; \, \bm s }\right|^2\nonumber\\
&= \sum_{\bm m} |c_{\bm m,  \, \bm s\rightarrow \bm s'} |^2,
\label{PAMPL}
\end{align}
where  $|n_{\bm j,\mu} \rangle$ is the single-mode Fock state with $n$ field excitations in mode $(\bm j,\mu)$ and all other modes in the vacuum state. Accordingly, the total transition probability is decomposed into probabilities induced by the corresponding $\bm m$-th subfield:
\begin{align}
\begin{split}
|c_{\bm m,  \, \bm s\rightarrow \bm s'}|^2 
= &\sum_{l,\mu}   \frac{\omega_{\bm j,\mu} e^2}{2 \varepsilon_0 \hbar}  \left| \int_\mathbb{R} \dd t  \chi(t) \e^{\ii(\omega_{\bm j,\mu} - \Omega_{\bm s\bm s'})t}\right. \\
&\left.\times\int_S \dd z \,  \bm u_{\bm j,\mu}(z) \cdot  \Smear(\ze(z)) \right|^2.
\end{split}
\label{CCCC}
\end{align}
Let us come back to our issue at hand: How many subfields of the reduced cavity are required? In particular: Do there exist special regimes where one can approximate the 3D model by a small number of subfields (perhaps one)?
We assume  that we restrict the number of subfields to some subset  $\bm N$ such that 
\begin{align}
\begin{split}
P_{\bm N,\, \bm s \rightarrow \bm s'} =  \sum_{\bm m \in \bm N}  |c_{\bm m,\, \bm s \rightarrow \bm s'}|^2. 
\label{NPROBRED}
\end{split}
\end{align}
Such a truncation can be analogously thought of as a modification in the atomic spatial profile, cf. Eq.~\eqref{KNMSS}. Intuitively, removing all but a few subfields is valid if the atomic shape is such that it couples only strongly to a few subfields. Of course, the field's mode functions are generally not of the shape of an  atomic wave function, and a truncation needs therefore careful analysis.
We can evaluate the validity of the truncation via the relative difference to the full transition probabilities (for a detailed discussion on different measures for the truncation error see~\cite{Lopp1}):
\begin{align}
\delta_{\bm N, \bm s\rightarrow \bm s'} = P^{-1}_{\bm s\rightarrow \bm s'}| P_{\bm s \rightarrow \bm s'}  - P_{\bm N,\bm s \rightarrow \bm s'}   |.
\label{delta}
\end{align}

\section{Example: Dimensional Reduction for a Cylindrical Cavity}
\label{NUMERICALESTIMATON}
\begin{figure}
\begin{center}
\includegraphics[width=8cm]{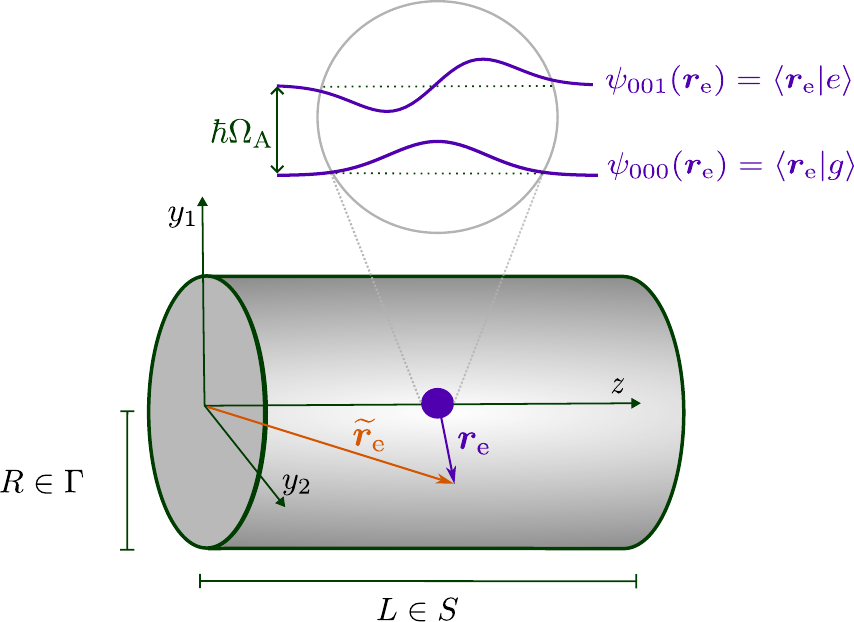}
\end{center}
\caption{Atom in the center of a cylindrical cavity with radius $R$ and length $L$. The electronic (cylindrical) coordinates are \mbox{$\bm r_\mathrm{e} = (\ye, z_\mathrm{e})$}  with respect to the atomic center of mass, and $\bm r = (\ye, z_\mathrm{e} + L/2)$ with respect to the field's quantization frame. The atom is modeled as a two-level system with the harmonic-oscillator eigenstates $\psi_{000} (\bm r_\mathrm{e}) = \psi_g (\bm r_\mathrm{e})$ and $\psi_{001} (\bm r_\mathrm{e}) = \psi_e (\bm r_\mathrm{e})$ with the energy gap $\hbar \Omega_\mathrm{A}$.}
\label{SetupFig}
\end{figure}

We now perform a numerical study as a working example of the dimensional reduction laid out in the previous sections and  consider a cylindrical cavity  of length $L$ in longitudinal direction and radius $R$ in transverse direction. A single two-level atom is placed in the center of the cavity at $z = L/2$ such that $\bm y = \bm y_\mathrm{e}$ (see Fig.~\ref{SetupFig}), with the field's quantization frame having its origin at $z =0$. For a discussion on transforming the interaction Hamiltonian to different frames see Refs.~\cite{Lopp2,Lopez}.

The atom is modeled as a 3D quantum harmonic-oscillator where we only consider the ground state and one excitation associated with the longitudinal ($z$) direction (the transverse directions being frozen out as is commonly assumed), expressed in the electric field's quantization frame:
\begin{subequations}
\begin{align}
\psi_{g} (r,z) = &\frac{1}{\pi^{3/4} \sigma^{3/2}}  \e^{ - \frac{r^2}{2\sigma^2}} \e^{ - \frac{(z-L/2)^2}{2 \sigma^2}},\\
\psi_{e} (r,z) =  &\frac{ 2^{1/2} }{\pi^{3/4} \sigma^{3/2}}  \e^{ - \frac{r^2}{2\sigma^2} } \e^{ - \frac{(z-L/2)^2}{2 \sigma^2}} \frac{z - L/2}{\sigma},
\end{align}   
\label{GAUSSIANWAVEFCTN}%
\end{subequations}
with harmonic-oscillator length
\mbox{$\sigma =  \sqrt{\hbar/(M \Omega_{\mathrm{A}})}$}~\cite{Foot}, $\Omega_\mathrm{A}$ being the frequency gap between excited and ground state, and $M$ being the mass of the atom.

We further consider two different time-dependent couplings in Eq.~\eqref{CCCC}. First a top-hat switching $\chi^{\mathrm{TS}}(t)$ and an adiabatic Gaussian switching $\chi^{\mathrm{GS}}(t)$:
\begin{align}
\chi^{\mathrm{TS}}(t) = \theta(t) \theta(t-T), \quad  \chi^\mathrm{GS}(t) =\exp\left( - \frac{t^2}{2 T^2}\right),
\label{ChiGauss}
\end{align}
respectively.
Both depend on the interaction time parameter $T$ which gives a characteristic time the atom is exposed to the field.

Considering the Hamiltonian~\eqref{HDEE}, due to the axial symmetry the $\bm e^{(\varphi)}$ component of the overlap from electric field and smearing function  vanishes in this setup. Moreover, for axially symmetric atomic wave functions centered on the symmetry axis the integration fixes the azimuth quantum number $m_2$ to zero. The vanishing $\bm e^{(\varphi)}$ and $\bm e^{(z)}$ component of the TE modes and the $\varphi$ derivative in $\bm e^{(r)}$ then leads to a vanishing interaction of the atom with the TE mode $\mu_1$. Thus we drop the polarization index for compactness of notation. 

For the sake of analytical results, we assume that the atom is sufficiently localized far from the cavity walls. That is,
$\sigma \ll R, L$.
In App.~\ref{GAUSSIANCYLCAVAPP} we calculate the overlap of the atomic smearing function~\eqref{KNMSS} with  the subfields.
We find the following transition probabilities to leading order~\footnote{Note that time $T$ is upper and lower bounded within perturbation theory. The upper limit is given by the convergence radius of the first order, dependent on switching, mode and atomic frequencies (\cite{Sakurai}, Chap. 5); roughly translating to $ \Omega_{\mathrm{A}} T \gg 1$. At the lower bound transitions to higher levels start getting relevant. Therefore, $T \to 0$ is excluded in the analysis.} 
\begin{align}
P_{(\pm)}\! &=  \frac{(e c \sigma)^2}{2 \pi \varepsilon_0 \hbar R^4 L}\sum_{m_1, l} \frac{\chi_{m_1}^2  \exp \left(-\frac{\chi_{m_1}^2 \sigma^2 }{2 R^2} -\frac{2 \pi^2 l^2 \sigma^2  }{ L^2}\right)}{\omega_{(m_1,0),2l} \, J_1^2(\chi_{m_1 })}\nonumber\\
&\times 
\begin{cases}
T^2 \mathrm{sinc}^2 ( \Delta_{(m_1 , 0), 2l,(\pm)}T ), \hspace{1.25cm} 
 \chi(t) \! =\! \chi^{\mathrm{TH}} (t), \\[5pt]
2\pi T^2 \exp \left(-  2 (\Delta_{(m_1, 0), 2l,(\pm)} T  )^2 \right), \chi(t) \! =\! \chi^{\mathrm{GS}} (t),
\end{cases}  \label{PREDTH}
\end{align}
where we defined $\chi_{m_1}$ as the $m_1$-th zero of the zeroth order Bessel function $J_0$, the frequencies \mbox{$\omega_{(m_1,0),2l}= c \, \sqrt{(\chi_{m_1}/R)^2 + (2 \pi l/L)^2 }$}, and the detuning \mbox{$\Delta_{\bm j,(\pm)} = ( \omega_{\bm j} \pm \Omega_\mathrm{A})/2$}, with $(-)$ denoting spontaneous emission ($e \rightarrow g$), and $(+)$ vacuum excitation ($g \rightarrow e$). 
We now examine different regimes for the subfield approximation reaching from the general waveguide regime with $\sigma \lessapprox R \ll L$ to the general optical resonator regime with $\sigma \ll R \lessapprox L$. By that notion, one would expect that the long-and-narrow waveguide is a much better regime for a small-number approximation in the subfields.

\begin{figure*}
\includegraphics[width=14.5cm]{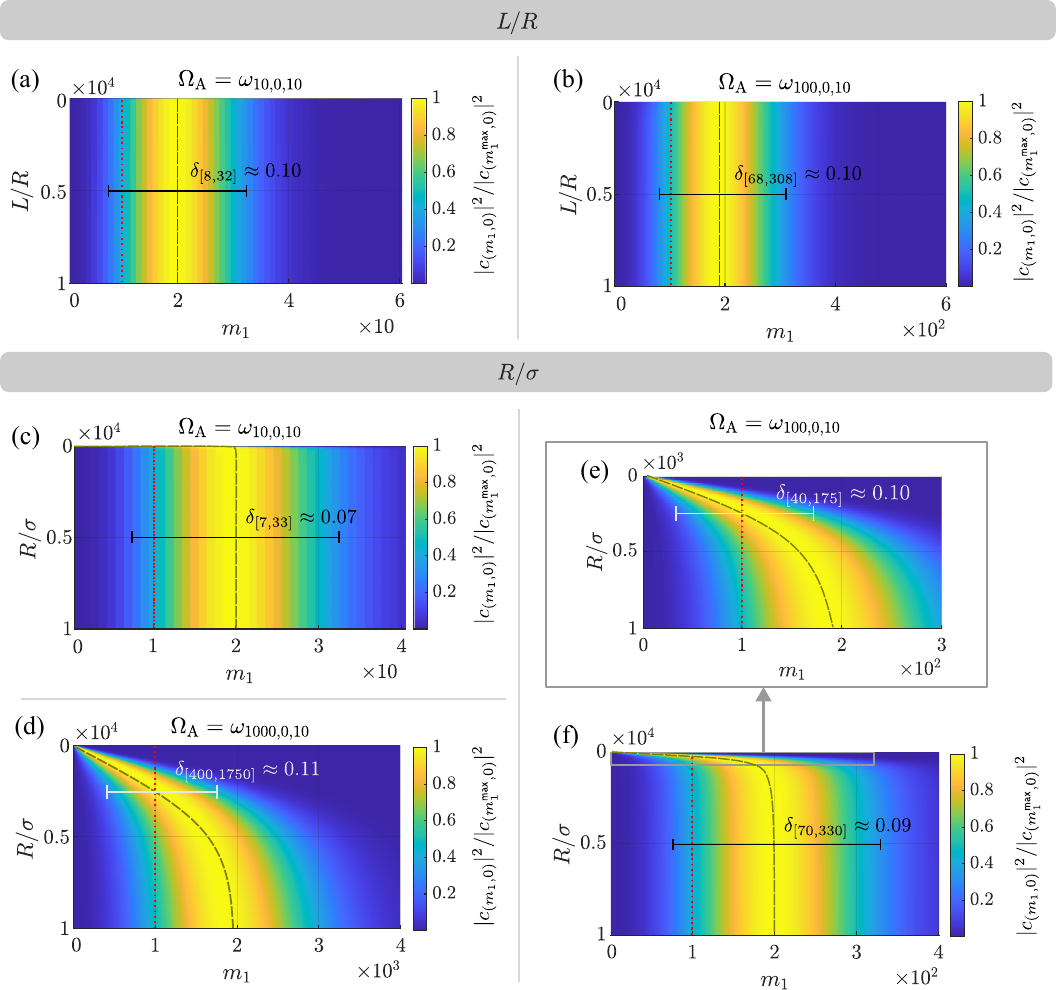}
\caption{Normalized subfield probabilities relative to the maximum transition probability \mbox{$|c_{(m_1,0)}|^2 /|c_{(m_1^\text{max},0)}|^2$} plotted as a function of subfield label $m_1$ and $L/R$  in (a)-(b) or $R/\sigma$ in (c)-(f) for different resonance conditions with $ \Omega_\mathrm{A}T = 1$ for spontaneous emission of the 2-level system in a cylindrical cavity, cf. Eq.~\eqref{PREDTH}. The resonant subfield  $m_1^\mathrm{res}$ is depicted as a red dotted line, the subfield with maximum transition probability $m_1^\text{max}$ as a green dashed line. Parameters in (a)-(b): $R/\sigma = 10^3$;  (c)-(f): $L/R = 10^3$. We further exemplarily show $\delta_{N}$ (cf. Eq.~\eqref{deltaCyl}) via the vertical bar indicating the set of subfields $\bm N$ needed for a relative error of about $10\,\%$. Note that the subscript $(-)$ has been dropped here for convenience.}
\label{Ratios}
\end{figure*}

\begin{figure*}
\includegraphics[width=14.5cm]{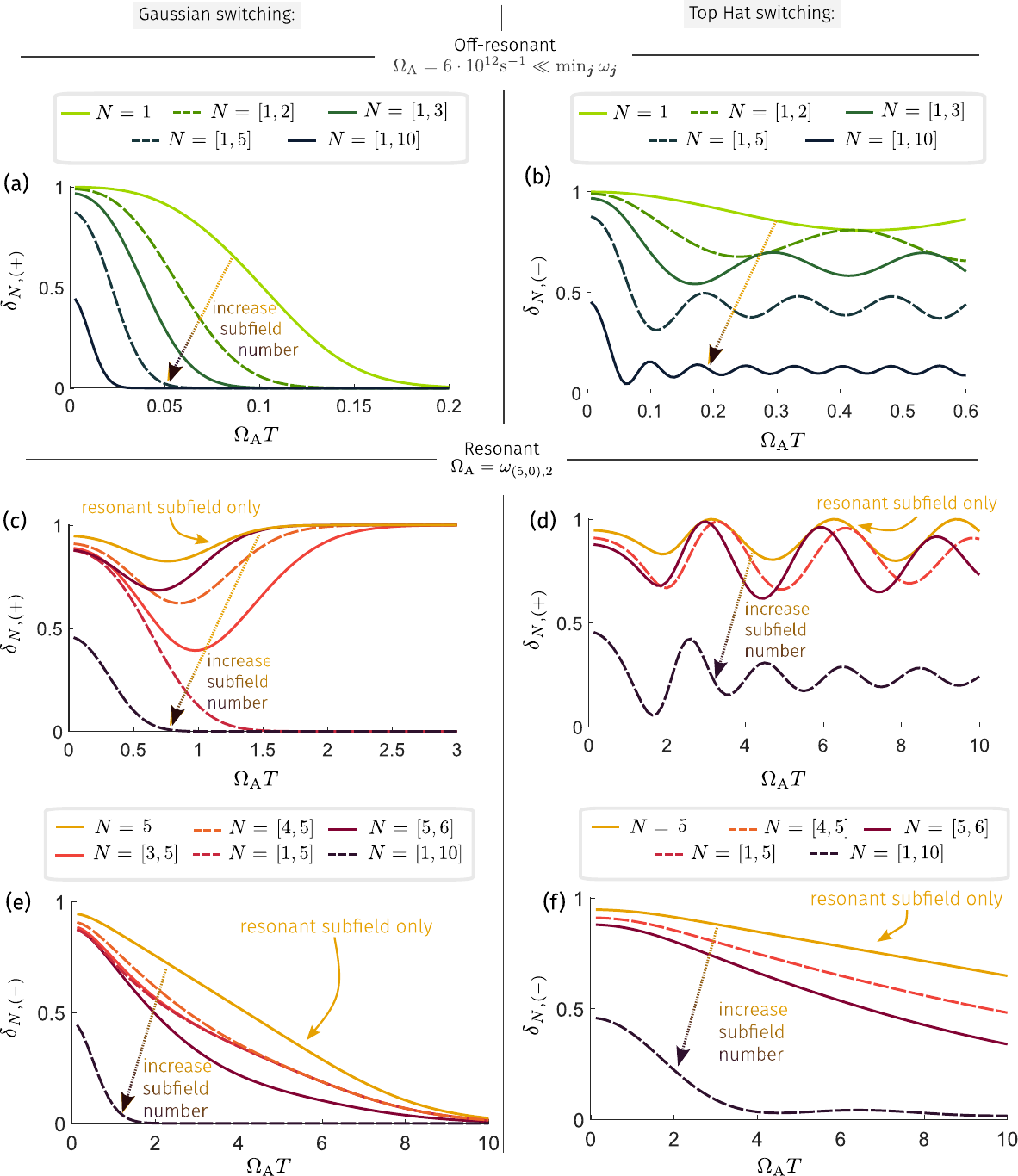}
\caption{Relative difference, Eq.~\eqref{deltaCyl}, between full and subfield-truncated (indexed by the set $N$) transition probabilities as a function of interaction time $\Omega_\mathrm{A} T$ for a two-level atom in a cylindrical cavity with $R/\sigma = 20$, $L/R = 10^5$, corresponding to a typical waveguide. 
The error for spontaneous emission is given by $\delta_{N,(-)}$ and for vacuum excitation by $\delta_{N,(+)}$.
The left column shows  Gaussian switching, the right top-hat switching. In (a-b):  Atom is strongly off-resonant with all subfields ($\Omega_\mathrm{A} = 6 \cdot 10^{12} \mathrm{s}^{-1} \ll \min_{\bm j} \omega_{\bm j}$); spontaneous emission and vacuum excitation coincide.  In (c-f): Atom is resonant with the 5-th subfield ($\Omega_\mathrm{A} = \omega_{(5,0),2}$). The (most-)resonant subfields are indicated. An arrow shows the direction of increasing number of subfields  from the resonant one.}
\label{MeasureWireFig}
\end{figure*}

\begin{figure*}
\includegraphics[width=14.2cm]{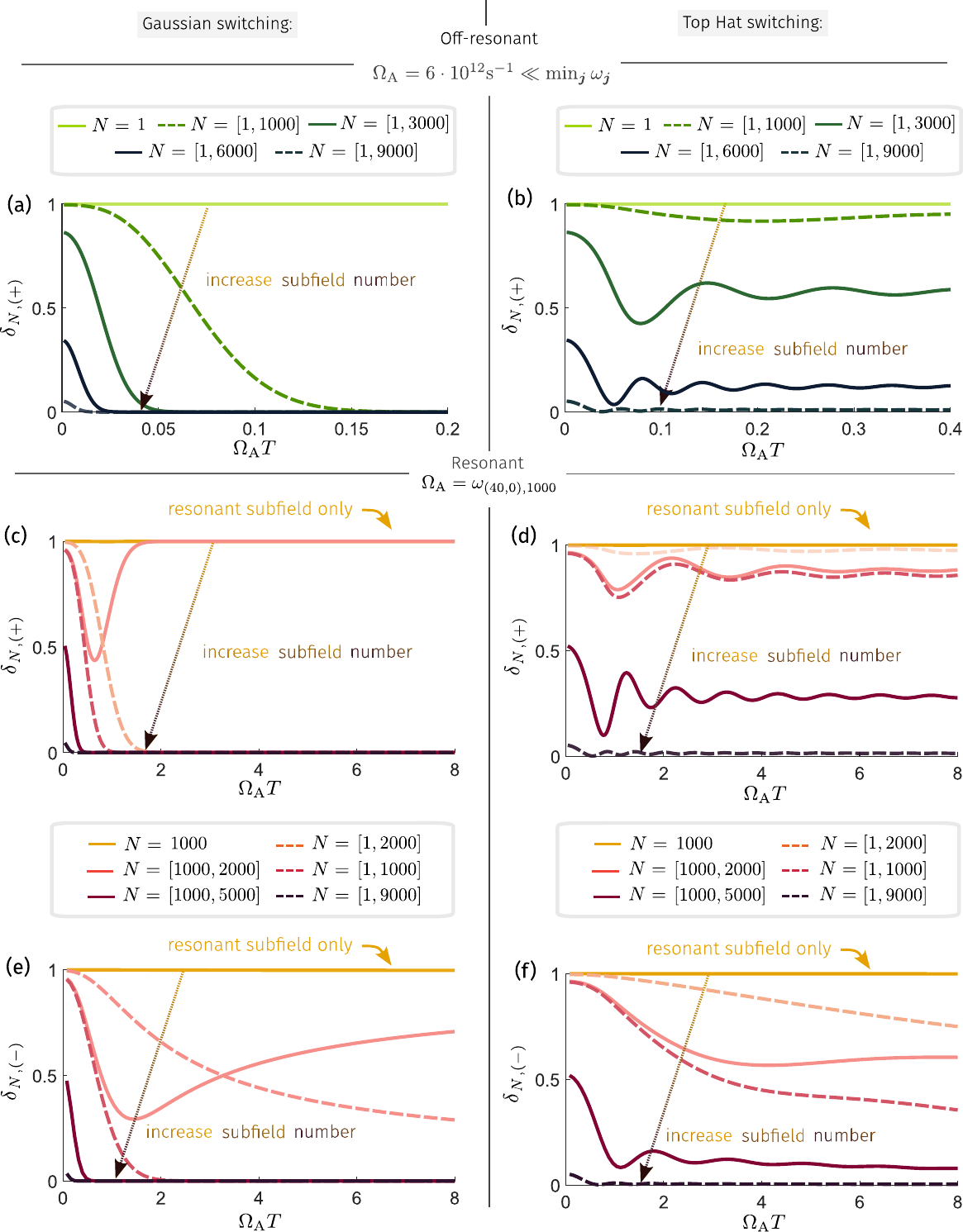}
\caption{Relative difference, Eq.~\eqref{deltaCyl}, between full and subfield-truncated (indexed by the set $N$) transition probabilities as a function of interaction time $\Omega T$ for a two-level atom in a cylindrical cavity with $R/\sigma = 10^4$, $L/R = 10^2$, corresponding to a typical optical resonator. 
The error for spontaneous emission is given by $\delta_{N,(-)}$ and for vacuum excitation by $\delta_{N,(+)}$.
The left column shows  Gaussian switching, the right top-hat switching. In (a-b):  Atom is strongly off-resonant with all subfields ($\Omega_\mathrm{A} = 6 \cdot 10^{12} s^{-1} \ll \min_{\bm j} \omega_{\bm j}$) ; spontaneous emission and vacuum excitation coincide.  In (c-f): Atom is resonant with the 40-th subfield ($\Omega_\mathrm{A} = \omega_{(40,0),1000}$). The (most-)resonant subfields are indicated. An arrow shows the direction of increasing number of subfields from the resonant one. }
\label{MeasureNotExactlyWireFig}
\end{figure*}

\subsection{Numerical Results}
\label{NUMEST}

We discuss the numerical results for the validity of the subfield approximation in terms of the four dimensionless parameters $L/R$, $R/\sigma$, $\Omega_{\mathrm{A}} T$, and $\omega_{\bm j}/\Omega_{\mathrm{A}}$. First, we investigate the geometric imprints (i.e., effects induced by varying the parameters $R/L$, $R/\sigma$, and consider different resonance conditions $\omega_{\bm j}/\Omega_{\mathrm{A}}$). 
Second, the impact of the dynamics, i.e., in terms of $\Omega_{\mathrm{A}}T$ and different switching functions, are studied.
Therefore, we particularize the relative error Eq.~\eqref{delta} to our example, reading 
\begin{align}
\delta_{N,(\pm)} =P^{-1}_{(\pm)}  | P_{(\pm)}  - P_{N,(\pm)}    |,
\label{deltaCyl}
\end{align}
where $P_{N,(\pm)}   = P_{(N,0),(\pm)}  $ is the transition probability defined in Eq.~\eqref{NPROBRED} with the azimuthal mode number being fixed at zero, i.e., $m_2=0$, and $m_1 \in N$.

\subsubsection*{Imprint of the Geometry}

Here, we only discuss spontaneous emission, where the impact of the geometry is particularly pronounced, and drop for simplicity the $(-)$ index in this section and  Fig.~\ref{Ratios}. 
First of all, see Fig.~\ref{Ratios}(a)-(b) for the example of Gaussian switching, the individual subfield probability amplitudes $|c_{(m_1,0)}|^2$ are insensitive to the (intuitively important) ratio $L/R$. This can be understood, cf. Eq.~\eqref{PREDTH}, by noting that the ratio $L/R$ appears only in the field frequencies. For the considered parameter space with $L/R \gg 1$, then, the most resonant subfield $m_1^\mathrm{res}$ has the property that $\chi^{\mathrm{res}}\gg 2\pi l^\mathrm{res} R/L$, where $\chi^{\mathrm{res}} \vcentcolon = \chi_{m_1^\mathrm{res},0}$. Note that the resonant mode is identical over the whole parameter regime from optical cavity to waveguide in these plots because it is only weakly affected by $L/R$. 
Nonetheless,  the number of subfields necessary for a proper subfield approximation are impacted by the chosen resonant subfield $m_1^\mathrm{res}$. To keep the error $\delta_{N}$ fixed, the number of subfields needed grows linearly with the resonant subfield  $m_1^\mathrm{res}$: for $\delta_{ N} \approx 0.1$, Fig.~\ref{Ratios}(a) with $m_1^\mathrm{res}=10$ requires $24$  and (b)  with $m_1^\mathrm{res}=100$ requires $240$ subfields.   

The parameter $R/\sigma$, on the other hand, is considerably more relevant to the truncation (see Fig.~\ref{Ratios}(c)-(f)). The subfield with the maximum contribution $m^\text{max}_1$ to the transition probabilities, which is distinct from the most resonant subfield $m_1^\mathrm{res}$, grows linearly with $R/\sigma$:  \mbox{$m^\text{max}_1 \approx 2R/(\pi \sqrt{2}\sigma)$} for \mbox{$R/\sigma  \,\le \, \pi m_1^{\mathrm{res}}/ \sqrt{2}$} . This is due to the geometrical parameter $R/\sigma$ dominating the energetic factor until the point of the resonance condition $\omega_{\bm j}=\Omega_{\mathrm{A}}$ is reached. From there the energetic part starts to dominate and, for interaction times $\Omega_\mathrm{A} T \approx 1$ and spontaneous emission,
\begin{align}
m^\text{max}_1 \approx \frac{2 \chi^{\mathrm{res}}}{\pi (\Omega_\mathrm{A} T)^2},
\label{LIMITI}
\end{align}
cf. App.~\ref{Asymptotics} for a derivation of $m^\text{max}_1$ for both vacuum excitation and spontaneous emission. 
In particular, due to the impact of the cavity geometry the maximum transition probability lies with a subfield higher in energy than predicted from the resonance condition; in our examples of Fig.~\ref{Ratios}(c)-(f) this amounts to about two times the resonant mode number.

On top of that, the number of significant subfields for a given error $\delta_{N}$ grows linearly with $R/\sigma$ until the resonance condition is reached. After that, the number of subfields needed stays approximately constant, being proportional to  $m_1^\mathrm{res}$, as was already found for Fig.~\ref{Ratios}(a)-(b). Moreover, the number of subfields needed for a given error $\delta_{N}$ and fixed $R/\sigma$ grows linearly with the resonance frequency. This can be seen, for instance, when comparing $\delta_{N}$ between Fig.~\ref{Ratios}(d) and (f). 
Thus in order to obtain the same accuracy for the number-of-subfield approximation, for the optical resonator ($R \gg \sigma$) a significantly higher number of subfields is required than for the waveguide ($R\gtrapprox \sigma$).

\subsubsection*{Imprint of the Dynamics}

Next, we investigate the influence of the dynamical process when the interaction time is varied.
We consider two regimes of the cylindrical cavity; first, a waveguide ($R=20\sigma$, \, $R \ll L$)~\cite{WG1,WG2,WG3,WG4,WG5} in Fig.~\ref{MeasureWireFig}, and secondly an optical resonator ($R = 10^4 \sigma, \, R < L$)~\cite{Kessler,Trupke,Herrmann} in Fig.~\ref{MeasureNotExactlyWireFig}. 
Even though the parameter $R/\sigma$ determines an upper bound of how many subfields generally might have to be included, as discussed in the previous section, the actual dynamical process can indeed result in a much improved accuracy for a lower number of subfields. When considering the case of a waveguide 
and studying the off-resonant coupling between atom and subfields with \mbox{$\Omega_\mathrm{A} \ll \min_{\bm j} \omega_{\bm j}$}, cf. Fig.~\ref{MeasureWireFig}(a) and~\ref{MeasureWireFig}(b), the processes of spontaneous emission and vacuum excitation are identical. For Gaussian switching, it is indeed possible to just choose  one subfield when considering interaction times that are sufficiently long. For shorter times, nonetheless, it is possible to chose just a few -- where of course more need to be taken into account for shorter times. For sudden switching, this is never the case and one is always required to choose several subfields for a low enough error. 

In the case of resonant atom-field coupling, cf. Fig.~\ref{MeasureWireFig}(c)-(f) (with $\Omega_{A} = \omega_{(5,0),2}$),  we see that again for spontaneous emission with  Gaussian switching one subfield may suffice in the long interaction time regime. Now, however, for vacuum excitation in Fig.~\ref{MeasureWireFig}(c) and~\ref{MeasureWireFig}(d) the error does not converge to zero with the one resonant subfield and may not even diminish when considering the near-resonant subfields (there emerges a unique interaction time where the error may assume a minimum). This can be understood from the exponential factor in Eq.~\eqref{PREDTH}. Therefore, the subfield truncation is generally better for spontaneous emission than for excitation processes. For small interaction times the impact of the resonant term is comparatively small leading to an increase of the number of subfields needed. 
As before, with sudden switching it is not possible to just choose the one resonant subfield to arrive at a reasonable approximation.

Studying now the optical resonator ($\sigma \ll R< L$), we see first of all that, in order to achieve the same error as for the waveguide, we need significantly more subfields. This is, as noted in the previous section (cf. Fig.~\eqref{Ratios}), related to the  ratio of $R/\sigma$ having increased. One can observe that the error has  a minimum for the case of spontaneous emission. This is due to the fact that the larger the ratio $R/\sigma$, the more the maximum subfield looses its unique status as term which predominantly influences the transition probability. 
Secondly, even for off-resonant Gaussian switching the single-subfield approximation may no longer be viable for long interaction times. The same holds for resonant Gaussian interactions; and only a sufficient number of subfields result in a diminishing truncation error. For sudden switching, again, we observe a generally oscillating error that does not reduce in time, both resonant and off-resonant.

\section{Lasers}
\label{LASERSEC}
Previously, we developed the formalism for dimensional reduction for closed cavities where the modes are compactly supported. In the following, we see that it can be extended to more general setups such as lasers where the modes decay sufficiently fast in transverse direction.
In the case of a laser beam mostly one or a few modes are being significantly excited~\cite{Lugiato}. This results in the increased coupling between the pumped modes and matter, which in general highly exceeds the coupling to the unpumped vacuum modes discussed in the previous section. 
Considering a predominant direction of propagation in $z$ and a sufficiently quickly decaying amplitude $\bm A_{\bm m,\mu}(\rr,k)$ in transverse direction, the modes take the form
\begin{align}
\bm u_{\bm m,\mu}(\rr,k) = \bm A_{\bm m,\mu}(\rr,k) \e^{\ii k z},
\label{PARA}
\end{align}
with a continuous wave vector $|\bm k| = k$ in the $z$ direction.
The amplitude satisfies a paraxial wave equation that can be achieved from the Helmholtz equation~\eqref{HELMHOLTZEQ} via a slowly varying envelope approximation~\cite{Pampaloni, Svelto, Kogelnik, Singh, Deutsch}:
\begin{align}
\left[ \Delta_{\Gamma}+ 2\ii k \partial_\mathrm{z} \right] \bm A_{\bm m,\mu}(\rr,k)  = 0,
\label{PARAXIALWAVEEQUATION}
\end{align}
where $\Delta_{\Gamma}$ is, again, the transverse Laplacian but now on the unbounded domain. Note that, in contrast to the  Helmholtz equation, the paraxial wave equation is no longer self-adjoint. 

The paraxial modes~\eqref{PARA} emerge from the zeroth order expansion in $(\sqrt{2}k w_0)^{-1}$ of the paraxial wave equation and thus solve Maxwell's equations only up to this order. By taking higher terms into account longitudinal polarization, e.g., to first order, and cross polarization, e.g., to second order, are introduced. A slightly different method to obtain higher orders is implemented by expanding the momentum-space wave function in terms of a small opening angle representing perturbations from the central momentum along the propagation axis~\cite{Deutsch}, leading to the same results of cross polarization and longitudinal fields. 
Unfortunately, going to second order the orthogonality of the modes is lost, causing couplings between laser modes of nonequal mode numbers~\cite{Singh}. However, for large cross sections compared with the beam's wavelength, a zeroth-order approximation, which is most commonly found in the literature~\cite{Grynberg,Svelto,Renk,Kogelnik,Pampaloni}, gives robust results.

\begin{figure}
\begin{center}
\includegraphics[width=7.5cm]{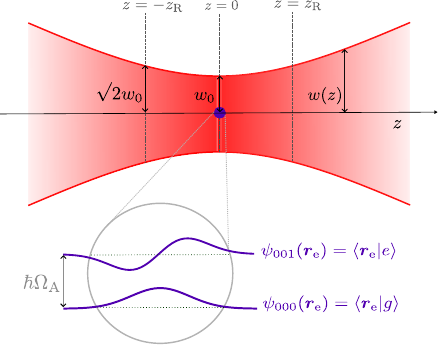}
\end{center}
\caption{Intersection in the $(x,y)$-plane of a Gaussian beam with focus at $z = 0$ interacting with matter modeled by a Gaussian two-level system with ground state $\psi_{000} (\bm r_\mathrm{e})$ and excited state $\psi_{001}(\bm r_\mathrm{e})$ in center of mass coordinates $\bm r_\mathrm{e}$ and an energy gap of $\hbar \Omega_\mathrm{A}$. The beam is symmetric in the $z = 0$-plane in which the beam radius $w_0$ is defined.}
\label{GAUSSIANBEAM}
\end{figure}

Most lasers generate electromagnetic waves with rectangular symmetry \cite{Singh} which are given by Hermite-Gaussian wave profiles. Then, the $L^2$ normalized solutions of~\eqref{PARAXIALWAVEEQUATION} can be expressed in Cartesian coordinates as
\begin{align}
\begin{split}
\bm A_{\bm m,\mu} (\bm r,k)  = &\frac{\mathrm{H}_{m_1}  \!\left( \frac{\sqrt{2}x}{w(z)} \right) \! \mathrm{H}_{m_2} \!\! \left( \frac{\sqrt{2}y}{w(z)} \right) \e^{- \frac{x^2+y^2}{w^2(z)}} \e^{\ii \theta_{\bm m}(\rr,k) }}{\sqrt{2^{m_1 +m_2-1} m_1! m_2! \pi w^2(z)}}\bm \epsilon_{\mu},
\end{split}
\label{MODEHERMITE}
\end{align}
with $H_{m_i}$ being the Hermite polynomial of order $m_i$ and $\bm \epsilon_\mu$ being the polarization. The beam contour (see Fig.~\ref{GAUSSIANBEAM})
\begin{align}
w(z) = w_0 \sqrt{1 + \left( \frac{z}{z_\mathrm{R}} \right)^2}
\label{BEAMWAIST}
\end{align}
is determined by the beam radius $w_0$
and the Rayleigh length 
$z_\mathrm{R} = k w_0^2/2.$
The phase $\theta_{\bm m} (\bm r,k)$ is given by
\begin{align}
\theta_{\bm m} (\rr,k) &=  k \frac{x^2+y^2}{2 R(z)}  - (m_1 + m_2 +1) \psi_\mathrm{G}(z),
\label{HERMITEPHASE}
\end{align}
where we define the radius of curvature of the phase front and the Gouy phase, respectively:
\begin{align}
R(z) &= z\left( 1 + \frac{z_\mathrm{R}^2}{z^2} \right), ~~~\psi_\mathrm{G}(z) = \arctan\left(\frac{z}{z_\mathrm{R}}\right).
\label{RADIUSOFCURVATURE}
\end{align}
Note that the following considerations are not restricted to our example but can also be applied to general laser modes such as Laguerre-Gaussian beams.

From these modes the quantized electric field of the laser can be constructed, reading
\begin{align}
    \hat{\E}(\bm r,t) &= \ii \sum_{\bm m,\mu} \int_\mathbb{R} \dd k \, \sqrt{\frac{\hbar \omega}{2\varepsilon_0}} \left[ \aop_{\bm m, \mu} (k)\bm u_{\bm m, \mu}(\rr) \mathrm{e}^{i \omega t} - \mathrm{H.c.} \right]\nonumber\\ 
    &= \sum_{\bm m} \left[\hat{\mathbfcal{E}}^{(+)}_{\bm m}(\bm r,t) + \hat{\mathbfcal{E}}^{(-)}_{\bm m}(\bm r,t)\right], \label{ELF}
\end{align}
where $\omega~= \omega (k)= c |k|$. Only at higher orders of the paraxial approximation is the frequencies' degeneracy in the transversal mode numbers $\bm m$ lifted~\cite{Deutsch}. Here, the frequencies observe an infinite level of degeneracy for each $k$, which will prevail throughout the dimensional reduction procedure.
Accordingly, the  field Hamiltonian reads 
\begin{align}
\hat{H}^\mathrm{field} = \sum_{\bm m,\mu}  \int_\mathbb{R} \dd k \,\hbar \omega \hat{a}^\dagger_{\bm m,\mu} (k) \aop^{\vphantom{\dagger}}_{\bm m,\mu}(k). 
\label{HAMINTFULLLASERFIELD}
\end{align}
With the quantized electromagnetic fields and the field Hamiltonian of the laser at hand, a reduction to 1D fields and the 1D dynamics can be performed.

\subsection*{Dimensional Reduction of a Hermite-Gaussian Beam}

Recall that the dimensional reduction requires (if one is interested in analytic results) the separability of modes, cf. Eq.~\eqref{IMPORTANTCONDITION} and Eq.~\eqref{IMPORTANTCONDITIONVMODES}. Due to the phase term~\eqref{HERMITEPHASE} and the $z$-dependence of the beam waist~\eqref{BEAMWAIST}, separability does not apply to the general form of the laser modes defined in Eq.~\eqref{MODEHERMITE}. However, we may consider the matter content to be strongly localized around the center of the beam, i.e., via a long-wavelength approximation: 
\begin{align}
\norm{\frac{\hat{x}_\mathrm{e}}{w_0}} \ll 1 \hspace{.2cm}  \land \hspace{.2cm} 
\norm{\frac{\hat{y}_\mathrm{e}}{w_0}} \ll 1 \hspace{.2cm} \land \hspace{.2cm}  \norm{\frac{\hat{z}_\mathrm{e}}{z_\mathrm{R}}} \ll 1.
\label{LARGEBEAMWAISTAPPROXIMATION}
\end{align}
with $\hat{x}_\mathrm{e}$, $\hat{y}_\mathrm{e}$ and $\hat{z}_\mathrm{e}$ being the electronic position operators, and $\norm{\hat{A}} = \expval{\hat{A}}{\psi}/\braket{\psi}{\psi}$.
Additionally, we require sufficiently low mode numbers,
\begin{align}
\max_{\bm m} (m_1 + m_2 +1) \ll \norm{\frac{\hat{z}_\mathrm{e}}{z_\mathrm{R}}}^{-1},
\label{APPRLAME}
\end{align}
such that the phase $\theta_{\bm m} (\rr,k)$ becomes independent of $z$. Note that higher-order corrections to the paraxial equation will have to be taken into account long before this condition is violated~\cite{Singh,Deutsch}.
Under the given assumptions, we get laser modes which separate in transversal scalar modes and a plane wave in the longitudinal direction, reading 
\begin{align}
\bm u_{\bm m,\mu} (\bm r,k) &= \varphi_{\bm m}(x,y) \e^{\ii kz} \bm \epsilon_\mu,
\label{PHIHERMITE}
\end{align}
where we defined 
\begin{align}
\varphi_{\bm m} (x,y) = &\frac{ \mathrm{H}_{m_1}  \left( \frac{\sqrt{2}x}{w_0} \right)  \mathrm{H}_{m_2}  \left( \frac{\sqrt{2}y}{w_0} \right) \e^{- \frac{x^2+y^2}{w_0^2}}}{\sqrt{2^{m_1 +m_2-1} m_1! m_2! \pi w_0^2}},
\label{TRANVMODEHERMITEAPPR}
\end{align}
which are $L^2$ orthonormalized on $\mathbb{R}^2$.
The magnetic field modes $\bm v_{\bm m,\mu} (\rr, k)$ of the laser are obtained fully analogously to Eq.~\eqref{PHIHERMITE} with the same functional form except for a polarization that is orthogonal to the electric mode to leading order in the paraxial approximation~\cite{Singh}.
 
We are now in the position to perform the dimensional reduction, which in contrast with the closed cavity has a simpler form due to the scalar nature of the longitudinal and transverse components. 
Following Eq.~\eqref{SUREANOTHER}, a map defined by the transversal modes~\eqref{TRANVMODEHERMITEAPPR} of the laser 
%\begin{align}
%\mathbfcal{P}^{(\mathrm{L})} [\bullet] = \sum_{\bm m} \int_{\mathbb{R}^2} \dd x \dd y \, \varphi_{\bm m}^\dagger (x,y) \, \bullet,
%\end{align}
yields dimensionally reduced plane-wave modes 
\begin{align}
\begin{split}
\bm u_{\mu} (z,k) &= \sum_{\bm m'} \inner{\varphi_{\bm m'} (x,y)}{\bm u_{\bm m,\mu}(\rr,k)}_{\mathbb{R}^2}
=  \mathrm{e}^{ikz} \bm \epsilon_\mu.
\end{split}
\label{SUREANOTHERLASER}
\end{align}
Analogously to~\eqref{DIMREDE} we map the complete transverse ancilla basis onto the 3D laser field 
\begin{align}
\hat{\E}(z,t) &= \!\sum_{\bm m} \! \int_{\mathbb{R}} \!\! \dd k \! \left(\!\sum_{\bm m'} \inner{\varphi_{\bm m'} (x,y)}{\hat{\mathbfcal{E}}^{(+)}_{\bm m}(\bm r,k,t)}_{\mathbb{R}^2} \! +\text{H.c.} \!\!\right) \nonumber\\
&= \sum_{\bm m} \hat{\E}_{\bm m}(z,t), 
\label{DECONPELASER}
\end{align}
with the subfields for the electric field of the laser being
\begin{align}
\hat{\E}_{\bm m}(z,t) = \! \ii \! \sum_\mu \! \int_\mathbb{R} \! \dd k \, \sqrt{\frac{\hbar \omega}{2 \varepsilon_0 }} \left[\aop_{\bm m,\mu}(k)  \e^{i(kz - \omega t)} \bm \epsilon_\mu \! - \mathrm{H.c.}\right].
\label{ELFMLASER}
\end{align}
Decomposing the  fields~\eqref{ELFMLASER} in positive and negative frequency parts, we can apply the same dimensional reduction method as already shown in Sec.~\eqref{DIMREDH}. Therefore, the dipolar interaction Eq.~\eqref{HDEE} decomposes into a sum of subfield Hamiltonians with the dimensionally reduced smearing functions (cf. Eq.~\eqref{KNMSS}) 
\begin{align}
\Smear(\ze) = 
\inner{\varphi^\dagger_{\bm m} (x_\mathrm{e},y_\mathrm{e})}{  \bm F_{\bm s\bm s'} (\re)}_{\mathbb{R}^2}.
\label{KNMSSLaser}
\end{align}

\subsection*{Example: Atom Interacting with a Laser}

We consider the electric dipole interaction of Eq.~\eqref{HDEE} (in the limit of $V \to \infty$) and choose  again a Gaussian two-level atom.
The calculation of the transition probabilities follows very closely the derivation from Sec.~\ref{MEASUREOFVALIDITY}.
Due to selection rules we assume that, for a linear polarization in the $x$ direction, the laser, with coherent-state amplitude $\alpha (k)$, is pumping into the $\text{TE}_{10}$ mode which we denote as $\bm \nu = (1,0,\epsilon_x)$, assuming  a sufficiently small bandwidth. 
Thus, for the atom being centered in the laser beam, the transition probabilities read to first order (see App.~\ref{LASERSMEARAPP} for details) 
\begin{align}
 &P_{\bm N,\epsilon_x,(\pm)} 
 \approx
 |c_{\bm \nu, (\pm)}|^2 +
 \sum_{\bm m \neq (1,0)}^{\bm N}  |c_{\bm m,\epsilon_x, (\pm)}  |^2 
 \label{ISTHATIT}
 \\
 &\quad= 
 g \left( |\alpha(k)|^2  \left| f_{(-)} (T)  +  \bar{f}_{(+)}(T) \right|^2 + \gamma_{\bm N}  \left|f_{(\pm)} (T)\right|^2\right), \nonumber
\end{align}
where $|c_{\bm \nu, (\pm)}  |^2 $ is the transition amplitude due to the laser mode and $|c_{\bm m,\epsilon_x, (\pm)}  |^2$  for the vacuum modes up to the mode numbers $\bm{N} = (N_1,N_2)$. The interaction-time-dependent function is 
\begin{align}
f_{(\pm)} (T)  =    
\begin{cases}
T \mathrm{sinc} ( \Delta_{(\pm)} T ),\hspace{.5cm} \, \chi(t) = \chi^{\mathrm{TS}}(t),\\
\sqrt{2\pi }T \e^{- \left(\Delta_{(\pm)} T\right)^2},  \, \chi(t) = \chi^{\mathrm{GS}}(t),\\
\end{cases}
\label{TIMEINTSECL}
\end{align}
with $\Delta_{(\pm)} = ( \omega \pm \Omega_\mathrm{A})/2$ being the detuning, which is independent of the mode numbers in the zeroth order paraxial wave approximation. 
Moreover, we defined the dimensionless coupling constant $g$ and the mode-number dependent function $\gamma_{\bm N}$: 
\begin{align}
g &=\frac{3 c^2 k^3 \sigma^6}{\hbar \varepsilon_0 \pi^4 \omega_0^4} \e^{-\frac{(k \sigma)^2}{2}}, \label{GCOUPL}
\\
\gamma_{\bm N} &= \frac{1}{3} \left(\frac{ 4\Gamma\left( \frac{5}{2} + N_1 \right)\Gamma\left( \frac{3}{2} + N_2 \right)  }{ 3\pi \Gamma\left( 1 + N_1 \right) \Gamma\left( 1 + N_2 \right)  }  - \frac{3}{4} \theta_1(N_1-1) \right),
\nonumber
\end{align}

where we employed the (nonstandard) Heaviside function with \mbox{$\theta_1(0)=1$}. 

The first term in Eq.~\eqref{ISTHATIT} originates from the laser, and the second  one comprises the contributions from the  vacuum modes.   To quantify their respective contribution, we define a measure $\zeta_{\bm N,(\pm)}$ which can be upper bounded: 
\begin{align}
\,\,\zeta_{\bm N, (\pm)} = \sum_{\bm m \neq (1,0)}^{\bm N}  \frac{|c_{\bm m,\epsilon_x, (\pm)}  |^2 }{|c_{\bm \nu, (\pm)}|^2 }  \le\frac{\gamma_{\bm N} }{4 |\alpha (k)|^2}.
\label{DELTALASER}
\end{align}
Consequently, a single-subfield approximation is indeed a very good approximation for strong lasers (cf. App.~\ref{LASERSMEARAPP}).
This is, for example, achieved in matter-wave interferometry where high-powered laser pulses (e.g., $6\,$W-$8\,$W~\cite{HerrmannLaser} or also up to $43\,$W~\cite{Kasevich}) in the Terahertz regime are applied producing average photon numbers of order $10^{20}$, exceeding significantly the coupling of the vacuum modes.

\section{Conclusion and Outlook}

In this paper we develop a systematic procedure to dimensionally reduce 3D quantum optical models; thereby yielding a precise measure of how the reduction  can be applied as an approximation. To that end, we decompose the electromagnetic modes into appropriate ancilla bases. The mapping of the modes onto an ancilla basis results in the elimination of the  corresponding spatial  dimension(s). 
The reduced modes obey independent, lower-dimensional Helmholtz equations which extend to the common interactions with matter. 
We show how lower-dimensional models emerge which comprise effective subfields yet constitute an exact reformulation of the 3D problem. An overview of the procedure can be found in Fig.~\ref{FIGFIELDS}.

By construction, this is generally not identical to the often \textit{ad hoc} applied approximation in the literature. However, by defining a measure with respect to some observable, for instance the relative difference to the statistics of an experiment, we are able to provide a handle on how good of an approximation the dimensional reduction should be; i.e., as a kind of number-of-mode approximation which is ubiquitously done in quantum optics. We show at the example of a two-level atom inside a cylindrical cavity that the number of subfields needed strongly depends on the parameters of the joint system and its dynamics. In general however a single subfield is not sufficient when considering the environment of the field's vacuum fluctuations. In particular, we find that the intuitive idea of a very long and narrow cavity as a 1D model is not easily justifiable.
We further apply the dimensional reduction to laser beams and see that a single-subfield approximation can be valid given the laser intensity is strong enough. This provides a justification for the standard approach.

Even though the framework established here applies to a wide range of setups commonly found in the literature, we only study a handful in detail.
For instance, we only consider  stationary atoms, yet motion (as was done for a toy model in~\cite{Lopp3}) or more generally the quantization of the center of mass of the atom may also be considered (the dimensional reduction of the electric dipole interaction for a fully quantized atom is derived in App.~\ref{MULTIPOLEAPP}). This extension could then be of interest to dynamical applications such as cavity-based atom interferometry in which high laser energies are focused into a very narrow area~\cite{Deutsch2}. 
Due to technical limitations in the stability of the modes one can excite cavity modes transverse to the laser propagation, leading to an adversarial impact on the experiments~\cite{CAVAI2,CAVAI3}. Consequently, one has to consider degrees of freedom that are not taken into account in the 1D case but could  be better understood by the procedure derived here. For that, one would have to combine the cavity and laser aspects discussed in this paper, i.e., by considering one or more lasers inside a cavity~\cite{BOOZER}. 

On the other hand, we see that the subfields still contain information about the full spectrum. Thus in metrological applications one could  imagine that by only measuring atoms one can reconstruct the cavity geometry or imperfections thereof~\cite{Grimmer}. 
Furthermore, the free choice of the transverse modes also suggests that the dimensional reduction can be applied to more general mode bases such as wavelets~\cite{Havu,Noikov,Unser}.  Additional setups might be of interest such as transmission lines interacting with superconducting qubits~\cite{McKay,Garcia}, or optomechanical interactions~\cite{Brenecke,ZHANG,Maas} where lower-dimensional models are commonly found. Lastly, treating leaky cavities within this formalism was beyond the scope of this paper and will undoubtedly be studied in the future.

\section*{Acknowledgements}
The authors would like to thank W. P. Schleich, M. Zych, H. M\"uller, A. Wolf, D. Fabian, M. Efremov and  A. Friedrich for stimulating and helpful discussions. Additionally the authors would like to thank the referee for careful reading and for helpful feedback and comments improving this paper.  
The QUANTUS and INTENTAS projects are supported by the German Space Agency at the German Aerospace Center (Deutsche Raumfahrtagentur im Deutschen Zentrum für Luft- und Raumfahrt, DLR) with funds provided by the Federal Ministry for
Economic Affairs and Climate Action (Bundesministerium
für Wirtschaft und Klimaschutz, BMWK) due to an enactment of the German Bundestag under Grant Nos.  50WM2250D (QUANTUS+) and 50WM2178 (INTENTAS).  
RL acknowledges the support by the Science Sphere Quantum Science of Ulm University.

\appendix

\section{Constructing the Electromagnetic Modes from the Scalar Helmholtz Equation}
\label{SecB}
\renewcommand{\theequation}{A.\arabic{equation}}

Starting from the eigenspectrum of the scalar Laplacian and the requirement of an orthonormal mode basis we show in a straightforward and simple fashion how expressions for the components of the mode decompositions of Eqs.~\eqref{IMPORTANTCONDITION} and~\eqref{IMPORTANTCONDITIONVMODES} can be obtained  in terms of the scalar Laplacian's eigenfunctions for arbitrary geometries. In particular, we verify Eq.~\eqref{LONGFIELDMODESSINCOS} for the longitudinal (recall, with respect to the cavity's symmetry axis) components of this decomposition, and find general expressions for the transversal degrees of freedom and polarization vectors.
To this end, let us consider the scalar Helmholtz equation 
\begin{align}
    (\Delta+ \bm k_{\bm j,\mu}^2) \psi_{\bm j,\mu} (\bm r)=0, \label{scalar_Helmholtz}
\end{align}
which is solved by the eigenfunctions $\psi_{\bm j,\mu}$ of the Laplacian  with discrete eigenvalues $-\bm k_{\bm j,\mu}^2$ (cf. Eq.~\eqref{SEPMODE}) corresponding to the electromagnetic theory. The index $\mu$, which corresponds to the polarization for the electromagnetic theory, indicates here the solutions to the scalar Helmholtz equation under (so far not specified) different boundary conditions.
From the definition of the cavity volume as a Cartesian product of  transversal and longitudinal space, i.e., $V=\Gamma\times S$, one can use the separation ansatz~\cite{Boyer} 
\begin{align}
\psi_{\bm j,\mu} (\rr) = \psi_{\bm m,\mu} (\bm y) \psi^{\vphantom{\dagger}}_{l,\mu}(z).
\label{Scalar_decomp}
\end{align}
Then one has two independent scalar Helmholtz equations; one being the \textit{transverse Helmholtz equation}
\begin{subequations}
\begin{align}
(\Delta_{\Gamma} + \bm k_{\bm m,\mu}^2 ) \psi_{\bm m,\mu} (\bm y) &= 0,
\label{ScalarHEHOTranv}
\end{align}
where $\Delta_{\Gamma}$ is the Laplacian with respect to the transverse coordinates $\bm y$  only, 
and the other being the \textit{longitudinal Helmholtz equation},
\begin{align}
(\partial_z^2 + k_l^2) \psi_{l,\mu} (z) &= 0,
\label{ScalarHEHOLong}
\end{align}
\end{subequations}\noindent
where we made use of the wave-vector separation~\eqref{SEPMODE}. Here we set, without loss of generality, the constant emerging from the separation of variables to zero. Since this constant  just adds an energy offset to the spectrum, the resulting energy spectrum is altered. 
To arrive at the correct boundary conditions for the electromagnetic modes,  cf. Eq.~\eqref{FULLHEHOEQ}, we preordain $\psi_{l,\mu_1}(z)$ to obey Dirichlet boundary conditions and $\psi_{l,\mu_2}(z)$ to obey Neumann boundary conditions. Moreover, in App.~\ref{DERVBOUNDSCALAR} we derive the boundary conditions of the  scalar modes which enable us to explicitly determine the scalar modes for a given geometry.
Granted this, we can write -- without specifying the cross section -- the two longitudinal solutions as
\begin{align}
\psi_{l,\mu_1}(z) =  \sqrt{\frac{2}{L}} \sin (k_l z),\quad
\psi_{l,\mu_2}(z) =  \sqrt{\frac{2}{L}}  \cos (k_l z),
\label{SCALMODES}
\end{align}
where  $k_l = \pi l/L$ is the longitudinal wave-vector component. From the $L^2$ normalization for the complete solution Eq.~\eqref{Scalar_decomp} it follows for the longitudinal part 
\begin{align}
\int_S \dd  z \,\,\psi_{l,\mu}^\dagger(z) \psi^{\vphantom{\dagger}}_{l',\mu'}(z) = 
\begin{cases}
\delta_{l,l'} \,\,\,\,\,\, ,\text{if}\,\, \mu = \mu',\\
\delta_{\mu,\mu'}\,\,\, ,\text{if} \,\, l = l',
\end{cases}
\label{ScalIntLong}
\end{align}
and additionally for the transversal part 
\begin{align}
\int_\Gamma \dd^2 y\,\, \psi_{\bm m,\mu}^\dagger (\bm y) \psi_{\bm m',\mu'} (\bm y) = \begin{cases}
\delta_{\bm m,\bm m'} \,\,\,\,\, ,\text{if}\,\, \mu = \mu',\\
\delta_{\mu,\mu'}\hspace{.5cm} ,\text{if} \,\, \bm m = \bm m',
\end{cases}
\label{ScalIntTranv}
\end{align}
Once the scalar modes are garnered -- requiring to know the transversal geometry --, one can construct, see e.g.,~\cite{Hanson, BUHMANN}, or~\cite{Gumerov} for a detailed derivation, the electric and magnetic modes (in the absence of charges and currents) via
\begin{subequations}
\label{FIELDMODES}
\begin{align}
    \bm u_{\bm j, \mu_1}(\bm r)&= \alpha_{\bm j,\mu_1} \, \bm \nabla\times \bm e \psi_{\bm j,\mu_1} (\bm r),
    \label{UMU1CON}\\ 
              \bm v_{\bm j, \mu_1}(\bm r)&=\frac{\beta_{\bm j,\mu_1}}{|\bm k_{\bm j,\mu_1}|} \bm \nabla\times\bm \nabla\times \bm e \psi_{\bm j,\mu_1} (\bm r),
    \label{VMU1CON}
    \intertext{for polarization $\mu_1$, and analogously for the $\mu_2$ polarization}
    \bm u_{\bm j, \mu_2}(\bm r)&=\frac{\alpha_{\bm j,\mu_2}}{|\bm k_{\bm j,\mu_2}|}\bm \nabla\times\bm \nabla\times \bm e \psi_{\bm j,\mu_2} (\bm r),
    \label{UMU2CON}\\
             \bm v_{\bm j, \mu_2}(\bm r)&= \beta_{\bm j,\mu_2} \, \bm \nabla\times \bm e \psi_{\bm j,\mu_2} (\bm r),
    \label{VMU2CON}
\end{align}
\label{RECTONSTRUCTION}%
\end{subequations}
satisfying the boundary conditions of Eq.~\eqref{FULLHEHOEQ}. Here, $\bm e$ is an \textit{a priori} arbitrary unit vector (or pilot vector) such that different choices yield different basis sets for $\bm u_{\bm j, \mu} (\rr)$ and $ \bm v_{\bm j, \mu} (\rr)$. In general, however, the symmetry of the problem gives a preferred choice for $\bm e$. For our purpose, and assuming cavities with axial symmetry, selecting the unit vector $\bm e^{(z)}$ in longitudinal direction is most natural. Note that the normalization constants $\alpha_{\bm j,\mu},\,\beta_{\bm j,\mu}$ depend on the form of the unit vector $\bm e$. For our choice, we require
\begin{align}
\alpha_{\bm j,\mu} = \beta_{\bm j,\mu} =  |\bm k_{\bm m,\mu}|^{-1}
\label{ALPHABETANORM}
\end{align}
such that the electromagnetic modes are normalized on $\textbf{L}^2(V)$. 
Accordingly, the electromagnetic modes can be expressed as a differential operator acting on the scalar solutions of the Helmholtz equation~\eqref{scalar_Helmholtz}, appended with a suitable pilot vector. Only the further decomposition of this operator in combination with the separability of the scalar modes defines a separable problem. This is in contrast to Ref.~\cite{Lopp1} where the dimensional reduction is applied to a scalar theory and thus only the separability of the scalar modes is required. We see in the following that the lower-dimensional modes are given by a reduced differential operator acting on the scalar solutions \mbox{$\psi_{l, \mu}(z)$} for the longitudinal part or on~\mbox{$\psi_{\bm m,\mu} (\bm y)$} for the transverse part, respectively.

We are now in the position to find the explicit expressions for the elements of the electromagnetic mode decompositions of Eqs.~\eqref{IMPORTANTCONDITION} and~\eqref{IMPORTANTCONDITIONVMODES} in terms of the scalar solutions. We define the components of the gradient in the transverse directions, reading
\begin{align}
\partial_{(y_1)} = \bm e^{(y_1)} \cdot \bm  \nabla,\, \text{and} \quad \partial_{(y_2)} = \bm e^{(y_2)} \cdot \bm \nabla.
\label{NOTATIONDERV}
\end{align}
We want to emphasize that for general geometries the derivatives $\partial_{(y_i)}$ include \textit{Lamé coefficients}~\cite{MORSE} due to the possible curvilinear nature of the coordinates, for instance in cylindrical coordinates $\partial_{(r)} = 
\partial_r$,\, $\partial_{(\varphi)} = r^{-1} 
\partial_\varphi$.
Employing then the Laplacian identity 
\begin{align}
\bm \nabla \times \bm \nabla \times \bm A = \bm \nabla (\bm \nabla \cdot \bm A) -  \Delta \bm A,
\label{LAPLACEIDENTITY}
\end{align}
for an arbitrary vector $\bm A$, we can write Eqs.~\eqref{RECTONSTRUCTION} as 
\begin{subequations}
\begin{align}
    \bm u_{\bm j, \mu_1}(\bm r)&= |\bm k_{\bm m,\mu_1}|^{-1}\begin{pmatrix}
    \ertwo\\ -\erone \\0   
    \end{pmatrix} \psi_{\bm j,\mu_1} (\bm r)
    ,
    \label{scalar_solAPP}
    \\
    \bm u_{\bm j, \mu_2}(\bm r) &= |\bm k_{\bm m,\mu_2}|^{-1} |\bm k_{\bm j,\mu_2}|^{-1}
    \begin{pmatrix}
    \erone \partial_z\\
    \ertwo \partial_z \\
    \bm k_{\bm m,\mu_2}^2
    \end{pmatrix}
    \psi_{\bm j,\mu_2} (\bm r),
    \nonumber
    \\
    \bm v_{\bm j, \mu_2}(\bm r) &= |\bm k_{\bm m,\mu_2}|^{-1}
    \begin{pmatrix}
    \ertwo\\ -\erone \\0   
    \end{pmatrix}  \psi_{\bm j,\mu_2} (\bm r)
    ,
    \label{scalar_solBAPP}
    \\  
    \bm v_{\bm j, \mu_1}(\bm r) &=|\bm k_{\bm m,\mu_1}|^{-1} |\bm k_{\bm j,\mu_1}|^{-1}  \begin{pmatrix}
    \erone \partial_z\\
    \ertwo \partial_z \\
    \bm k_{\bm m,\mu_1}^2
    \end{pmatrix} \psi_{\bm j,\mu_1} (\bm r),
    \nonumber
\end{align}
\label{RECONSTRUCTION}
\end{subequations}
where we used the transversal Helmholtz equation~\eqref{ScalarHEHOTranv}.
Inserting the scalar modes defined in Eq.~\eqref{ScalarHEHOLong}, with 
\begin{align}
\begin{split}
k_l \psi_{\bm j,\mu_2} (\bm r) &= \partial_z \psi_{\bm j,\mu_1} (\rr),\\ 
- k_l \psi_{\bm j,\mu_1} (\bm r) &= \partial_z \psi_{\bm j,\mu_2} (\rr),
\end{split}
\label{CONDSPSI}
\end{align}
Eqs.~\eqref{scalar_solAPP} and~\eqref{scalar_solBAPP} can be cast as a product of individual elements via Eq.~\eqref{IMPORTANTCONDITION} for the electric field modes and via Eq.~\eqref{IMPORTANTCONDITIONVMODES} for the magnetic field modes, respectively.
Therefore, we define the matrices of the transverse degrees of freedom as a superposition of the two polarization contributions (so that they may serve as ancilla basis later on for the dimensional reduction)
\begin{subequations}
\begin{align}
\mathbfcal{S}_{\bm m} (\bm y) &= 
\mathbfcal{S}_{\bm m,\mu_1} (\bm y) + \mathbfcal{S}_{\bm m,\mu_2} (\bm y)
\label{WHYMOREMATRICES1}\\
&=  
\begin{pmatrix}
&\ertwo & - \ertwo  &0 \\  &- \erone &\erone &0 \\ &0 &0 &0 
\end{pmatrix} \frac{\psi_{\bm m,\mu_1} (\bm y)}{\sqrt{2} |\bm k_{\bm m,\mu_1}|}\nonumber\\ 
&\hspace{.5cm}+ 
\begin{pmatrix}
&\erone & \erone &0 \\ &\ertwo &\ertwo &0  \\ &0 &0 & \sqrt{2}|\bm k_{\bm m,\mu_2}|  
\end{pmatrix} \frac{\psi_{\bm m,\mu_2} (\bm y)}{\sqrt{2} |\bm k_{\bm m,\mu_2}|}, \nonumber
\\
\mathbfcal{T}_{\bm m} (\bm y) 
&= \mathbfcal{T}_{\bm m,\mu_1} (\bm y) + \mathbfcal{T}_{\bm m,\mu_2} (\bm y) \label{WHYMOREMATRICES2}\\
&= 
\begin{pmatrix}
&\erone & \erone &0 \\ &\ertwo &\ertwo &0  \\ &0 &0 & \sqrt{2} |\bm k_{\bm m,\mu_1}|  
\end{pmatrix} \frac{\psi_{\bm m,\mu_1} (\bm y)}{\sqrt{2} |\bm k_{\bm m,\mu_1}|} \nonumber\\
&\hspace{.5cm}+ 
\begin{pmatrix}
&\ertwo & - \ertwo  &0 \\  &- \erone &\erone &0 \\ &0 &0 &0 
\end{pmatrix} \frac{\psi_{\bm m,\mu_2} (\bm y)}{\sqrt{2} |\bm k_{\bm m,\mu_1}|},\nonumber
\end{align}
\label{STDECOMPOSUITIONMATRICES}%
\end{subequations}
for the electric and magnetic field, respectively. Furthermore, the polarization vectors read
\begin{align}
\bm \epsilon_{\bm j,\mu_1}  &= 
\begin{pmatrix}
 &1/\sqrt{2}\\
 -&1/\sqrt{2}\\
 &0
\end{pmatrix},
\quad
\bm \epsilon_{\bm j,\mu_2}   = | \bm k_{\bm j,\mu_2}|^{-1}  \begin{pmatrix}
-k_l/\sqrt{2}\\
-k_l/\sqrt{2}\\
|\bm k_{\bm m,\mu_2}|
\end{pmatrix}\nonumber,\\
\bm \kappa_{\bm j,\mu_1}  &= | \bm k_{\bm j,\mu_1}|^{-1}  \begin{pmatrix}
k_l/\sqrt{2}\\
k_l/\sqrt{2}\\
|\bm k_{\bm m,\mu_1}|    
\end{pmatrix},
\quad
\bm \kappa_{\bm j,\mu_2} =
\begin{pmatrix}
 &1/\sqrt{2}\\
 -&1/\sqrt{2}\\
 &0
\end{pmatrix},
\label{WHYMOREMATRICES3}
\end{align}
and the matrices for the longitudinal degrees of freedom become (utilizing Eq.~\eqref{SCALMODES})
\begin{align}
\begin{split}
\mathbfcal{U}_{l} (z) &= \mathrm{diag} \left(
\psi_{l,\mu_1} (z),\,
\psi_{l,\mu_1} (z), \,
\psi_{l,\mu_2} (z)\right),\\
\mathbfcal{V}_{l} (z) &= \mathrm{diag} \left(
\psi_{l,\mu_2} (z),\,
\psi_{l,\mu_2} (z),\,
\psi_{l,\mu_1} (z)
\right).
\end{split}
\label{UPMATRIX}
\end{align}

Since the electromagnetic modes obey Eq.~\eqref{ID1}, the components of the magnetic field decomposition, Eq.~\eqref{IMPORTANTCONDITIONVMODES}, can be realized via simple replacement in the electric mode components. In particular, the transverse-mode matrices $\mathbfcal{T}_{\bm m}$ can be found by interchanging the polarization index, i.e., \mbox{$\mathbfcal{T}_{\bm m} (\bm y) =  \mathbfcal{S}_{\bm m} (\bm y)\Big|_{\mu_1 \leftrightarrow \mu_2}$}.
The polarization vectors $\bm \epsilon_{\bm j,\mu}$, however, depend not only on the transverse-mode numbers $\bm m$, but also on the longitudinal-mode number $l$. Hence, the component-wise orthogonality~\eqref{COMPONENTWISEULZ} of the longitudinal elements $\mathbfcal{U}_{l}(z)$ and $\mathbfcal{V}_{l}(z)$  must also be taken into account. For this reason one has to not only switch the polarization index of the transverse wave vectors $\bm k_{\bm m,\mu}$ to arrive at the corresponding polarizations $\bm \kappa_{\bm j,\mu}$ but also  the sign of the longitudinal wave vector $k_l$. Then one obtains $\bm \kappa_{\bm j,\mu_1}$ from $\bm \epsilon_{\bm j,\mu_2}$ by the substitution \mbox{$\{\bm k_{\bm m,\mu_2}, k_l\} \rightarrow \,\{\bm k_{\bm m,\mu_1}, - k_l\}$}  and vice versa.
The explicit electromagnetic modes can then be easily given in terms of the scalar Helmholtz modes by particularizing to a specific cavity geometry.

\section{Mapping of the Electromagnetic Modes and Orthonormality}
\renewcommand{\theequation}{B.\arabic{equation}}
\label{SEPPELFIELDMODES}

Here, we verify the orthonormality of Eqs.~\eqref{STID1} of the transversal matrices  $\mathbfcal{S}_{\bm m} (\bm y)$ that serve as ancilla modes (the results follow analogously for $\mathbfcal{T}_{\bm m} (\bm y)$) with which we arrive at the dimensionally reduced 1D modes of Eqs.~\eqref{PROJECTIONSUANDV}. We further show the orthonormality conditions for these reduced modes. The same will then be executed for the reduction to the 2D modes.

\subsection{Reduction to 1D Field Modes}
\renewcommand{\theequation}{B.1.\arabic{equation}}
\label{APPREDTO1D}

First, recall the decomposition of $\mathbfcal{S}_{\bm m} (\bm y)$  into the two polarization contributions in Eq.~\eqref{STDECOMPOSUITIONMATRICES}. 
With help of the scalar mode approach discussed in App.~\ref{SecB} and the use of the boundary conditions of the transverse scalar modes as shown subsequently in App.~\ref{DERVBOUNDSCALAR} we obtain by means of Green's first identity
\begin{align}
\begin{split}
&\int_\Gamma \dd^2 y\, \partial_{(y_2)} \bar{\psi}_{\bm m,\mu} (\bm y) \partial_{(y_1)} \psi_{\bm m',\mu'} (\bm y)\\ 
&\quad=\oint_{\partial_\Gamma} \dd y \, n_\Gamma^{(y_2)} \bar{\psi}_{\bm m,\mu} (\bm y)  \partial_{(y_1)}  \psi_{\bm m',\mu'} (\bm y)\\ 
&\quad\quad- \int_{\Gamma} \dd^2 y\,\bar{\psi}_{\bm m,\mu} (\bm y)  \partial_{(y_1)} \partial_{(y_2)}  \psi_{\bm m',\mu'} (\bm y) \\
&\quad=\oint_{\partial_\Gamma} \dd y \, n_\Gamma^{(y_2)} \bar{\psi}_{\bm m,\mu} (\bm y)  \partial_{(y_1)}  \psi_{\bm m',\mu'} (\bm y)\\ 
&\quad\quad- 
\oint_{\partial_\Gamma} \dd y \, n_\Gamma^{(y_1)} \bar{\psi}_{\bm m,\mu} (\bm y)  \partial_{(y_2)}  \psi_{\bm m',\mu'} (\bm y)\\
&\quad\quad+ \int_\Gamma \dd^2 y\, \partial_{(y_1)} \bar{\psi}_{\bm m,\mu} (\bm y) \partial_{(y_2)} \psi_{\bm m',\mu'} (\bm y) \\
&\quad =\int_\Gamma \dd^2 y\, \partial_{(y_1)} \bar{\psi}_{\bm m,\mu} (\bm y) \partial_{(y_2)} \psi_{\bm m',\mu'} (\bm y), 
\label{CANYOUFEELIT}
\end{split}
\end{align}
where we used for the infinitesimal surface vector $\dd \bm y = \bm n_\Gamma \dd y$, with the surface vector of the cavity cross section \mbox{$\bm n_\Gamma = (n_\Gamma^{(y_1)}, \, n_\Gamma^{(y_2)},\,0)$}. For identical polarizations it follows analogously by applying integration by parts and the orthogonality and boundary conditions
\begin{align}
\begin{split}
\sum_{\bm m'} &\int_\Gamma \dd^2 y \, \left( \partial_{(y_1)} \bar{\psi}_{\bm m',\mu} (\bm y) \partial_{(y_1)} \psi_{\bm m,\mu} (\bm y) \right.\\ 
&+ \left. \partial_{(y_2)} \bar{\psi}_{\bm m',\mu} (\bm y) \partial_{(y_2)} \psi_{\bm m,\mu} (\bm y)  \right) - \bm k_{\bm m,\mu}^2 = 0.
\end{split}
\label{NOICANT}
\end{align}
Then, we can write
\begin{subequations}
\begin{align}
&\int_\Gamma \dd^2 y \, \mathbfcal{S}^\dagger_{\bm m,\mu_1} (\bm y)  \mathbfcal{S}^{\vphantom{\dagger}}_{\bm m',\mu_1} (\bm y) =
\frac{\delta_{\bm m, \bm m'}}{2} 
\begin{pmatrix}
\,\,\,1 &-1 &\,\,\,0 \\ -1&\,\,\,\,1&\,\,\,0\\ \,\,\,0&\,\,\,\,0&\,\,\,0
\end{pmatrix}, \nonumber\\
&\int_\Gamma \dd^2 y \mathbfcal{S}^\dagger_{\bm m,\mu_2} (\bm y)  \mathbfcal{S}^{\vphantom{\dagger}}_{\bm m',\mu_2} (\bm y) = 
\frac{\delta_{\bm m, \bm m'}}{2}
\begin{pmatrix}
1 & 1 &0 \\ 1&1&0\\ 0&0&2
\end{pmatrix}. \quad
\end{align}
For distinct polarizations with $\mu \neq \mu'$, we find by means of Eq.~\eqref{CANYOUFEELIT}:
\begin{align}
\int_\Gamma \dd^2 y\, \mathbfcal{S}^\dagger_{\bm m,\mu} (\bm y)  \mathbfcal{S}^{\vphantom{\dagger}}_{\bm m',\mu'} (\bm y) = 0.
\end{align}
\label{ORTHGONALITYSMATRICES}%
\end{subequations}
These results imply the orthonormality relations~\eqref{STID1}.
%Via identity~\eqref{NMATRICES} the relations~\eqref{ORTHGONALITYSMATRICES} can be shown analogously for the matrices $\mathbfcal{T}_{\bm m,\mu} (\bm y)$ defined in Eq.~\eqref{WHYMOREMATRICES2}, yielding a proof for the orthogonality relation~\eqref{ORTHOGTT}. 
Consequently, the dimensionally reduced 1D modes defined in Eq.~\eqref{PROJECTIONSUANDV} are achieved in a straightforward manner:
\begin{subequations}
\begin{align}
    \bm u_{\bm j,\mu_1} (z) &= \sqrt{L}^{-1} 
    \begin{pmatrix}
        &\sin(k_l z) \\ 
        -&\sin( k_l z) \\ 
        &0
    \end{pmatrix}, \\
    \bm u_{\bm j,\mu_2} (z) &= (\sqrt{L}|\bm k_{\bm j,\mu_2}|)^{-1}
    \begin{pmatrix}
            -k_l \sin(k_l z) \\ 
            -k_l \sin(k_l z) \\ 
            \, \sqrt{2} |\bm k_{\bm m,\mu_2}| \cos(k_l z) 
    \end{pmatrix}, \nonumber \\
        \bm v_{\bm j,\mu_2} (z) &= \sqrt{L}^{-1} 
    \begin{pmatrix}
        &\cos(k_l z) \\ 
        -&\cos( k_l z) \\ 
        &0
    \end{pmatrix}, \\
    \bm v_{\bm j,\mu_1} (z) &= (\sqrt{L} |\bm k_{\bm j,\mu_1}|)^{-1}
    \begin{pmatrix}
            k_l \cos(k_l z) \\ 
            k_l \cos(k_l z) \\ 
            \,  \sqrt{2} |\bm k_{\bm m,\mu_1}| \sin(k_l z) \nonumber 
    \end{pmatrix}.
\end{align}
\label{UVEQUATIONSCOMPLETE}%
\end{subequations}
Importantly, for this choice of the polarization basis or equivalently of the pilot vector $\bm e$, one of the sets of 1D electric and magnetic modes still depends on the transverse-mode numbers $\bm m$ and polarization $\mu$ after dimensional reduction. 

For $\bm m = \bm m'$, i.e., for modes corresponding to the same subspace $\textbf{L}^2_{\bm m}(S)$, and via the normalization of $\mathbfcal{U}_l (z)$  and $\mathbfcal{V}_l (z)$ in Eqs.~\eqref{ORTHONORMALITYUANDVMATRICES} the orthonormality of $\bm u_{\bm j,\mu} (z)$ and $\bm v_{\bm j,\mu} (z)$ follows directly (cf. Eq.~\eqref{normCondlongmodespace}). For the orthogonality between the electric and magnetic modes we have 
\begin{align}
\begin{split}
&\int_S \dd z \,\,\bm u_{\bm j,\mu}^\dagger (z) \cdot \bm v_{\bm j,\mu}^{\vphantom{\dagger}} (z)\\ 
&\quad= \left( \bm \epsilon_{\bm j.\mu}^\dagger \!\!\! \cdot \bm \kappa_{\bm j,\mu}^{\vphantom{\dagger}} \right) \int_S \dd z \sin (k_l z) \cos(k_l z) = 0,  
\end{split}
\end{align}
whereby in the last step we used that the integral vanishes on $S$, proving Eq.~\eqref{UVSTR}.

\subsection{Reduction to 2D Field Modes}
\renewcommand{\theequation}{B.2.\arabic{equation}}
\label{APPREDTO2D}

The same can be done now for the dimensional reduction to 2D. 
Here the normalization of the longitudinal matrices (that now serve as ancilla modes) $\mathbfcal{U}_l (z) $ and $\mathbfcal{V}_l (z) $ in Eqs.~\eqref{ORTHONORMALITYUANDVMATRICES} follows directly from their definition in Eqs.~\eqref{LONGFIELDMODESSINCOS}.
Then, the normalized modes solving the transverse Helmholtz Eq.~\eqref{INITM1} on $\Gamma$ can be constructed via Eq.~\eqref{NONONOTANOTHER}:
\begin{subequations}
\begin{align}
    \bm s_{\bm j,\mu_1} (\bm y) &= |\bm k_{\bm m,\mu_1}|^{-1} 
    \begin{pmatrix}
        &\ertwo \\ 
        -&\erone \\ 
        &0
    \end{pmatrix}\psi_{\bm m,\mu_1} (\bm y),\label{CURLYMMATRICES}\\ 
\bm s_{\bm j,\mu_2} (\bm y) &= |\bm k_{\bm m,\mu_2}|^{-1} |\bm k_{\bm j,\mu_2}|^{-1}
    \begin{pmatrix}
            -k_l \erone \\ 
            -k_l \ertwo \\ 
            \,  \bm k_{\bm m,\mu_2}^2 
    \end{pmatrix}
    \psi_{\bm m,\mu_2} (\bm y).\nonumber
\end{align}
Likewise, we have for the 2D solutions of the magnetic transverse Helmholtz equation, cf. Eq.~\eqref{INITN1}, in terms of the transverse scalar modes
\begin{align}
    \bm t_{\bm j,\mu_1} (\bm y) &= |\bm k_{\bm m,\mu_1}|^{-1} |\bm k_{\bm j,\mu_1}|^{-1} 
    \begin{pmatrix}
        k_l \erone \\ 
        k_l \ertwo \\
      \bm k_{\bm m,\mu_1}^2 
    \end{pmatrix}
    \psi_{\bm m,\mu_1} (\bm y) \nonumber, \\
        \bm t_{\bm j,\mu_2} (\bm y) &= |\bm k_{\bm m,\mu_2}|^{-1} 
    \begin{pmatrix}
        &\ertwo \\ 
        -&\erone \\
        &0
    \end{pmatrix}
    \psi_{\bm m,\mu_2} (\bm y).
\label{CURLYNMATRICES}
\end{align}
\end{subequations}

The orthonormality of the electric field modes $\bm s_{\bm j,\mu} (\bm y)$ (cf. Eq.~\eqref{COND15}) and thus also of the magnetic field modes $\bm t_{\bm j,\mu} (\bm y)$ follows directly from Eq.~\eqref{STID1} and the orthogonality of the polarization vectors, i.e., for $l = l'$ which corresponds to the smae transversal mode space $\textbf{L}_l^2(\Gamma)$. 
Furthermore, the orthogonality of the 2D electric and magnetic modes, cf. Eq.~\eqref{COND15EXTENDED}, follows directly from the identity~\eqref{CANYOUFEELIT}:
\begin{align}
&\int_\Gamma \dd^2 y\, \bm s_{\bm j,\mu}^\dagger \cdot \bm t_{\bm j,\mu}^{\vphantom{\dagger}} \label{CALCD2} \\
&\quad = \pm  \frac{k_l}{|\bm k_{\bm m,\mu}|^2 |\bm k_{\bm j,\mu}|} \int_\Gamma \dd^2 y\, \left[ \partial_{(y_2)}  \bar{\psi}_{\bm m,\mu} (\bm y) \partial_{(y_1)} \psi_{\bm m,\mu} (\bm y) \right.\nonumber\\
&\quad \quad \left.- \partial_{(y_1)} \bar{\psi}_{\bm m,\mu} (\bm y) \partial_{(y_2)} \psi_{\bm m,\mu} (\bm y)  \right]= 0.\nonumber
\end{align}

\section{Boundary Conditions of the lower-dimensional Helmholtz Equations}
\setcounter{equation}{0}
\renewcommand{\theequation}{C.\arabic{equation}}
\label{DERVBOUND}

In the following we derive the boundary conditions for the lower-dimensional electromagnetic modes, i.e., the boundary conditions stated in Eqs.~\eqref{INITU1}, \eqref{INITV1},~\eqref{INITM1} and~\eqref{INITN1}, as well as for the transverse scalar modes $\psi_{\bm m,\mu}$, which allow us to easily find explicit expressions for the electromagnetic modes for a given geometry.
Therefore, we divide the boundary conditions of the cavity into longitudinal boundary conditions on $\partial S$ with corresponding normal vector $\pm \bm e^{(z)}$, and transverse boundary conditions on  $\partial \Gamma$ with normal vector $\bm n^{(\Gamma)}$ (cf. Fig.~\ref{CAVLOPP}). 
If both boundary conditions (transversal and longitudinal boundary conditions) are satisfied, the boundary conditions at the cavity edges, i.e., the case $\bm y \in \partial \Gamma \land z \in \partial S$, are trivially satisfied as well.

\subsection{Boundary Conditions for the lower-dimensional EM Modes}
\label{BOUNDARYCONDITIONSFIELDMODES}
\setcounter{equation}{0}
\renewcommand{\theequation}{C.1.\arabic{equation}}

\subsubsection*{Electric Boundary Conditions}
Starting with the lower-dimensional electric modes, we plug the mode decomposition~\eqref{IMPORTANTCONDITION} into the boundary condition~\eqref{CONDEFIELD}. Then one obtains on the transversal boundary 
\begin{align}
\bm n_{\Gamma}\times \bm u_{\bm j,\mu} (\rr) \Big|_{\bm y \in \partial \Gamma } = 
\bm n_{\Gamma}\times \mathbfcal{U}_{l}(z) \bm s_{\bm j,\mu} (\bm y)  \Big|_{\bm y \in \partial \Gamma }=\bm  0.
\label{EQLAT1}
\end{align}
Recall that in App.~\ref{APPREDTO2D} we found expressions for $\bm s_{\bm j,\mu} (\bm y)$ and $\bm t_{\bm j,\mu} (\bm y)$ in terms of the scalar modes (cf. Eq.~\eqref{CURLYMMATRICES} and Eq.~\eqref{CURLYNMATRICES}).
Via a decomposition in the basis spanning the cavity volume~\eqref{BASIS}, one can write Eq.~\eqref{EQLAT1} in terms of a superposition of the two different longitudinal scalar solutions $\psi_{l,\mu_1}(z)$ and $\psi_{l,\mu_2}(z)$ (see Eq.~\eqref{UPMATRIX}): 
\begin{align}
\begin{split}
&\quad\left(n_{\Gamma}^{(y_2)} s^{(z)}_{\bm j,\mu} (\bm y), \, -n_{\Gamma}^{(y_1)} s^{(z)}_{\bm j,\mu} (\bm y), \, 0  \right)^\mathrm{T}\Big|_{\bm y \in \partial \Gamma } \psi_{l,\mu_1}(z) \\
&+\!\left(0, \, 0, \, n_{\Gamma}^{(y_1)} s^{(y_2)}_{\bm j,\mu} (\bm y) -n_{\Gamma}^{(y_2)} s^{(y_1)}_{\bm j,\mu} (\bm y)  \right)^\mathrm{T} \!\Big|_{\bm y \in \partial \Gamma } \!\psi_{l,\mu_2}(z) \! = \! \bm 0.
\end{split}
\label{INCOMPONENTSWETRUST}
\end{align}
Each of these terms decouples again into longitudinal and transversal degrees of freedom. Since the two vectors depending on the transversal degrees of freedom are independent of each other, one attains the first boundary condition of Eq.~\eqref{INITM1}:
\begin{align}
&\left(n_{\Gamma}^{(y_2)} s^{(z)}_{\bm j,\mu} (\bm y), \, -n_{\Gamma}^{(y_1)} s^{(z)}_{\bm j,\mu} (\bm y), \, n_{\Gamma}^{(y_1)} s^{(y_2)}_{\bm j,\mu} (\bm y)\right. \label{IMDONE}\\ 
&\quad\left.-n_{\Gamma}^{(y_2)} s^{(y_1)}_{\bm j,\mu} (\bm y)  \right)^\mathrm{T}\Big|_{\bm y \in \partial \Gamma } = 
\bm n_{\Gamma}\times   \bm s_{\bm j,\mu} (\bm y)  \Big|_{\bm y \in \partial \Gamma } = \bm 0.\nonumber
\end{align}
To arrive at the longitudinal boundary conditions in Eq.~\eqref{INITU1}, we evaluate boundary condition~\eqref{CONDEFIELD} on $S$ and make use of the matrix components $\mathcal{U}_{l}^{(y_1,y_1)}(z)=\mathcal{U}_{l}^{(y_2,y_2)}(z)=\psi_{l,\mu_1}(z)$ such that 
\begin{align}
\begin{split}
&\bm e^{(z)} \times  \mathbfcal{U}_{l}(z) \bm s_{\bm j,\mu} (\bm y)  \Big|_{z \in \partial S}\\
&\quad = \left(s^{(z)}_{\bm j,\mu} (\bm y), \,  s^{(z)}_{\bm j,\mu} (\bm y), \, 0  \right)^\mathrm{T} \psi_{l,\mu_1}(z) \Big|_{\bm z \in \partial S } =   \bm 0.
\end{split}
\label{EQLAT1xx}
\end{align}
It follows directly that this boundary condition acts solely on the longitudinal scalar mode $\psi_{l,\mu_1} (z)$. Thus, multiplying the diagonal matrix $\mathbfcal{U}_{l} (z)$ by the appropriate polarization vector (i.e., either $\bm \epsilon_{\bm j,\mu_1}$ for $\mu_1$-polarization or $\bm \epsilon_{\bm j,\mu_2}$ for $\mu_2$-polarization) yields the longitudinal boundary condition for the 1D modes
\begin{align}
\left(u_{\bm j,\mu}^{(y_2)} (z) ,\, - u_{\bm j,\mu}^{(y_1)} (z) , \, 0\right)^\mathrm{T} \!\Big|_{z \in \partial S } \!\! = \bm e^{(z)} \times  \bm u_{\bm j, \mu} (z)  \Big|_{z \in \partial S } \!\! = \bm 0.
 \label{IMDONEZ}
\end{align}
For the set of boundary conditions originating from Eq.~\eqref{CONDBFIELD}, we decompose the nabla operator in the basis spanned by the cavity~\eqref{BASIS} into its longitudinal and cross sectional part via 
\begin{align}
\bm \nabla = \bm\nabla_{\Gamma} + \bm e^{(z)} \partial_z.
\label{NABLADECOMP}
\end{align}
Thus, we obtain on the transversal boundary
\begin{align}
\bm \nabla_\Gamma  \mathbfcal{U}_{l} (z)  \bm s_{\bm j,\mu} (\bm y) \Big|_{\bm y\in \partial \Gamma} = - \bm e^{(z)} \partial_z  \mathbfcal{U}_{l} (z)  \bm s_{\bm j,\mu} (\bm y) \Big|_{\bm y\in \partial \Gamma}  .
\label{BC5}
\end{align}
From boundary condition~\eqref{IMDONE} it follows that for $\bm y \in \partial \Gamma: \,\bm s_{\bm j,\mu} \cdot \bm e^{(z)}= 0$. Thus, the right-hand side of Eq.~\eqref{BC5} vanishes. Using the identity $\mathcal{U}_{l}^{(y_1,y_1)}(z) = \mathcal{U}_{l}^{(y_2,y_2)}(z)$ (cf. Eq.~\eqref{LONGFIELDMODESSINCOS}), one finds the second boundary condition of Eq.~\eqref{INITM1}, i.e.,
\begin{align}
\begin{split}
&\partial_{(y_1)} s_{\bm j,\mu}^{(y_1)} (\bm y) \Big|_{\bm y \in \partial \Gamma} + \partial_{(y_2)} s_{\bm j,\mu}^{(y_2)} (\bm y) \Big|_{\bm y \in \partial \Gamma} \\
&\quad=
\bm  \nabla_\Gamma \cdot  \bm s_{\bm j,\mu} (\bm y)  \Big|_{\bm y \in \partial \Gamma } = 0.
\end{split}
\label{WHATABOUND}
\end{align}
To derive the longitudinal boundary condition from Eq.~\eqref{CONDBFIELD}, one uses  \mbox{$\mathcal{U}_{l}^{(y_1,y_1)}(z) \Big|_{z \in \partial S} = \mathcal{U}_{l}^{(y_2,y_2)}(z) \Big|_{z \in \partial S} = 0$ } from Eq.~\eqref{IMDONEZ} such that
\begin{align}
\partial_z u_{\bm j,\mu}^{(z)} (z) \Big|_{z\in \partial S} =  
\bm e^{(z)}  \partial_z \cdot \bm u_{\bm j,\mu} (z) \Big|_{z\in \partial S}= 0,
\label{SECONDBOUNDUZ}
\end{align}
which corresponds to the first boundary condition of~\eqref{INITU1}.

\subsubsection*{Magnetic Boundary Conditions}

Analogously to the derivation of Eq.~\eqref{WHATABOUND}, one can show via the matrix components $\mathcal{V}_{l}^{(y_1,y_1)}(z) = \mathcal{V}_{l}^{(y_2,y_2)}(z) $ of $\mathbfcal{V}_l(z)$ that the boundary condition~\eqref{CONDBBFIELD} splits into 
\begin{align}
\begin{split}
&n_{\Gamma}^{(y_1)} t_{\bm j,\mu}^{(y_1)} (\bm y) \Big|_{\bm y \in \partial \Gamma} + n_{\Gamma}^{(y_2)} t_{\bm j,\mu}^{(y_2)} (\bm y) \Big|_{\bm y \in \partial \Gamma} \\ 
&\quad = \bm  n_{\Gamma} \cdot \bm t_{\bm j,\mu} (\bm y)  \Big|_{\bm y \in \partial \Gamma } = 0,
\label{1STNBBOUND}
\end{split}
\\
&v_{\bm j,\mu}^{(z)} (z) \Big|_{z\in \partial S}  
= \bm e^{(z)}  \cdot \bm v_{\bm j,\mu} (z) \Big|_{z\in \partial S}= 0,
\label{2NDBBOUND}
\end{align}
corresponding to the first boundary condition in Eq.~\eqref{INITN1} and the second boundary condition in Eq.~\eqref{INITV1}, respectively.
The second boundary condition of the 3D magnetic modes~\eqref{CONDBBBFIELD} reads
\begin{align}
\bm n \times \left( \bm \nabla \times \mathbfcal{V}_{l}(z) \bm t_{\bm j,\mu} (\bm y)  \right)  \Big|_{\bm r\in \partial V} = \bm 0.
\end{align}
Focusing first on the transverse boundary gives by means of the vector identity~\cite{Hanson}
\begin{align}
\bm A \times (\bm \nabla \times \bm B) = \bm \nabla_{\bm B}(\bm A \cdot \bm B) - (\bm A \cdot \bm \nabla) \bm B,
\label{HOWMANYIDENTITIESARELEFTOUTTHERE}
\end{align}
with $\bm \nabla_{\bm B}$ being the nabla operator acting only the vector $\bm B$:
\begin{align}
\bm \nabla \! \left(  \bm n_\Gamma^\dagger \mathbfcal{V}_{l}(z) \bm t_{\bm j,\mu} (\bm y) \right)\! \Big|_{\bm y \in \partial \Gamma} \!\!\! - (\bm n_\Gamma \! \cdot \! \bm \nabla )  \mathbfcal{V}_{l} (z) \bm t_{\bm j,\mu} (\bm y)  \Big|_{\bm y \in \partial \Gamma} \!\!\!= \bm 0. 
\end{align}
The gradient term on the left-hand side vanishes due to boundary condition~\eqref{1STNBBOUND}.
Equivalently to the calculation performed in Eq.~\eqref{INCOMPONENTSWETRUST}, we find an expression purely in terms of the transversal degrees of freedom:
\begin{align}
&\left(
n_{\Gamma}^{(y_2)} \left[ \partial_{(y_1)} t_{\bm j,\mu}^{(y_2)} (\bm y) - \partial_{(y_2)} t_{\bm j,\mu}^{(y_1)} (\bm y) \right],\right. \nonumber \\
&\quad
n_{\Gamma}^{(y_1)} \left[ \partial_{(y_1)} t_{\bm j,\mu}^{(y_2)} (\bm y) - \partial_{(y_2)} t_{\bm j,\mu}^{(y_1)} (\bm y) \right], \\
&\quad
\left.\left[n_{\Gamma}^{(y_1)} \partial_{(y_1)} + n_{\Gamma}^{(y_2)}\partial_{(y_2)} \right] t^{(z)}_{\bm j,\mu} (\bm y) 
\right)\Big|_{\bm y \in \partial \Gamma} 
= \bm 0. \nonumber
\end{align}
Writing this equation in a more compact form, one obtains with $\bm n_{\Gamma}\cdot \bm \nabla = \bm n_{\Gamma}\cdot \bm \nabla_\Gamma$ the second boundary condition of Eq.~\eqref{INITN1}, reading
\begin{align}
\bm n_{\Gamma}\times \left[\bm \nabla_\Gamma \times  \bm t_{\bm j,\mu} (\bm y)\right] \Big|_{\bm y \in \partial \Gamma} = \bm 0.
\label{BOUNDN2APP}
\end{align}
Finally, for the longitudinal boundary we have 
\begin{align}
\bm e^{(z)} \times \left(\bm \nabla \times \mathbfcal{V}_{l}(z) \bm t_{\bm j,\mu} (\bm y) \right) \Big|_{z\in \partial S} 
= \bm 0.
\label{ITS12HOURSLATERANDSTILLNOPROGRESSMADE}
\end{align}
Applying identity~\eqref{HOWMANYIDENTITIESARELEFTOUTTHERE} yields
\begin{align}
\begin{split}
&\bm \nabla \left( \bm e^{(z)\dagger} \mathbfcal{V}_{l}(z) \bm t_{\bm j,\mu} (\bm y)  \right)\Big|_{ z \in \partial S}  \\ 
&\quad- (\bm e^{(z)} \cdot \bm \nabla ) \mathbfcal{V}_{l} (z)  \bm t_{\bm j,\mu} (\bm y) \Big|_{z \in \partial S} = \bm 0.
\end{split}
\end{align}
By use of boundary condition~\eqref{2NDBBOUND} the gradient term vanishes.  
Thus, when decomposing the nabla operator into transverse and longitudinal components again (cf. Eq.~\eqref{NABLADECOMP}) and applying boundary condition~\eqref{2NDBBOUND} one arrives at
\begin{align}
\begin{split}
&\quad- \left( \partial_z v_{\bm j,\mu}^{(y_1)} (z) ,\, \partial_z v_{\bm j,\mu}^{(y_2)} (z) , \, 0\right) \Big|_{z \in \partial S }\\ 
&= 
\bm e^{(z)} \times \left[ \bm e^{(z)} \partial_z \times \bm v_{\bm j,\mu} (z) \right]\Big|_{z \in S} = \bm 0.
\end{split}
\label{BOUNDVL}
\end{align}
Equivalently, the first boundary condition presented in~\eqref{INITV1} is realized, reading 
\begin{align}
\bm e^{(z)}  \partial_z \times \bm v_{\bm j,\mu} (z) \Big|_{z \in S} = \bm 0.
\label{THIRDBOUND}
\end{align}

Thus, we found that both the boundary conditions for the electric modes and the boundary conditions for the magnetic modes separate on the underlying geometry into two groups of boundary conditions. One group entails the transverse degrees of freedom and one group entails the longitudinal degrees of freedom, corresponding each to  lower-dimensional dynamics after dimensional reduction.

\subsection{Boundary Conditions for the Scalar Modes}
\setcounter{equation}{0}
\renewcommand{\theequation}{C.2.\arabic{equation}}
\label{DERVBOUNDSCALAR}

Here we reexpress the boundary conditions we found in the previous section for the lower-dimensional electromagnetic modes in terms of the scalar solutions; enabling us to find the explicit electromagnetic modes for a given geometry in terms of the scalar modes.
For a cavity of length $L$, i.e., $\partial S \in \{0,L\}$, we verify that the boundary conditions are in accordance with our initial choice~\eqref{SCALMODES}. Indeed, by making use of boundary condition~\eqref{IMDONEZ} (cf. Eq.~\eqref{UPMATRIX}) or analogously boundary condition~\eqref{2NDBBOUND}  we find 
\begin{align}
\psi_{l,\mu_1} (z) \Big|_{z \in \partial S} = 0. 
\label{bc:scalar0}
\end{align}
From boundary condition~\eqref{SECONDBOUNDUZ} and boundary condition~\eqref{THIRDBOUND} we get similarly
\begin{align}
\partial_z \psi_{l,\mu_2} (z) \Big|_{z \in \partial S} = 0.
\end{align}

For the transverse scalar modes, we have for $\psi_{\bm m,\mu_1} (\bm y)$ a Neumann boundary condition from boundary conditions~\eqref{IMDONE}:
\begin{align}
(n_{\Gamma}^{(y_1)} \partial_{(y_2)} + n_{\Gamma}^{(y_2)} \partial_{(y_1)} ) \psi_{\bm m,\mu_1} (\bm y)\Big|_{\bm y \in \partial \Gamma} 
%=  (\bm n_{\Gamma}\times \nabla_\Gamma) \psi_{\bm m, \mu_1}(\bm y) \Big|_{\bm y \in \partial\Gamma}
=  0.
\label{bc:scalar1}
\end{align}
This condition can also be found from boundary condition~\eqref{1STNBBOUND} and~\eqref{BOUNDN2APP}, whereas boundary condition~\eqref{WHATABOUND} yields no additional information for the boundary condition of the scalar modes $\psi_{\bm m,\mu_1} (\bm y)$ whatsoever.
For the scalar modes $\bm \psi_{\bm m,\mu_2}(\bm y)$, the boundary conditions~\eqref{IMDONE},~\eqref{WHATABOUND} and~\eqref{1STNBBOUND} yield boundary conditions of Dirichlet kind
\begin{align}
\psi_{\bm m,\mu_2} (\bm y) \Big|_{\bm y \in \partial \Gamma} = 0.
\label{bc:scalar2}
\end{align}

%Thus we have shown that if the longitudinal scalar solution $\psi_{l,\mu} (z)$ is chosen to satisfy Dirichlet boundary conditions, the corresponding transversal solution $\psi_{\bm m,\mu} (\bm y)$ is constrained by Neumann boundary conditions and vice versa. 

In summary, we find a duality  in the boundary conditions of the scalar modes: Dirichlet ($\psi_{l,\mu_1} (z)$, $\psi_{\bm m,\mu_2} (\bm y)$) to Neumann ($\psi_{l,\mu_2} (z)$, $\psi_{\bm m,\mu_1} (\bm y)$) and vice versa by interchanging the polarization index, with complementary behavior for longitudinal and transverse components. Let us briefly connect this to the electromagnetic theory. This duality is due to the construction of the field modes from the scalar modes in Eq.~\eqref{RECONSTRUCTION} and Faraday's law of Eq.~\eqref{ID1} connecting electric and magnetic fields, which by construction need to have complementary (either Dirichlet or Neumann) boundary conditions for normal/transverse field components (see Eqs.~\eqref{CONDEFIELD},~\eqref{CONDBFIELD}  and Eqs.~\eqref{CONDBBFIELD},~\eqref{CONDBBBFIELD}). Therefore, the curl operation can be seen as switching the boundary conditions of the fields and the scalar modes, explaining the complementary boundary conditions for scalar components of distinct polarization indices. Furthermore, the complementary boundary conditions between transverse and longitudinal scalar components for fixed polarization index arise due to the normal electric field component vanishing at the boundary and analogously for the perpendicular magnetic field component.

%More precisely: The opposing boundary conditions for scalar modes of different polarization ensure that field modes of the same polarization and that electric and magnetic field keep their orthogonality, the opposing boundary conditions in transverse and longitudinal scalar modes of same polarization ensure the correct conditions account directly to the field mode's boundary conditions.}    

\section{Self-Adjointness of the Dimensionally Reduced Helmholtz Equations}
\renewcommand{\theequation}{D.\arabic{equation}}
\label{DERVSELFADJOINTNESS}
In the following we prove self-adjointness of the lower-dimensional Helmholtz equations (cf. Eqs.~\eqref{INITU1},~\eqref{INITV1} for the 1D problem and Eqs.~\eqref{INITM1},~\eqref{INITN1} for the 2D problem) after dimensional reduction, and thus the validity of the decomposition in an orthonormal eigenmode basis spanning the Hilbert space of the lower-dimensional problems. For a detailed discussion on self-adjoint problems in cavities and also for a proof of the self-adjointness of the initial 3D problem see Ref.~\cite{Hanson}. 

\subsection{1D Helmholtz Equations}
\setcounter{equation}{0}
\renewcommand{\theequation}{D.1.\arabic{equation}}
\label{SELFADJOINTNESS1D}

We first show self-adjointness of the 1D electric and magnetic Helmholtz equations~\eqref{INITU1} and~\eqref{INITV1} on their respective longitudinal-mode space $\textbf{L}^2_{\bm m}(S)$ for arbitrary transverse-mode numbers $\bm m$. Performing two times integration by parts on the respective inner product gives
\begin{align}
&\hspace{.5cm}\int_S \!\dd z \, \bm u_{\bm m l,\mu}^\dagger(z)\cdot \partial_z^2 \bm u_{\bm m l',\mu}^{\vphantom{\dagger}}(z) \label{norm1Dselfadj1}\\
&= \bm u_{\bm m l,\mu}^\dagger(z)\cdot \partial_z \bm u_{\bm m l',\mu}^{\vphantom{\dagger}}(z)\Big|_{z \in \partial S}\! - \left[\partial_z \bm u_{\bm m l,\mu}^\dagger(z) \right]\nonumber\\ 
&\quad\times \bm u_{\bm m l',\mu}^{\vphantom{\dagger}}(z)\Big|_{z \in \partial S} + \!\int_S\! \dd z \, \left[\partial_z^2 \bm u_{\bm m l,\mu}^\dagger(z) \right] \cdot  \bm u_{\bm m l',\mu}^{\vphantom{\dagger}}(z).\nonumber
\end{align}
The boundary terms can be written component-wise and vanish as
\begin{align}
&\bm u_{\bm m l,\mu}^\dagger \!(z)\cdot \partial_z \bm u_{\bm m l',\mu}^{\vphantom{\dagger}}\!(z)\Big|_{z \in \partial S}  \!\!\!\!- \left[ \partial_z \bm u_{\bm m l,\mu}^\dagger \!(z) \right] \cdot  \bm u_{\bm m l',\mu}^{\vphantom{\dagger}} \!(z) \Big|_{z \in \partial S}\nonumber
\\&= \! \sum_i \! \left[ \bar{u}^{(i)}_{\bm m l,\mu} \! (z) \partial_z u^{(i)}_{\bm m l',\mu} \! (z) \!  -  \!  u^{(i)}_{\bm m l,\mu} \! (z) \partial_z \bar{u}^{(i)}_{\bm m l',\mu} \! (z)  \right] \! \!\Big|_{z \in \partial S} \!\!\!\! = 0.
\label{HOWLONGTHISWILLLAST}
\end{align}
Due to boundary condition~\eqref{IMDONEZ} we have that the components $u^{(y_1)}_{\bm j,\mu} (z)$ and $u^{(y_2)}_{\bm j,\mu} (z)$ vanish at the boundary; the remaining component with $i = z$ vanishes due to boundary condition~\eqref{SECONDBOUNDUZ}, which gives \mbox{$\partial_z u^{(z)}_{\bm j,\mu} (z) = 0$} for $z \in \partial S$. Thus, we verified the self-adjointness of the Laplace operator $\partial_z^2$ for the dimensionally reduced dynamics of the longitudinal electric modes. 
Analogously for the longitudinal magnetic modes, the self-adjointness of the boundary-value problem~\eqref{INITV1} can be seen since 
\begin{align}
\begin{split}
&\bm v_{\bm m l,\mu}^\dagger \!(z) \cdot\partial_z \bm v_{\bm m l',\mu}^{\vphantom{\dagger}}\!(z)\Big|_{z \in \partial S} \!\!\!\! - \left[ \partial_z  \bm v_{\bm m l,\mu}^\dagger \!(z) \right] \cdot \bm v_{\bm m l',\mu}^{\vphantom{\dagger}} \!(z)\Big|_{z \in \partial S}\\ 
&= \! \sum_i \! \left[ \bar{v}^{(i)}_{\bm m l,\mu}  \! (z) \partial_z v^{(i)}_{\bm m l',\mu} \! (z) \! -  \! v^{(i)}_{\bm m l,\mu} \! (z) \partial_z \bar{v}^{(i)}_{\bm m l',\mu} \! (z)  \right]\! \! \Big|_{z \in \partial S} \!\!\!\! = 0,
\end{split}
\label{HOWLONGTHISWILLLASTV}
\end{align}
where we used boundary condition~\eqref{BOUNDVL}, yielding \mbox{$\partial_z v^{(y_1)}_{\bm j,\mu} (z) = \partial_z v^{(y_2)}_{\bm j,\mu} (z)=0$} at the boundary. The remaining terms associated with $i = z$, likewise,  vanish at the boundary via boundary condition~\eqref{2NDBBOUND}.  
Hence we have shown that the dimensionally reduced longitudinal modes' dynamics associated with the electric and magnetic field, respectively, correspond to a self-adjoint Laplace operator $\partial_z^2$ with appropriate boundary conditions. It is obvious that also the terms of the Helmholtz equations~\eqref{INITU1} and~\eqref{INITV1} associated with $k_l^2$ are self-adjoint; thereby guaranteeing that the reduced modes, being an orthonormal basis for given transverse-mode numbers, reconstruct their respective dimensionally-reduced longitudinal Hilbert space $\textbf{L}^2_{\bm m} (S)$.

\subsection{2D Helmholtz Equations}
\setcounter{equation}{0}
\renewcommand{\theequation}{D.2.\arabic{equation}}
\label{SELFADJOINTNESS2D}

Next, we show explicitly the self-adjointness of the 2D, i.e., transverse, Helmholtz equation after dimensional reduction. This is shown for the boundary-value problem of the electric modes~\eqref{INITM1} and the magnetic modes~\eqref{INITN1} on their respective transverse-mode space $\textbf{L}^2_l (\Gamma)$, separately. In both cases, one can rewrite the transverse Laplacian acting on the transverse modes via Eq.~\eqref{LAPLACEIDENTITY} in terms of a Helmholtz decomposition 
\begin{align}
  \Delta_{\Gamma} \bm s_{\bm j,\mu} &=  \bm \nabla_\Gamma \left[\bm \nabla_\Gamma \cdot  \bm s_{\bm j,\mu} (\bm y) \right] + \bm \nabla_\Gamma \times  \left[ \bm \nabla_\Gamma \times   \bm s_{\bm j,\mu} (\bm y)\right]. 
\label{2DHEHOADJEQ1}
\end{align}
Making use of the (anti-)linearity of the $L^2$ inner product (cf. Eq.~\eqref{COND15}), we start with the gradient operator contribution  in Eq.~\eqref{2DHEHOADJEQ1}. From the product rule we have
\begin{align}
 &\quad\quad\bm s_{\bm m l,\mu}^\dagger (\bm y) \bm \nabla_\Gamma \left[\bm \nabla_\Gamma \cdot  \bm s_{\bm m' l,\mu}^{\vphantom{\dagger}} (\bm y) \right]\nonumber\\   
&= \bm \nabla_\Gamma \cdot \left( \bm s_{\bm m l,\mu}^\dagger (\bm y)  \left[ \bm \nabla_\Gamma \cdot  \bm s_{\bm m'l,\mu}^{\vphantom{\dagger}} (\bm y)  \right]  \right)\label{PLEASEHELPME}\\
&\quad
- \left[\bm \nabla_\Gamma \cdot \bm s_{\bm ml,\mu}^\dagger (\bm y) \bm \right]
\left[\bm \nabla_\Gamma \cdot \bm s^{\vphantom{\dagger}}_{\bm m'l,\mu} (\bm y) \right]\nonumber.
\end{align}
Performing the integration over the transverse cavity domain $\Gamma$ and using the divergence theorem on the first term on the right-hand side, 
we get for the inner product by defining the infinitesimal surface vector $\dd \bm y_{\Gamma} =  \bm n_{\Gamma}\dd y \in \partial \Gamma$:
\begin{align}
&\quad\quad\int_\Gamma \!\dd^2 y \,  \bm s_{\bm ml,\mu}^\dagger
(\bm y) \cdot \!  \bm \nabla_\Gamma \left[\bm \nabla_\Gamma \cdot  \bm s^{\vphantom{\dagger}}_{\bm m' l,\mu} (\bm y) \right]\nonumber\\
&=  \oint_{\partial \Gamma}\! \dd \bm y_{\Gamma} \cdot \bm s^{\dagger}_{\bm m l,\mu} (\bm y)  \left[\bm \nabla_\Gamma \cdot \bm s_{\bm m'l,\mu}^{\vphantom{\dagger}} (\bm y)  \right] \label{TRANVAPP1}\\
&\quad -\!\int_{\Gamma} \dd^2 y \left[\bm \nabla_\Gamma \cdot  \bm s_{\bm ml,\mu}^\dagger (\bm y)  \right]\!
\left[\bm \nabla_\Gamma \cdot   \bm s^{\vphantom{\dagger}}_{\bm m'l,\mu} (\bm y) \right].\nonumber
\end{align}
The contour integral on the boundary $\partial \Gamma$ vanishes due to boundary condition Eq.~\eqref{WHATABOUND}. Using this boundary condition again the remaining term gives by means of the divergence theorem 
\begin{align}
&\quad \quad \int_\Gamma \!\dd^2 y   \bm s_{\bm m l,\mu}^\dagger
(\bm y)  \cdot \bm \nabla_\Gamma\! \left[\bm \nabla_\Gamma \cdot  \bm s^{\vphantom{\dagger}}_{\bm m' l,\mu} (\bm y) \right]\nonumber\\   
&= -\oint_{\partial \Gamma}\! \dd \bm y_{\Gamma}   \cdot  \bm s^{\vphantom{\dagger}}_{\bm m'l,\mu} (\bm y) \left[ \bm \nabla_\Gamma  \bm s_{\bm ml,\mu}^\dagger (\bm y) \right]\label{TRANVAPP2}\\ 
&\quad+ \!\int_{\Gamma} \!\dd^2 y  \bm \nabla_\Gamma \left[ \bm \nabla_\Gamma \bm s_{\bm ml,\mu}^\dagger (\bm y) \right]  \cdot   \bm s_{\bm m'l,\mu} (\bm y). \nonumber
\end{align}
Here, the contour integral does, similarly to the contour integral in Eq.~\eqref{TRANVAPP1}, vanish due to boundary condition~\eqref{WHATABOUND}, yielding the self-adjointness of the gradient term of the Laplacian $\Delta_{\Gamma}$.
For the rotation term in Eq.~\eqref{2DHEHOADJEQ1} we obtain by applying the vector analog of the second Greens theorem (cf.~\cite{Stratton}, Eq. 5) 
\begin{align}
&\quad\quad\int_\Gamma \dd^2 y  \, \bm s_{\bm m l,\mu}^\dagger   (\bm y)   \cdot \left\{ \bm \nabla_\Gamma \times  \left[ \bm \nabla_\Gamma \times   \bm s_{\bm m' l,\mu}^{\vphantom{\dagger}} (\bm y)\right]  \right\}\nonumber \\ 
&=  \int_\Gamma \dd^2 y \, \left\{ \bm \nabla_\Gamma \times  \left[ \bm \nabla_\Gamma \times   \bm s_{\bm m l,\mu}^{\dagger} (\bm y)\right] \right\} \cdot \bm s^{\vphantom{\dagger}}_{\bm m' l,\mu} (\bm y)\nonumber\\
&\quad+ \oint_{\partial \Gamma} \dd \bm y_{\Gamma} \cdot  \left\{  \bm s_{\bm ml,\mu}^\dagger (\bm y)  \times   \left[ \bm \nabla_\Gamma \times  \bm s_{\bm m' l,\mu}^{\vphantom{\dagger}}  (\bm y) \right] \right\}\label{2DHEHOADJEQ3}  \\
&\quad- \oint_{\partial \Gamma} \dd \bm y_{\Gamma} \cdot \left\{ \bm s_{\bm m' l,\mu}  (\bm y)   \times \left[ \bm \nabla_\Gamma \times  \bm s_{\bm ml,\mu}^\dagger (\bm y)\right] \right\}.\nonumber
\end{align}
With the vector identity $\bm A \cdot (\bm B \times \bm C) = (\bm A \times \bm B) \cdot \bm C$, the terms involving the contour integral in Eq.~\eqref{2DHEHOADJEQ3} vanish, i.e., 
\begin{align}
&\quad\quad\oint_{\partial \Gamma} \dd  \bm y_{\Gamma} \cdot \left\{ \bm s_{\bm ml,\mu}^\dagger (\bm y) \times   \left[ \bm \nabla_\Gamma \times    \bm s_{\bm m' l,\mu}^{\vphantom{\dagger}}  (\bm y) \right]\right\}\label{2DHEHOADJEQ4}\\
&=\oint_{\partial \Gamma} \dd  y \left[\bm n_{\Gamma}\times  \bm s_{\bm ml,\mu}^\dagger (\bm y)   \right] \cdot \left[ \bm \nabla_\Gamma \times    \bm s_{\bm m' l,\mu}^{\vphantom{\dagger}}  (\bm y) \bm \right]= 0,\nonumber
\end{align}
where from the second to the third line we used boundary condition~\eqref{IMDONE}. 

The calculation for the self-adjointness of the boundary-value problem of the transverse magnetic field modes, cf. Eq.~\eqref{INITN1}, is achieved fully analogously by substituting the transverse electric field modes $\bm s_{\bm j,\mu} (\bm y)$ by the transverse magnetic field modes $\bm t_{\bm j,\mu} (\bm y)$ in the calculations performed in Eq.~\eqref{2DHEHOADJEQ1}-\eqref{2DHEHOADJEQ3}. In this case, self-adjointness of the gradient term, i.e., $\bm \nabla_\Gamma \cdot [ \bm \nabla_\Gamma \cdot \bm t_{\bm j,\mu}(\bm y)]$, is obtained by the use of condition~\eqref{1STNBBOUND} in Eq.~\eqref{TRANVAPP1} and Eq.~\eqref{TRANVAPP2}. 
Furthermore, for the rotation term, i.e., $\bm \nabla_\Gamma \times [ \bm \nabla_\Gamma \times \bm t_{\bm j,\mu}(\bm y)]$, the boundary terms in Eq.~\eqref{2DHEHOADJEQ4} vanish by the following calculation 
\begin{align}
&\quad\quad\oint_{\partial \Gamma} \dd \bm y_\Gamma \cdot  \left\{ \bm t_{\bm ml,\mu}^\dagger (\bm y) \times   \left[ \bm \nabla_\Gamma \times   \bm t_{\bm m' l,\mu}^{\vphantom{\dagger}}  (\bm y) \right] \right\}
\label{2DHEHOADJEQ5}\\
&= \oint_{\partial \Gamma} \dd  y \,\bm t_{\bm ml,\mu}^\dagger (\bm y)  \cdot \left[ n^{(\Gamma)} \times \bm \nabla_\Gamma \times    \bm t_{\bm m' l,\mu}^{\vphantom{\dagger}}  (\bm y) \right]= 0,\nonumber
\end{align}
where from the second to the third step we used boundary condition~\eqref{2NDBBOUND}.
Finally, the terms of the Helmholtz equations~\eqref{INITM1} and~\eqref{INITN1} associated with  $\bm k_{\bm m,\mu}^2$ are self-adjoint as well. To this end, we have verified self-adjointness of the boundary-value problems for the transverse electric modes defined in Eq.~\eqref{INITM1} and the transverse magnetic modes defined in Eq.~\eqref{INITN1} on their respective transverse domain $\textbf{L}^2_l (\Gamma)$ after dimensional reduction.

\section{ Example -- Dimensional Reduction for  a Cylindrical Cavity}
\renewcommand{\theequation}{E.\arabic{equation}}
\label{EXAMPLEAPP}
In the following a working example for reducing the dimensions of a 3D cavity to 2D as well as 1D is provided.
Therefore, let us consider an ideal cylindrical cavity, of length $L$ in direction $\bm e^{(z)}$ and radius $R$. Following App.~\ref{SecB}, both the 2D and 1D modes can be constructed solely by the scalar eigenmodes of the Helmholtz equation under the appropriate boundary conditions derived in App.~\ref{BOUNDARYCONDITIONSFIELDMODES}. Since the longitudinal scalar solutions $\psi_l (z)$ are independent of the cross section (cf. Eq.~\eqref{SCALMODES}), the dimensional reduction reduces to a single task: The construction of the scalar solutions $\psi_{\bm m,\mu} (\bm y)$ of the transverse Helmholtz equation.

\subsection{Construction of the Transverse Scalar Modes}
\label{SecE1}

The transverse Helmholtz equation~\eqref{ScalarHEHOTranv} in cylindrical coordinates, i.e.,
\begin{align}
\left[r^{-1}\partial_r (r \partial_r) + r^{-2} \partial_\varphi^2 + \bm k_{\bm m,\mu}^2 \right] \psi_{\bm m,\mu} (r,\varphi) = 0,
\label{C1}
\end{align}
 with a so far unspecified wave vectors $\bm k_{\bm m, \mu}$ can be solved by
\begin{align}
\psi_{\bm m,\mu} (r,\varphi) = c_{\bm m,\mu}  J_{m_2} \left( |\bm k_{\bm m,\mu}| r \right)\e^{\ii m_2\varphi},
\label{CC2}
\end{align}
where $c_{\bm m,\mu}$ is a normalization constant and $J_{m_2}$ the $m_2$-th Bessel function of first kind. Recall that we use the shorthand notation $\bm m = (m_1,m_2)$ and neglected unbounded solutions of Eq.~\eqref{C1}. Applying the boundary conditions, cf. Eq.~\eqref{bc:scalar1} and Eq.~\eqref{bc:scalar2}, yields 
\begin{align}
\partial_r \psi_{\bm m,\mu_1} (r,\varphi) \Big|_{r = R} = 0,\quad
\psi_{\bm m, \mu_2} (r,\varphi) \Big|_{r = R} = 0.
\end{align}
Under these boundary conditions a unique $|\bm k_{\bm m,\mu}|$ can be determined for each polarization yielding: $|\bm k_{\bm m,\mu_1}| = \chi_{\bm m,\mu_1}/R$, with $\chi_{\bm m,\mu_1}$ being the $m_1$-th zero of the derivative of the $m_2$-th Bessel function, and $|\bm k_{\bm m,\mu_2}| = \chi_{\bm m,\mu_2}/R$ with
$\chi_{\bm m,\mu_2}$ being the $m_1$-th zero of the  $m_2$-th Bessel function.
We conclude from normalization conditions~\eqref{ScalIntLong} and~\eqref{ScalIntTranv}
\begin{align}
c_{\bm m,\mu_1} &= \left( \pi R^2 [J_{m_2}^2(\chi_{\bm m,\mu_1})- J_{m_2+1}^2(\chi_{\bm m,\mu_1})] \right)^{-1/2},\nonumber\\
c_{\bm m,\mu_2} &= \left( \pi R^2 J_{m_2+1}^2(\chi_{\bm m,\mu_2})  \right)^{-1/2}.
\label{C4}
\end{align}
One can use these transverse scalar modes on the disk in combination with the longitudinal scalar modes~\eqref{SCALMODES} to reconstruct the 3D modes of the cylinder via Eq.~\eqref{RECONSTRUCTION}. 
With the 3D modes and thus the 3D electric and magnetic fields of the cylinder at hand one now can start to reduce the electric and magnetic fields to lower dimensions by the approach presented in Sec.~\ref{DIMREDELFIELDS}.

\begin{widetext}
\subsection{Dimensional Reduction to a 1D cavity -- ``Thin Fiber Limit"}
\renewcommand{\theequation}{E.2.\arabic{equation}}

Mapping from the 3D to the 1D problem for a cylindrical cavity is achieved via the  transverse ancilla basis~\eqref{WHYMOREMATRICES1} in terms of  the scalar modes defined in Eq.~\eqref{CC2}:
\begin{align}
\begin{split}
\mathbfcal{S}_{\bm m} (\bm y) =  &\frac{c_{\bm m, \mu_1}}{\sqrt{2} |\bm k_{\bm m,\mu_1}|} 
\begin{pmatrix}
r^{-1} \partial_\varphi & - r^{-1} \partial_\varphi & 0 \\
- \partial_r &   \partial_r & 0 \\
0 & 0 & 0 \\ 
\end{pmatrix}
J_{m_2} \left( |\bm k_{\bm m,\mu_1}| r \right)\e^{-\ii m_2\varphi}\\
&+  \frac{c_{\bm m, \mu_2}}{\sqrt{2} |\bm k_{\bm m,\mu_2}|} 
\begin{pmatrix}
\partial_r & \partial_r & 0 \\
r^{-1} \partial_\varphi & r^{-1} \partial_\varphi & 0 \\
0 & 0 & \sqrt{2} |\bm k_{\bm m,\mu_2}| \\ 
\end{pmatrix}
J_{m_2} \left( |\bm k_{\bm m,\mu_2}| r \right)\e^{-\ii m_2\varphi}.
\end{split}
\label{PCYLINDER}
\end{align}
The $\mathbfcal{T}_{\bm m} (\bm y)$ for the magnetic field can be obtained from Eq.~\eqref{PCYLINDER} by the substitutions discussed in  App.~\ref{SecB}. By means of Eqs.~\eqref{PROJECTIONSUANDV} the longitudinal 1D modes as already given in Eqs.~\eqref{UVEQUATIONSCOMPLETE} are obtained.
Plugging these modes into the expressions for the 1D fields, i.e.,  Eq.~\eqref{RQEF} and Eq.~\eqref{RQBF}, provides the 1D electric and magnetic subfields, respectively.

\subsection{Dimensional Reduction to a 2D cavity -- ``Large Mirror Limit"}
\renewcommand{\theequation}{E.3.\arabic{equation}}

The dimensional reduction to 2D is executed via the longitudinal ancilla basis $\mathbfcal{U}_{l} (z)$ and $\mathbfcal{V}_l (z)$ as shown in Eq.~\eqref{NONONOTANOTHER} and Eq.~\eqref{SEEMSMYWORDHASNOWEIGHTHERE}, with the explicit forms given in Eqs.~\eqref{LONGFIELDMODESSINCOS}.
By means of the transverse scalar modes derived in Eq.~\eqref{C1} to Eq.~\eqref{C4} the dimensionally reduced 2D electric field modes read via Eq.~\eqref{CURLYMMATRICES}: 
\begin{align}
\begin{split}
    \bm s_{\bm j,\mu_1} (\bm y) &= 
    \frac{c_{\bm m,\mu_1}}{|\bm k_{\bm m,\mu_1}|}
    \begin{pmatrix}
        \frac{\ii m_2}{r} J_{m_2} \left( \frac{\chi_{\bm m,\mu_1}}{R} r\right) \e^{\ii m_2 \varphi}\\ 
        -\frac{|\bm k_{\bm m,\mu_1}|}{2} \left[ J_{m_2-1} \left( \frac{\chi_{\bm m,\mu_1}}{R} r\right) - J_{m_2+1} \left( \frac{\chi_{\bm m,\mu_1}}{R} r\right) \right] \e^{\ii m_2 \varphi}\\ 
        0
    \end{pmatrix},
    \label{CURLYMMATRICESCYLINDER}
    \\
    \bm s_{\bm j,\mu_2} (\bm y) &
    =   - \frac{c_{\bm m,\mu_2}}{|\bm k_{\bm m,\mu_2}||\bm k_{\bm j,\mu_2}|} \begin{pmatrix}
        \frac{k_l|\bm k_{\bm m,\mu_2}|}{2} \left[ J_{m_2-1} \left( \frac{\chi_{\bm m,\mu_2}}{R} r\right) - J_{m_2+1} \left( \frac{\chi_{\bm m,\mu_2}}{R} r\right) \right] \e^{\ii m_2 \varphi}  \\
        k_l\frac{\ii m_2}{r} J_{m_2} \left( \frac{\chi_{\bm m,\mu_2}}{R} r\right) \e^{\ii m_2 \varphi}\\ 
        -|\bm k_{\bm m,\mu_2}|^2 J_{m_2} \left( \frac{\chi_{\bm m,\mu_2}}{R} r\right) \e^{\ii m_2 \varphi}
    \end{pmatrix}. 
\end{split}
\end{align}
The 2D magnetic modes likewise are obtained via Eq.~\eqref{CURLYNMATRICES}:
\begin{align}
\begin{split}
    \bm t_{\bm j,\mu_1} (\bm y) &= 
    \frac{c_{\bm m,\mu_1}}{|\bm k_{\bm m,\mu_1}||\bm k_{\bm j,\mu_1}|} \begin{pmatrix}
        \frac{k_l|\bm k_{\bm m,\mu_1}|}{2} \left[ J_{m_2-1} \left( \frac{\chi_{\bm m,\mu_1}}{R} r\right) - J_{m_2+1} \left( \frac{\chi_{\bm m,\mu_1}}{R} r\right) \right] \e^{\ii m_2 \varphi}  \\
        k_l\frac{\ii m_2}{r} J_{m_2} \left( \frac{\chi_{\bm m,\mu_1}}{R} r\right) \e^{\ii m_2 \varphi}\\ 
        |\bm k_{\bm m,\mu_1}|^2 J_{m_2} \left( \frac{\chi_{\bm m,\mu_1}}{R} r\right) \e^{\ii m_2 \varphi}
    \end{pmatrix}, 
    \\
    \bm t_{\bm j,\mu_2} (\bm y) &= 
    \frac{c_{\bm m,\mu_2}}{|\bm k_{\bm m,\mu_2}|}
    \begin{pmatrix}
        \frac{\ii m_2}{r} J_{m_2} \left( \frac{\chi_{\bm m,\mu_2}}{R} r\right) \e^{\ii m_2 \varphi}\\ 
        -\frac{|\bm k_{\bm m,\mu_2}|}{2} \left[ J_{m_2-1} \left( \frac{\chi_{\bm m,\mu_2}}{R} r\right) - J_{m_2+1} \left( \frac{\chi_{\bm m,\mu_2}}{R} r\right) \right] \e^{\ii m_2 \varphi}\\ 
        0
    \end{pmatrix}.
    \label{CURLYNMATRICESCYLINDER}
\end{split}
\end{align}
\end{widetext}
These lower-dimensional modes still depend on the mode number $l$ associated with the integrated-out dimension which is due to the polarization surviving the reduction. 
The 2D quantum fields are then obtained in the same manner as the 1D fields of Eq.~\eqref{RQEF} and Eq.~\eqref{RQBF} but now via  Eq.~\eqref{NONONOTANOTHER} and Eq.~\eqref{SEEMSMYWORDHASNOWEIGHTHERE} as 
\begin{subequations}
\begin{align}
\begin{split}
\hat{\E}(\bm y,t) &=\sum_{\bm j,l'} \inner{\mathbfcal{U}_{l'} (z)}{ \hat{\mathbfcal{E}}^{(+)}_{\bm j, \mu} (\rr,t)}_S+ \mathrm{H.c.} = 
\sum_l \hat{\E}_{l}(\bm y,t),\\
\hat{\B}(\bm y,t) &=  \sum_{\bm j,l'}  \inner{\mathbfcal{V}_{l'} (z)}{ \hat{\mathbfcal{B}}^{(+)}_{\bm j, \mu} (\rr,t)}_S+ \mathrm{H.c.} = 
\sum_l \hat{\B}_{l}(\bm y,t),
\end{split}
\label{DIMREDE2D}
\end{align}
where we define the subfields in the 2D case
\begin{align}
\begin{split}
&\hat{\E}_l(\bm y,t) =
\ii \sum_{\bm m,\mu} 
 \sqrt{\frac{\hbar \omega_{\bm j,\mu} }{2 \varepsilon_0 }} \Big( \aop_{\bm j, \mu}  \e^{-\ii \omega_{\bm j,\mu} t} \bm s_{\bm j,\mu}(\bm y) - \mathrm{H.c.}\Big),\\
 &\hat{\B}_l(\bm y,t) =
\sum_{\bm m,\mu} 
 \sqrt{\frac{\hbar \omega_{\bm j,\mu} }{2 \varepsilon_0 c^2}} \Big( \aop_{\bm j, \mu}  \e^{-\ii \omega_{\bm j,\mu} t} \bm t_{\bm j,\mu}(\bm y) + \mathrm{H.c.}\Big),
 \end{split}
 \label{TRANVSUB}
\end{align}
\end{subequations}
with each subfield describing the dynamics on its respective subspace $\textbf{L}_{l}^2(\Gamma)$.

\section{Dimensionally Reduced Dynamics}
\label{DETERMINATIONSUBFIELDHAMILTONIAN}

\subsection{Free Field Hamiltonian}
\setcounter{equation}{0}
\renewcommand{\theequation}{F.1.\arabic{equation}}
\label{FIELDHAMAPP}

In the following we derive the decomposition of the original free electromagnetic Hamiltonian into an infinite sum of 1D subfield Hamiltonians~\eqref{h1D}.
It is straightforward to show that for fixed $l$, i.e., after integrating out the $z$ direction in the Hamiltonian~\eqref{Hh}, all off-diagonal terms vanish since 
\begin{align}
\int_\Gamma \dd y^2 \left( \bm s_{\bm j,\mu}(\bm y) \cdot \bm s_{\bm m'l,\mu'} (\bm y) - \bm t_{\bm j,\mu} (\bm y) \cdot \bm t_{\bm m' l,\mu'} (\bm y) \right) = 0.
\end{align}
On the other hand we obtain for the diagonal terms by means of the identities~\eqref{STID1} by splitting the 1D fields into positive and negative frequency components (cf. Eq.~\eqref{BACKTRANSFORM})
\begin{align}
\hat{H}^{\mathrm{field}} 
&= \frac{\varepsilon_0}{2} \sum_{\bm j,\bm j'} \int_V \dd^3 r \left[   \hat{\mathbfcal{E}}_{\bm j'}^{(-)}(z,t) \mathbfcal{S}_{\bm m'}^\dagger (\bm y) \cdot \mathbfcal{S}_{\bm m} (\bm y) \hat{\mathbfcal{E}}_{\bm j}^{(+)}(z,t)\right.\nonumber\\   
&\quad+ \left. c^2  \hat{\mathbfcal{B}}_{\bm j'}^{(-)} (z,t) \mathbfcal{T}_{\bm m'}^\dagger (\bm y) \cdot  \mathbfcal{T}_{\bm m} (\bm y) \hat{\mathbfcal{B}}_{\bm j}^{(+)}(z,t) + \mathrm{H.c.}  \right]\nonumber\\ 
&= 
\frac{\varepsilon_0}{2} \sum_{\bm m} \sum_{l, l'} \int_S \dd z \left[  \hat{\mathbfcal{E}}_{\bm m l}^{(-)} (z,t) \cdot  \hat{\mathbfcal{E}}_{\bm m l'}^{(+)}(z,t)\right.\nonumber\\  
&\quad \left.+ c^2  \hat{\mathbfcal{B}}_{\bm m l}^{(-)} (z,t) \cdot  \hat{\mathbfcal{B}}_{\bm m l'}^{(+)}(z,t) + \mathrm{H.c.} \right]
\label{FIELDHAMAPPEQ2}
\end{align}
Since the polarization vectors satisfy for all modes \mbox{$\bm \epsilon_{\bm m l,\mu} \cdot \bm \epsilon_{\bm m l',\mu'} - \bm \kappa_{\bm m l,\mu} \cdot \bm \kappa_{\bm m l',\mu'} = 0$}, the terms with $l\neq l'$ vanish equally to the 3D case. Thus Eq.~\eqref{FIELDHAMAPP} recasts to  Eq.~\eqref{h1D}:
\begin{align}
\hat{H}^{\mathrm{field}} &= 
\frac{\varepsilon_0}{2} \sum_{\bm m}\int_V \dd^3 r \left[ |\E_{\bm m} (z,t)|^2  + c^2   |\B_{\bm m} (z,t)|^2\right]\nonumber\\
&= \sum_{\bm m} \hat{h}_{\bm m}^{\mathrm{field}},
\label{FIELDHAMAPPEQ3}
\end{align}

\subsection{Electric Dipole Hamiltonian with Quantized Atomic COM}
\setcounter{equation}{0}
\renewcommand{\theequation}{F.2.\arabic{equation}}

\label{MULTIPOLEAPP}

Here we consider general hydrogen-like atoms with all their degrees of freedom being quantized; in particular the center of mass (COM) is no longer assumed to follow a classical trajectory but can exhibit quantum delocalization~\cite{Lopp2,Stritzelberger}.
We show that the dimensional reduction can likewise be implemented for the electric dipole interaction with quantized COM, and, following the procedure analogously, the same is expected to hold for general multipole interactions or interactions with the magnetic field. Note that this also includes diamagnetic or R\"ontgen terms where potentially more involved components arise due to the cross product.
In the COM variables, the electric dipole Hamiltonian reads
\begin{align}
\hat{H}_{\mathrm{COM}}^\mathrm{I} &= -\chi(t) e  \int_V \dd^3 \mathcal{R} \,\dyad{\mathbfcal{R} }{\mathbfcal{R}} \hat{\bm r}  \cdot \hat{\E} (\mathbfcal{R},t),
\label{HAMAPPHO}
\end{align}
where $\mathbfcal{R} = (\mathbfcal{Y}, \mathcal{Z})$ denotes the atomic COM position and $\bm r$ is the relative position. Considering the transition elements with respect to $L^2$-normalized COM position distributions $\ket{\bm s}=\int \dd^3 \mathcal{R}  \, \psi_s(\mathbfcal{R})\ket{\mathbfcal{R}}$, we have
\begin{align}
\matrixel{\bm s}{\hat{H}_{\mathrm{COM}}^\mathrm{I}} {\bm s '} \! &= \! - \chi(t) e \!\!
\int_V \!\!\!\mathrm{d}^3 \mathcal{R} \,
%\psi^*_s(\mathbfcal{R})  \psi_{s'}(\mathbfcal{R}) \hat{\bm r} \cdot 
\hat{\bm F}_{\bm s \bm s'} (\mathbfcal{R}) \!\cdot 
\!\hat{\E} (\mathbfcal{R},t), \\
\hat{\bm F}_{\bm s \bm s'} (\mathbfcal{R}) &= \hat{\bm r} \psi_{\bm s}^\ast (\mathbfcal{R}) \psi_{\bm s'} (\mathbfcal{R}),
\end{align}
where $\hat{\bm F}_{\bm s \bm s'} (\mathbfcal{R})$ denotes, as in the semiclassical case, the atomic smearing functions.
%Whereas the set of fields differ from to the fields investigated in Sec.~\ref{DIMREDQF} only by their different coordinate frame, the smearing functions (cf. compare Eq.~\eqref{SMEARNORMAL}) additionally become operator valued when transforming to the COM picture.
Integrating out the transverse COM degrees of freedom $\mathbfcal {Y}$ gives the collection of dimensionally reduced transition matrix elements 
\begin{align}
\matrixel{\bm s}{\hat{H}_{\mathrm{COM}}^\mathrm{I}} {\bm s '}  \! &= \! - \chi(t) e  \! \sum_{\bm j} \!
\int_S \!\! \mathrm{d} \mathcal{Z} \, 
\hat{\bm F}_{\bm m,\bm s\bm s'} (\mathcal{Z}) \!\cdot \!\hat{\mathbfcal{E}}^{(+)}_{\bm j} (\mathcal{Z},t)\nonumber\\
&\quad+ \mathrm{H.c.}, 
\end{align}

where the dimensionally reduced spatial smearing function, obtained by a mapping onto the transversal ancilla mode associated with the $\bm m$-th subfield (as introduced in Eq.~\eqref{KNMSS}), reads
\begin{align}
\begin{split}
\hat{\bm F}_{\bm m,\bm s \bm s'} (\mathcal{Z})  &= 
\int_{\Gamma} \dd^2 \mathcal{Y}\,   \mathbfcal{S}_{\bm m}(\mathbfcal{Y})  \hat{\bm F}_{\bm s \bm s'} (\mathbfcal{R}) \\
&= 
\int_{\Gamma} \dd^2 \mathcal{Y}\,   \mathbfcal{S}_{\bm m}(\mathbfcal{Y})   \hat{\bm r} \psi^\ast_{\bm s} (\mathbfcal{R}) \psi_{\bm s'} (\mathbfcal{R}). 
\end{split}
\label{KNMSSRELCORD}
\end{align}

\begin{widetext}

\section{Determination of the Transition Probabilities for the Subfield Truncation}

\subsection{Gaussian Wave Packet in a Cylindrical Cavity}
\setcounter{equation}{0}
\renewcommand{\theequation}{G.1.\arabic{equation}}
\label{GAUSSIANCYLCAVAPP}

Recall that the transition probabilities~\eqref{PAMPL} for a two-level system decompose into the transition amplitudes 
\begin{align}
|c_{\bm m, (\pm)}|^2 
= \sum_{l,\mu}   \frac{\omega_{\bm j,\mu} e^2}{2 \varepsilon_0 \hbar}  \left| \int_\mathbb{R} \dd t  \chi(t) \e^{\ii(\omega_{\bm j,\mu} \pm \Omega_{\mathrm{A}})t}\int_S \dd z \,  \bm u_{\bm j,\mu} (z) \cdot  \Smear(\ze(z)) \right|^2.
\label{CCCCAPP}
\end{align}
Performing the time integration yields (using the detuning $\Delta_{\bm j,\mu,(\pm)} = ( \omega_{\bm j,\mu} \pm \Omega_\mathrm{A})/2$) 
\begin{align}
f_{\bm j,\mu,(\pm)} (T) = \int_{\mathbb{R}} \dd t  \chi (t) \e^{\ii(\omega_{\bm j,\mu} \pm \Omega_\mathrm{A})t} =    
\begin{cases}
T \mathrm{sinc} ( \Delta_{\bm j,\mu,(\pm)} T ),\hspace{.5cm} \text{if}\, \chi(t) = \chi^{\mathrm{TS}}(t),\\
\sqrt{2\pi }T \e^{- \left(\Delta_{\bm j,\mu,(\pm)} T\right)^2}, \,\,\,\, \text{if}\, \chi(t) = \chi^{\mathrm{GS}}(t).\\
\end{cases}
\label{TIMEINT}
\end{align}
For clarity, we separately consider the terms (corresponding to the basis vectors of $\mathcal{C}$) of the inner product for the spatial integration.
While the azimuth integration in $\Smear (z_\mathrm{e}(z))$  fixes the azimuth quantum number $m_2$ to zero, the radial integration of the radial component in~\eqref{CCCCAPP} gives by the use of integration by parts (cf. App.~\ref{SecB} and App.~\ref{SecE1} to obtain an explicit expression for the field modes)
\begin{align}
\begin{split}
&\hspace{0.4cm} \int_0^L \dd  z \, u_{\bm j,\mu_2}^{(r)}(z) F_{\bm m}^{(r)} (z - L/2) \\
&= -\frac{2 c_{\bm m,\mu_2}k_l}{ \sigma^4 |\kk_{\bm j,\mu_2}||\bm k_{\bm m,\mu_2}| \sqrt{\pi^3L}}  \int_0^L \int_0^R \int_0^{2 \pi} \dd z \dd r \dd \varphi  \,\,   r^2 \e^{-r^2/\sigma^2}  \partial_\mathrm{r} J_{m_2} \left( |\bm k_{\bm m,\mu_2}|r\right) \e^{i m_2 \varphi}\e^{- \frac{(z-L/2)^2}{\sigma^2}}(z - L/2) \sin(k_l z) \\
%&= \frac{ 8  c_{n0,\mu_2}k_l}{\sigma^4 | \kk_{(m_1,0),l,\mu_2}||\bm  \kk_{(m_1,0),\mu_2}|\sqrt{\pi L}} \int_0^L  \int_0^R \dd z \dd r \,\,   \left(r - \frac{r^3}{\sigma^2}\right) \e^{-r^2/\sigma^2}  J_0 \left( |\bm  \kk_{(m_1,0),\mu_2}|r\right) \e^{- \frac{(z-L/2)^2}{\sigma^2}} (z - L/2) \sin(k_l z)\\
%&=  \frac{4 c_{n0,\mu_2}k_l}{\sqrt{2 \pi}\sigma^2 | \kk_{(m_1,0),l,\mu_2}|} \sqrt{\frac{2}{L}} \int_0^L  \int_0^{\sqrt{2}R/\sigma} \dd z \dd u \,\, \left(u - \frac{u^3}{2}\right) \e^{-u^2 / 2} J_0\left(\frac{\chi_{n0}\sigma}{\sqrt{2}R} u\right) (z - L/2) \sin(k_l z)\\
&\approx - \frac{4 c_{(m_1,0),\mu_2}k_l}{\sigma^2 |\kk_{(m_1,0),l,\mu_2}||\bm k_{(m_1,0),\mu_2}|\sqrt{\pi L}}  \int_0^L \int_0^\infty \dd z \dd v \,\, \left( \frac{v^3}{2}-v\right) \e^{-v^2 / 2} J_0\left(\frac{|\bm k_{(m_1,0),\mu_2}|\sigma}{\sqrt{2}} v\right) \e^{- \frac{(z-L/2)^2}{\sigma^2}}(z - L/2) \sin (k_l z),
\end{split}
\label{FIRSTINTEGRAL}
%Stimmt mit MA überein
\end{align}
where we used in the last step the substitution $v = \sqrt{2} r/\sigma$. 
Note that the wave functions  are real-valued for the transitions considered here with \mbox{$\bm F_{\bm m,(+)} (z) = \bm F_{\bm m,(-)} (z)$} such that we dropped the transition-process indices.
Since the dominating Gaussian  decays sufficiently fast for $v > \sqrt{2}R/\sigma$, we  expanded the upper limit of the $r$-integration in Eq.~\eqref{FIRSTINTEGRAL} to infinity. The integral over $[0, \infty)$ here is solved analytically by Hankel transformations (\cite{BATE}, Sec. 8.2, Eq.~(21)) which results in a sum of hypergeometric functions ${}_1F_1$, reading 
\begin{align}
\begin{split}
&\hspace{0.4cm} \int_0^L \dd z \, u_{\bm j,\mu_2}^{(r)}(z) F_{\bm m}^{(r)} (z - L/2) \\
&= - \frac{4 c_{(m_1,0),\mu_2}k_l}{\sigma^2 |\kk_{(m_1,0),l,\mu_2}||\bm k_{(m_1,0),\mu_2}|\sqrt{\pi L}}\left( {}_1 F_1 \left[2; \, 1; \, - \left( \frac{|\bm k_{(m_1,0),\mu_2} | \sigma}{2}\right)^2\right] - \,{}_1 F_1 \left[1; \, 1; \, -\left( \frac{|\bm k_{(m_1,0),\mu_2} | \sigma}{2}\right)^2\right]   \right) \\ 
&\quad \times \int_0^L  \dd z \e^{- \frac{(z-L/2)^2}{\sigma^2}}(z - L/2) \sin(k_l z)  \\
%&=  \frac{4 c_{n0,\mu_2}k_l}{\sqrt{2 \pi}\sigma^2 | \kk_{(m_1,0),l,\mu_2}||\bm  \kk_{(m_1,0),\mu_2}|} \frac{\chi_{n0}^2 \sigma^2}{4 R^2}  \e^{ - \frac{\chi_{n0}^2 \sigma^2}{4 R^2}}\e^{- \frac{(z-L/2)^2}{\sigma^2}} (z - L/2)\\
&=  \frac{ k_l |\bm  \kk_{(m_1,0),\mu_2}|}{|J_1(\chi_{n0})  \kk_{(m_1,0),l,\mu_2}| \pi R \sqrt{L}}   \e^{ - \left( \frac{|\bm k_{(m_1,0),\mu_2} | \sigma}{2}\right)^2}  \int_0^L \dd z  \e^{- \frac{(z-L/2)^2}{\sigma^2}} (z - L/2) \sin (k_l z ).
\end{split}
%Stimmt mit MA überein
\label{HINTR}
\end{align}
For the $z$-integration we also use integration by parts and expand the limits of the integral to infinity
\begin{align}
 \int_0^L \!\dd z  u_{\bm j,\mu_2}^{(r)}(z) F_{\bm m}^{(r)} (z - L/2) 
%&= - \frac{\sigma^2 |\bm  \kk_{(m_1,0),\mu_2}| k_l}{2 |J_1(\chi_{n0})  \kk_{(m_1,0),l,\mu_2} | \pi R \sqrt{L}}  \e^{ - \left( \frac{|\bm k_{(m_1,0),\mu_2} | \sigma}{2}\right)^2}  \int_{0}^{L} \dd z\, \sin ( k_l z) \partial_{z} \e^{- \frac{(z-L/2)^2}{\sigma^2}} \nonumber \\
&\approx\frac{\sigma^2 |\bm  \kk_{(m_1,0),\mu_2}| k_l^2}{2 |J_1(\chi_{n0})  \kk_{(m_1,0),l,\mu_2}| \pi R \sqrt{L}}    \e^{ - \left( \frac{|\bm k_{(m_1,0),\mu_2} | \sigma}{2}\right)^2} \int_{-\infty}^{\infty} \dd z    \e^{-(z-L/2)^2/\sigma^2} \cos (k_l z).
%&= - \frac{F_{n0,\mu_2}^{(r)}}{| \kk_{(m_1,0),l,\mu_2}|} \sqrt{\frac{\pi^5  \sigma^4}{2 L^5}} l^2 \cos \left( \frac{\pi l}{2} \right) \e^{-\frac{\pi^2 l^2 \sigma^2 }{4 L^2}},
%&= (-1)^{l+2} (2l+1) \frac{F_{n0,\mu_2}^{(r)}  \pi}{| \kk_{(m_1,0),l,\mu_2}|} \sqrt{\frac{2}{L^3}} \e^{-  \frac{(2l+1)^2\pi^2 \sigma^2 }{4 L^2} }, 
\label{FUR1}
\end{align}
For odd $l$, the argument in the integral becomes an odd function and vanishes. For even $l$, the longitudinal component gives (cf. ~\cite{BATE}, Sec. 1.3. Eq.~(11))
\begin{align}
\int_0^L \dd z \, u_{\bm j,\mu_2}^{(r)}(z) F_{\bm m}^{(r)} (z - L/2) 
\approx\frac{\sigma^3 |\bm  \kk_{(m_1,0),\mu_2}| k_l^2}{2|J_1(\chi_{n0})  \kk_{(m_1,0),l,\mu_2}| R \sqrt{\pi L} } \e^{-\frac{\ii \pi l}{2}}  \e^{ - \left( \frac{|\bm  \kk_{(m_1,0),\mu_2}|\sigma}{2}\right)^2} \e^{-\left( \frac{k_l \sigma }{2} \right)^2}.
\label{FUR2}
\end{align}
Since the atom is located in the center of the cylinder, the component in the Hamiltonian~\eqref{HDEE} emerging from the $\bm e^{(\varphi)}$-component of the scalar product of the reduced smearing function and the longitudinal-mode vector vanishes due to the rotational invariance. Finally for the $z$-component, we get by performing the azimuthal integration -- which again fixes the azimuthal mode number $m_2$ to zero -- and extending the upper limit of the radial integration to infinite 
\begin{align}
\begin{split}
\int_0^L \!\! \dd z \, u_{\bm j,\mu_2}^{(z)}(z) F_{\bm m}^{(z)} (z - L/2) %\\
&\approx  \!\!  \frac{ 4 c_{(m_1,0),\mu_2}|\bm k_{(m_1,0),\mu_2}|}{\sigma^4 |\kk_{(m_1,0),l,\mu_2}|\sqrt{\pi L} } \!\!  \int_0^L \!\! \!\!  \int_0^\infty  \!\! \!\!  \dd z \dd r  \,\,   r  \e^{-r^2/\sigma^2} \!\! \!\!  J_0 \left( |\bm  \kk_{(m_1,0),\mu_2}| r\right) \e^{- \frac{(z-L/2)^2}{\sigma^2}} (z - L/2)^2 \cos(k_l z). 
\end{split}
\label{HINTZ1}
\end{align}
The radial integration can also be solved by Hankel transformations (cf.~\cite{THEGODFATHER} Sec. 8.6. Eq.~(10)):
\begin{align}
\int_0^L \dd z \, u_{\bm j,\mu_2}^{(z)}(z) F_{\bm m}^{(z)} (z - L/2) 
\approx  
\frac{ 2 |\bm  \kk_{(m_1,0),\mu_2}|}{\sigma^2 | J_1 (\chi_{n0})  \kk_{(m_1,0),l,\mu_2}|\pi R \sqrt{L}} \e^{ - \left( \frac{|\bm  \kk_{(m_1,0),\mu_2}|\sigma}{2}\right)^2} \int_0^L  \dd z  \,\,   \e^{- \frac{(z-L/2)^2}{\sigma^2}} (z - L/2)^2 \cos(k_l z). 
\label{HINTZ2}
\end{align}
For the cosine term we thus have $\cos( k_l z + \pi l /2)$. Then, the integral is non-zero only for even $l$ (odd $l$ results in an odd function)
\begin{align}
\begin{split}
\hspace{.7cm}\int_0^L \dd z \, u_{\bm j,\mu_2}^{(z)}(z) F_{\bm m}^{(z)} (z - L/2) 
\approx  \frac{ |\bm  \kk_{(m_1,0),\mu_2}| \sigma }{2 | J_1 (\chi_{n0})  \kk_{(m_1,0),l,\mu_2}| R \sqrt{\pi L}} \left(1 - \frac{k_l^2 \sigma^2}{2} \right) \e^{-\frac{\ii \pi l}{2}} \e^{ - \left( \frac{|\bm  \kk_{(m_1,0),\mu_2}|\sigma}{2}\right)^2}  \e^{-\left(\frac{k_l \sigma}{2 }\right)^2}, 
\end{split}
\label{FUZEND}
\end{align}
where we used in the last step Eq.~(14) in Sec. 1.4 of~\cite{THEGODFATHER}.
Combining Eq.~\eqref{FUR2} and Eq.~\eqref{FUZEND} yields for the total overlap between longitudinal modes and smearing functions
\begin{align}
\hspace{.7cm}\int_0^L \dd z \, \bm u_{\bm j,\mu_2} (z) \cdot \bm F_{\bm m} (z - L/2) 
&\approx  \frac{ |\bm  \kk_{(m_1,0),\mu_2}| \sigma}{| J_1 (\chi_{n0})  \kk_{(m_1,0),l,\mu_2}| R \sqrt{\pi L}}  \e^{-\frac{\ii \pi l}{2}}\e^{ - \left( \frac{|\bm  \kk_{(m_1,0),\mu_2}|\sigma}{2}\right)^2} \e^{-\left(\frac{k_l \sigma}{2 }\right)^2}.
\label{LASTEQ}
\end{align}
Combining this with Eq.~\eqref{TIMEINT} yields the transition probabilities in Eq.~\eqref{PREDTH}.

\subsection{Determining the Maximum Subfield Probability}
\setcounter{equation}{0}
\renewcommand{\theequation}{G.2.\arabic{equation}}
\label{Asymptotics}

To find the transverse-mode number $m_1^\text{max}$  for the example of Sec.~\ref{NUMEST} which corresponds to the subfield with maximum transition probability \mbox{$|c_{m_1^\text{max}, 0, (\pm)}|^2$} of Eq.~\eqref{NPROBRED}, we consider
\begin{align}
f_{m_1,(\pm)} = \frac{|c_{(m_1,0)+(1,0), (\pm)}|^2 - |c_{(m_1,0), (\pm)}|^2}{\chi_{m_1+1} - \chi_{m_1}} \rightarrow 0,
\label{CONDASYMPTOTIC}
\end{align}
Since Eq.~\eqref{CONDASYMPTOTIC} contains an infinite sum over the longitudinal-mode numbers $l$, it is useful employ the Euler-Maclaurin formula (cf.~\cite{Abramowitz} Eq.~(23.1.30)) which yields in the case of Gaussian switching 
\begin{align}
 |c_{(m_1,0), (\pm)}|^2 =  &\int_{\mathbb{R}^+} \!\!\dd l  \,\, \frac{\chi_{m_1}^2  \exp \left(-\frac{\chi_{m_1}^2 \sigma^2 }{2 R^2} -\frac{2 \pi^2 l^2 \sigma^2  }{ L^2} -  2 (\Delta_{(m_1, 0), 2l,(\pm)} T  )^2\right) }{\omega_{(m_1,0),2l} \,\,J_1^2(\chi_{m_1})} + \frac{\chi_{m_1}  \exp \left( -\frac{\chi_{m_1}^2 \sigma^2 }{2 R^2} -  \frac{(\chi_{m_1} \pm \Omega_\mathrm{A})^2}{2}T^2 \right) }{2 J_1^2(\chi_{m_1})}\nonumber\\   
  &+ \mathcal{O}\left(\frac{R^2}{L^2}\right),
\end{align}
where we neglect higher-order terms due to $R/L \ll 1$.
The integral allows for an analytical solution in the range of $T \approx \Omega_\mathrm{A}^{-1}$ for spontaneous emission (for vacuum excitation a larger range of interaction times is allowed)
in terms of the modified Bessel function of the second kind in zeroth order $K_0$ reading 
\begin{align}
 |c_{(m_1,0), (\pm)}|^2 \approxprop \, \frac{\chi_{m_1}^2}{2 J_1^2(\chi_{m_1})} \left( \frac{L}{\pi R} K_0 (q \chi_{m_1}^2) + \frac{\exp (- q\chi_{m_1}^2)}{\chi_{m_1}}  \right)\exp \left( - q \chi_{m_1}^2  \pm \frac{\tau^2}{\widetilde{\Omega}_\mathrm{A}}\chi_{m_1} - \frac{\tau^2}{2}  \right), 
 \label{ASYMPTOT}
\end{align}
with the dimensionless parameters 
\begin{align}
q = \left( \frac{\sigma}{2 R}  \right)^2 + \left( \frac{\tau}{2 \widetilde{\Omega}_\mathrm{A}}  \right)^2 = \left( \frac{\sigma}{2 R}  \right)^2 + \left( \frac{cT}{2 R}  \right)^2 , \quad \tau = \Omega_\mathrm{A} T, \quad \widetilde{\Omega}_\mathrm{A} = \sqrt{ \chi_{m_1^\mathrm{res}}^2 + \left(\frac{2 \pi l^{\mathrm{res}} R}{L} \right)^2}. 
\label{CPM}
\end{align}
Equation~\eqref{CONDASYMPTOTIC} becomes then 
\begin{align}
f_{m_1,(\pm)} \approxprop \frac{\frac{\chi_{m_1+1}}{J_1^2(\chi_{m_1 +1})}\exp \left( - q \chi_{m_1 +1}^2  \pm \frac{\tau^2}{\widetilde{\Omega}_\mathrm{A}}\chi_{m_1 + 1} - \frac{\tau^2}{2}  \right) -  \frac{\chi_{m_1}}{J_1^2(\chi_{m_1})} \exp \left( - q \chi_{m_1}^2  \pm \frac{\tau^2}{\widetilde{\Omega}_\mathrm{A}}\chi_{m_1} - \frac{\tau^2}{2}  \right) }{\chi_{m_1+1} - \chi_{m_1}} \rightarrow 0,
\end{align}
where we approximated the Bessel function to first order as \mbox{$K_0 \approx \sqrt{\pi / (2 q \chi_{m_1})^2} \exp ( - q \chi_{m_1}^2)$} for $q \chi_{m_1}^2 > \pi$ (cf.~\cite{Abramowitz}, Eq.~(9.7.2)), corresponding to the parameter regime of Fig.~\eqref{Ratios}.
In this regime, the zeros of the zeroth order Bessel function can be approximated by $\chi_{m_1} \approx \pi (m_1 -1/4)$ and therefore $J_1^2(\chi_{m_1}) \approx 2/[\pi (m_1 -1/4)] $ (cf.~\cite{Abramowitz}, Eq.~(9.5.12) and Eq.~(9.2.1), respectively). Finally, solving for $m_1$ and approximating the logarithm for large arguments to first order yields a quadratic equation which is solved by 
\begin{align}
m_{1,(\pm)}^\mathrm{max} \approx \frac{4}{2 \pi^2 q \pm \frac{\pi \tau^2}{\widetilde{\Omega}_\mathrm{A}} + \sqrt{32 \pi^2 q  + \left(2 \pi^2 q \pm \frac{\pi \tau^2}{\widetilde{\Omega}_\mathrm{A}} \right)^2 } }. 
\end{align}
For large ratios $R/\sigma$ (optical resonator limit) in the case of spontaneous emission we have $q \approx [\tau/(2 \widetilde{\Omega}_\mathrm{A})]^2$. Additionally, for the resonance frequencies considered in Fig.~\ref{Ratios}, we can use $q \ll  \pi \tau^2\widetilde{\Omega}_\mathrm{A}$, and therefore for spontaneous emission
\begin{align}
m_{1,(-)}^\mathrm{max}  \approx \frac{2 \widetilde{\Omega}_\mathrm{A}}{\pi \tau^2},   
\end{align}
corresponding to Eq.~\eqref{LIMITI} for the ratios $l_{\mathrm{res}}R/L$  considered in the plots.

\subsection{Gaussian Wave Packet in a Laser Beam}
\setcounter{equation}{0}
\renewcommand{\theequation}{G.3.\arabic{equation}}
\label{LASERSMEARAPP}

For fields polarized in $\bm \epsilon_x$, we have only a nonvanishing overlap with the $x$-component of the smearing function. Therefore, we obtain (for the Gaussian states defined in~\eqref{GAUSSIANWAVEFCTN})
\begin{align}
F_{\bm m}^{(x)} (z) %= \iint_{\mathbb{R}^2} \ddx \ddy \, x^2  \varphi_{\bm m}(x,y) \psi_{000}(\bm r)   \psi_{001}(\bm r) \bm \epsilon_x  
= \sqrt{\frac{2}{ \pi^7 \sigma^8 2^{m_1+m_2} m_1! m_2! w_0^2}} \iint_{\mathbb{R}^2} \dd x \dd y \, x z  \, \mathrm{H}_{m_1}  \left( \frac{\sqrt{2}x}{w_0} \right) \mathrm{H}_{m_2}  \left( \frac{\sqrt{2}y}{w_0} \right)\e^{- \frac{x^2+y^2}{w^2_0}} \e^{- \frac{x^2+y^2 + z^2}{\sigma_{\vphantom{0}}^2}}. 
\label{FRR1}
\end{align}
To perform the integration we define the constant $a = w_0^{-2} + \sigma^{-2}$.
%\end{align}
For the integration of~\eqref{FRR1} in $x$ one has an odd function and thus a vanishing integration for even $m_1$. For odd $m_1$, the integral can be solved in terms of the \textit{probabilist's} Hermite polynomials, $\mathrm{He}_{m_1}(x) = 2^{-m_1/2} \mathrm{H}_{m_1}(x/\sqrt{2})$, as
\begin{align}
\int_{\mathbb{R}} \dd x  x   \mathrm{H}_{2 m_1+1} \left( \frac{\sqrt{2}x}{w_0}\right) \!\e^{-ax^2} \!= 
 2^{(2m_1+3)/2} \!\int_{\mathbb{R}^+} \dd x  x  \mathrm{He}_{2 m_1+1} \left( \frac{2x}{w_0}\right) \e^{-ax^2} \!= \frac{w_0^2}{4} \sqrt{\frac{\pi}{2}} \frac{(2m_1+1)!}{m_1!} \frac{\left(\frac{1}{2} - \frac{w_0^2 a}{4}\right)^{m_1}}{\left(\frac{w_0^2 a}{4}\right)^{m_1+3/2}}.
\end{align}
For solving the integral we used an integral transformation (\cite{THEGODFATHER}, p.~172, Eq.~(12)).
For the integration in $y$ one has a vanishing integral for odd $m_2$. Using the same approach as above one obtains
\begin{align}
\int_{\mathbb{R}} \dd y \, \mathrm{H}_{2 m_2} \left( \frac{\sqrt{2}y}{w_0}\right) \e^{-ay^2}=
2^{m_2+1}\int_{\mathbb{R}^+} \dd y \, \mathrm{He}_{2m_2} \left( \frac{2y}{w_0}\right) \e^{-ay^2} = \frac{w_0}{2} \sqrt{\pi} \frac{(2m_2)!}{m_2!} 
\frac{\left(\frac{1}{2} - \frac{w_0^2 a}{4}\right)^{m_2}}{\left(\frac{w_0^2 a}{4} \right)^{m_2 + 1/2}}.
\end{align}
With these results we find for the reduced smearing function~\eqref{FRR1}:
\begin{align}
F_{\bm m}^{(x)} (z) 
=&   \sqrt{\frac{(2m_1+1)! (2m_2)!}{ \pi^5 4^{m_1+m_2-1} (m_1!m_2!)^2}} \frac{w_0^2}{\sigma^4}  \left(\frac{\sigma^2}{w_0^2 + \sigma^2} \right)^2 \left( \frac{\sigma^2 - w_0^2}{\sigma^2 + w_0^2} \right)^{m_1 + m_2}  z \e^{- \frac{ z^2}{\sigma_{\vphantom{0}}^2}}. \nonumber\\
\approx&  (-1)^{m_1+m_2} \sqrt{\frac{(2m_1+1)! (2m_2)!}{\pi^5 4^{m_1+m_2-1} (m_1!m_2!)^2}}   \frac{z}{w_0^2}  \e^{- \frac{ z^2}{\sigma_{\vphantom{0}}^2}}, 
\label{FRR2}
\end{align}
where we used Eq.~\eqref{LARGEBEAMWAISTAPPROXIMATION} to approximate $
\sigma^2 + w_0^2 \approx w_0^2$.
% This combined with condition~\eqref{APPRLAME} allows to write the $x$-component of the reduced smearing function~\eqref{FRR2} in a more compact form, reading
% \begin{align}
% F_{\bm m}^{(x)} (z) 
% \approx  (-1)^{m_1+m_2} \sqrt{\frac{(2m_1+1)! (2m_2)!}{\pi^5 4^{m_1+m_2-1} (m_1!m_2!)^2}}   \frac{z}{w_0^2}  \e^{- \frac{ z^2}{\sigma_{\vphantom{0}}^2}}. 
% \label{FRR22}
% \end{align}
One can now calculate the longitudinal overlap of the lower-dimensional smearing function with the 1D modes for a beam polarized in $\bm \epsilon_x$:
\begin{equation}
\int_\mathbb{R} \dd z \, \bm F_{\bm m} (z) \cdot
\begin{Bmatrix}
\bm u_\mu^\ast (z) \\
\bm u_\mu (z)
\end{Bmatrix}\Bigg|_{\mu = \bm \epsilon_x}
\approx  (-1)^{m_1+m_2} \sqrt{\frac{(2m_1+1)! (2m_2)!}{w_0^4 \pi^5 4^{m_1+m_2-1} (m_1!m_2!)^2}} \int_\mathbb{R} \dd z \, z  \e^{- \frac{ z^2}{\sigma_{\vphantom{0}}^2}}  \begin{Bmatrix}
\e^{-\ii k z}\\
\e^{\ii k z}
\end{Bmatrix}. 
\label{FRR3}
\end{equation}
Completing the square we obtain for the integral
\begin{align}
\int_{\mathbb{R}} \dd z \, z  \e^{- \frac{ z^2}{\sigma_{\vphantom{0}}^2}}  \e^{\pm \ii k z} = - \frac{\sigma^2}{2} \e^{-\frac{k^2\sigma^2}{4}} \left[    \int_{\mathbb{R}} \dd z \, \partial_z    \e^{- \sigma^{-2} ( z \mp \ii k \sigma^2 /2)^2} \mp\ii  k  \int_{\mathbb{R}} \dd z \,  \e^{- \sigma^{-2} ( z \mp \ii k \sigma^2 /2 )^2 }    \right]
= \pm i \frac{k \sigma^3 \sqrt{\pi}}{2} \e^{-\frac{k^2\sigma^2}{4}},
\label{COMPLETINGTHESQUARE}
\end{align}
where we used in the last step that the derivative of a Gaussian gives a odd function in the integral and thus vanishes when integrated over $\mathbb{R}$. 
To find the transition amplitudes defined in Eq.~\eqref{ISTHATIT} we have to consider two different possible final states: $
\ket{\kappa_{\bm \nu} (k) } = \ket{ \alpha(k) }_{\bm \nu} \otimes\ket{ \Omega  }_{\neg \bm \nu},
$
with $|\Omega \rangle_{\neg \bm \nu}$ being the vacuum for the complement modes, and 
$
|\theta_{\bm m,\mu}(k) \rangle =
|\alpha(k) \rangle_{\bm \nu} \otimes |1 (k) \rangle_{\bm m,\mu} \otimes |\Omega (k) \rangle_{\neg \bm \nu \neg (\bm m,\mu)}
$ for $(\bm m,\mu) \neq \bm \nu$.
To find the transition amplitudes we need to evaluate the following matrix elements for the annihilation and creation operators:
\begin{subequations}
\begin{align}
&\langle \kappa_{\bm \nu} (k') | \hat{a}^{\vphantom{\dagger}}_{\bm m,\mu} (k) |\kappa_{\bm \nu} (k') \rangle = \delta_{(\bm m,\mu),\bm \nu}  \, \delta(k-k') \alpha(k'),\quad
\langle \kappa_{\bm \nu} (k') | \hat{a}_{\bm m,\mu}^\dagger (k) |\kappa_{ \bm \nu} (k') \rangle = \delta_{(\bm m,\mu),\bm \nu} \delta(k - k') \bar{\alpha} (k'),
\intertext{
whereas for the other final state $|\theta_{\bm m,\mu}(k) \rangle$:
}
&\langle \theta_{\bm m,\mu} (k') | \hat{a}^{\vphantom{\dagger}}_{\bm m,\mu} (k) |\kappa_{\bm \nu} (k') \rangle = 0,\hspace{3.4cm}
\langle \theta_{\bm m, \mu} (k') | \hat{a}_{\bm m,\mu}^\dagger (k) |\kappa_{ \bm \nu} (k') \rangle = (1 -  \delta_{(\bm m,\mu),\bm \nu} ) \delta (k - k').
\label{OPIDENT}
\end{align}
\end{subequations}
Thus, we have for the pumped mode in the case of stimulated emission
\begin{align}
|c_{\bm \nu, g \rightarrow e} (t) |^2 
%\approx &\left| \langle \kappa_{\bm \nu} (k); \,  e|  \left[  \openone - \frac{i}{\hbar}  \int_{\mathbb{R}} \dd t \,\, \hat{H}^\mathrm{I}(t)  \right]  | \kappa_{\bm \nu} (k); \,  g \rangle \right|^2,\\
\approx & \frac{ c e^2}{ \hbar  \varepsilon_0} \left| \langle \kappa_{\bm \nu} (k); \,  e|  \int_\mathbb{R} \dd t  \int_\mathbb{R} \dd k' \int_\mathbb{R} \dd z \, \sqrt{k'} \chi(t) \sum_{\bm m,\mu} \left(  \hat{a}^{\vphantom{\dagger}}_{\bm m, \mu} (k') \e^{- \ii c k' t} \bm u_{\mu} (k',z) -   \hat{a}^{\dagger}_{\bm m,\mu} (k') \e^{\ii c k' t} \bm u_{\mu}^\dagger (k',z)  \right) \right.\nonumber\\
&\times \left. \bm F_{\bm m,g \rightarrow e} (z)  \left( \dyad{e}{g} \e^{\ii \Omega_\mathrm{A} t} +  \dyad{g}{e} \e^{-\ii \Omega_\mathrm{A} t}   \right)| \kappa_{\bm \nu} (k); \,  g \rangle \right|^2\nonumber\\
\approx &\frac{3 c e^2 \sigma^6 }{4 \hbar  \varepsilon_0 \pi^4 w_0^4} \left|   \int_\mathbb{R} \dd t  \int_\mathbb{R} \dd k' \, \sqrt{{k'}^3} \e^{- \frac{{k'}^2}{2 \sigma^2}} \chi(t) \left(  \alpha_{\bm \nu} (k') \e^{- \ii c k' t} +    \bar{\alpha}_{\bm \nu} (k') \e^{\ii c k' t}   \right)   \e^{\ii \Omega_\mathrm{A} t} \delta(k' - k) \right|^2   \nonumber \\
= &\frac{3 c e^2 k^3 \sigma^6 }{\hbar  \varepsilon_0 \pi^4 w_0^4} \e^{- \frac{k^2 \sigma^2}{2}}  \left|  \int_\mathbb{R} \dd t\, \chi(t) \mathrm{Re} \left[  \alpha_{\bm \nu} (k) \e^{- \ii \omega t}   \right]   \e^{\ii \Omega_\mathrm{A} t}\right|^2.
\label{NPTP}
\end{align}
For the nonpumped modes, i.e., for $(\bm m,\mu) \neq \bm \nu$, it yields 
\begin{align}
|c_{\bm m, \mu , g \rightarrow e} (t) |^2 
%\approx &\left| \langle \theta_{\bm \nu} (k); \,  e|  \left[  \openone - \frac{i}{\hbar}  \int_{\mathbb{R}} \dd t \,\, \hat{H}^\mathrm{I}(t)  \right]  | \kappa_{\bm \nu} (k); \,  g \rangle \right|^2,\\
\approx & \frac{c e^2}{2 \hbar  \varepsilon_0} \left| \langle \theta_{\bm \nu} (k); \,  e|  \int_\mathbb{R} \dd t  \int_\mathbb{R} \dd k' \int_\mathbb{R} \dd z \, \sqrt{k'} \chi(t) \sum_{\bm m,\mu}\left(  \hat{a}^{\vphantom{\dagger}}_{\bm m, \mu} (k') \e^{- \ii c k' t} \bm u_{\mu} (k',z) -   \hat{a}^{\dagger}_{\bm m,\mu} (k') \e^{\ii c k' t} \bm u_{\mu}^\dagger (k',z)  \right) \right.\nonumber\\
&\times \bm F_{\bm m,g \rightarrow e} (z)  \left( \dyad{e}{g} \e^{\ii \Omega_\mathrm{A} t} +  \dyad{g}{e} \e^{-\ii \Omega_\mathrm{A} t}   \right)| \kappa_{\bm \nu} (k); \,  g \rangle \Bigg|^2\nonumber\\
\approx &\frac{c e^2 k^3 \sigma^6 }{8 \hbar  \varepsilon_0 \pi^4 w_0^4} \e^{- \frac{k^2 \sigma^2}{2}}\frac{ (2m_1+1)! (2m_2)!}{4^{m_1+m_2-1} (m_1!m_2!)^2}  \left|  \int_\mathbb{R} \dd t\, \chi(t)  \e^{- \ii (\omega  + \Omega_\mathrm{A})t}\right|^2,
\label{VMTA}
\end{align}
where we used the operator identities defined in Eq.~\eqref{OPIDENT} and the overlap from reduced smearing function and longitudinal modes (Eq.~\eqref{FRR3} and Eq.~\eqref{COMPLETINGTHESQUARE}). 
Recall that the transition probabilities~\eqref{ISTHATIT} are in the paraxial wave approximation to zeroth order, and by restriction of the number of vacuum modes to $\bm N$. Thus, by adding the vacuum transition amplitudes~\eqref{VMTA} to the laser transition amplitude~\eqref{NPTP}, two couplings can be identified. One mode-number independent coupling $g$, coupling laser and vacuum modes to the two-level system equally, and a mode-number dependent coupling $\gamma_{\bm N}$, yielding an additional modulation of the coupling to the set of vacuum modes:   
%Summing over all transition amplitudes~\eqref{VMTA} up to a tuple of mode numbers $\bm N$, defining a coupling constant $g$ and the function $\gamma_{\bm N}$, i.e.,
\begin{align}
\begin{split}
g =\frac{3 c e^2 k^3 \sigma^6 }{\hbar  \varepsilon_0 \pi^4 w_0^4} \e^{- \frac{k^2 \sigma^2}{2}}, 
%\gamma_{\bm N}=xxx
%p_{\bm m} = \frac{ (2m_1+1)! (2m_2)!}{ 2 \pi^5 4^{m_1+m_2} (m_1!m_2!)^2},
\gamma_{\bm N} =  \frac{1}{3} \sum_{\bm m \neq (1,0)}^{\bm N}  \frac{ (2m_1+1)! (2m_2)!}{ 4^{m_1+m_2+1/2} (m_1!m_2!)^2} = \frac{1}{3} \left(\frac{ 4\Gamma\left( \frac{5}{2} + N_1 \right)\Gamma\left( \frac{3}{2} + N_2 \right)  }{ 3\pi \Gamma\left( 1 + N_1 \right) \Gamma\left( 1 + N_2 \right)  }  - \frac{3}{4} \theta_1(N_1-1) \right),
\end{split}
\label{GCOUPLAPP}
\end{align}
where we assume here that the Heaviside function obeys (nonconventionally) $\theta_1(0)=1$. By means of Eq.~\eqref{TIMEINTSECL}
%\begin{align}
%f_{(\pm)} (T) = \int_\mathbb{R} \dd t \, \chi(t,T) \e^{- \ii (\omega \pm \Omega_{\mathrm{A}}) t},
%\label{FUNCP}
%\end{align}
%and the coupling constants~\eqref{GCOUPLAPP}
one arrives at expression for the transition probabilities Eq.~\eqref{ISTHATIT}). 
In particular, the time-integral contribution (Eq.~\eqref{NPTP}) of the laser mode to the transition probabilities reads 
\begin{align}
\begin{split}
&\hspace{.3cm}\left| \int_\mathbb{R} \dd t \left( f_{(-)} (t) + \bar{f}_{(+)} (t) \right) \right|^2\\ 
%&\hspace{.3cm}\left| \int_\mathbb{R} \chi(t) \left( \e^{\ii  2 \Delta_{(\pm)} t} + \e^{ - \ii  2 \Delta_{(\mp)} t}\right) \right|^2\\ 
&=\begin{cases}
\mathrm{sinc}^2 \left( \Delta_{(+)} T \right) + \mathrm{sinc}^2 \left( \Delta_{(-)} T\right)   +   \mathrm{sinc} \left( \Delta_{(+)} T \right) \mathrm{sinc} \left( \Delta_{(-)} T \right) \left[  2 \cos^2 \left( \frac{\omega}{2} T \right) +1 \right] 
\!, &\text{$\chi(t) = \chi^{\mathrm{TH}} (t)$,} \\[2pt]
8 \pi T^2 \exp \left( -  \frac{\omega^2 + \Omega_\mathrm{A}^2}{2} T^2 \right) \mathrm{cosh}^2 \left( \frac{\omega \Omega_{\mathrm{A}}}{2} T^2\right), &\text{$\chi(t) = \chi^{\mathrm{GS}} (t)$.}
\end{cases}
\end{split}
\label{TIMEDEPAMPLLASER}
\end{align}
%To investigate the ratio of the transition probability given only by the pumped compared and the correction by emerging from the vacuum mode to some maximum mode number $\bm N$ (cf. section~\ref{LASERSEC} for a detailed discussion) a measure of how strongly the vacuum modes influence the mode pumped by the laser has to be defined. Therefore, the transition amplitudes calculated in Eqs.~\eqref{NPTP} and Eq.~\eqref{VMTA} and the integration of the time domain in Eq.~\eqref{TIMEDEPAMPLLASER} yield

The measure $\zeta_{\bm N,(\pm)}$ quantifies the ratio of the contributions from the vacuum~\eqref{VMTA} versus the laser mode~\eqref{NPTP}:
\begin{align}
\,\,\zeta_{\bm N, (\pm)} &= \sum_{\bm m}^{\bm N}  \frac{|c_{\bm m,\epsilon_x, (\pm)}  |^2 }{|c_{\bm \nu, (\pm)}|^2 } 
=
\frac{\gamma_{\bm N} }{|\alpha (k)|^2}\!
\begin{cases}
 \frac{\mathrm{sinc}^2 \left( \Delta_{(\pm)} T \right) }{\mathrm{sinc}^2 \left( \Delta_{(+)} T \right) + \mathrm{sinc}^2 \left( \Delta_{(-)} T \right)   +   \mathrm{sinc} \left( \Delta_{(+)} T \right) \mathrm{sinc} \left( \Delta_{(-)} T \right) \left[  2 \cos^2 \left( \frac{\omega}{2} T \right) +1 \right]} 
\!, &\text{$\chi(t) = \chi^{\mathrm{TH}} (t)$,} 
\\[6pt]
\frac{1}{4}\exp \left(\pm  \omega \Omega_\mathrm{A} T^2 \right) \mathrm{sech}^2 \left( \frac{\omega \Omega_{\mathrm{A}}}{2} T^2\right), &\text{$\chi(t) = \chi^{\mathrm{GS}} (t)$.}
\end{cases}
\label{DELTALASERAPP}
\end{align}
Since all parameters of $\zeta_{\bm N,(\pm)}$ except the interaction time are assumed fixed, maximizing the right-hand side of this equation in interaction time $T$ gives the following upper bound:
\begin{align}
\zeta_{\bm N,(\pm)} \le \frac{\gamma_{\bm N} }{|\alpha (k)|^2},
\end{align}
which  only depends on the mean photon number of the laser $|\alpha(k)|^2$ and the range $\bm N$ of vacuum modes considered. In Fig.~\ref{Contourplot} we plot $\gamma_{\bm N}$, showing that for strong laser intensities the vacuum modes' contribution is highly suppressed. 

\begin{figure}[H]
\begin{center}
\includegraphics[width=5.9cm]{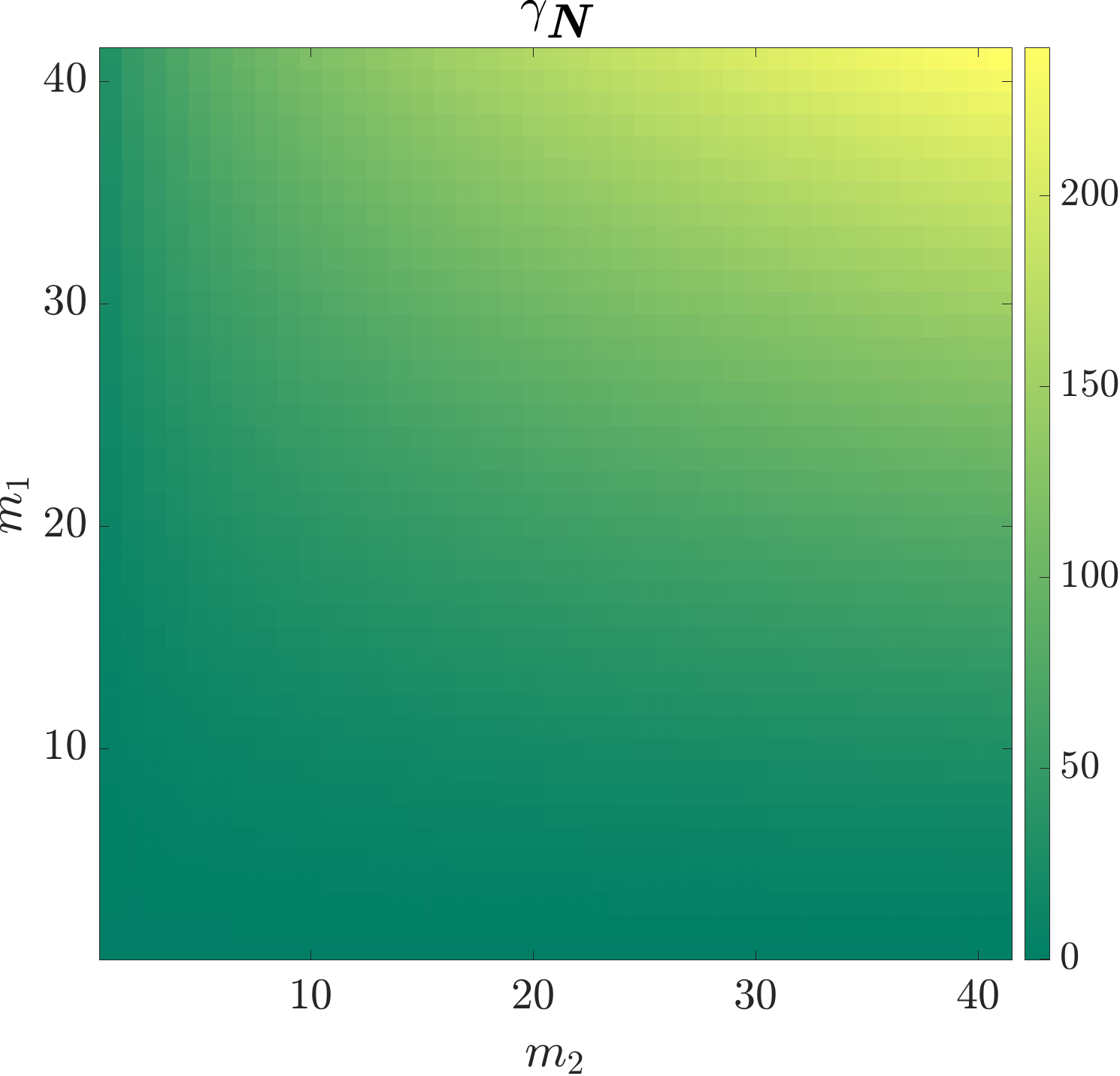}
\end{center}
\caption{The mode-number dependent coupling $\gamma_{\bm N}$ versus mode numbers $m_1$ and $m_2$.}
\label{Contourplot}
\end{figure}

\end{widetext}
\clearpage

\bibliography{Bib}

\end{document}